\documentclass[12]{article}

\usepackage{amsmath,amssymb}
\usepackage{bbm}
\usepackage{pstricks,pst-plot}
\usepackage{graphicx}
\usepackage{array}
\psset{unit=1cm,linewidth=1pt}
\newrgbcolor{paleyellow}{1 1 0.8}
\newrgbcolor{paleblue}{0.8 0.9 1}

\usepackage[left=1in,right=1in,top=1in,bottom=1.25in]{geometry}
\usepackage{setspace}
\usepackage[bookmarks=true]{hyperref}

\newcommand\tr{\mathop{\rm tr}\nolimits}
\newcommand\Tr{\mathop{\rm Tr}\nolimits}
\newcommand\Det{\mathop{\rm Det}\nolimits}
\newcommand\be{\begin{equation}}
\newcommand\ee{\end{equation}}

\newcommand\comment[1]{\langle\!\langle\,\mbox{#1}\,\rangle\!\rangle}
\newcommand\vev[1]{\left\langle#1\right\rangle}

\newcommand\ctanh{\mathop{{\rm ctanh}}\nolimits}

\def\d{{\rm d}}
\def\CF{\mathcal{F}}
\def\CA{\mathcal{A}}

\def\bp{\vec{\bf p}}
\def\bq{\vec{\bf q}}
\def\bn{\vec{\bf n}}
\def\bv{\vec{\bf v}}

\newcommand\Trule{\rule{0pt}{2.0ex}}
\newcommand\Brule{\rule[-1.2ex]{0pt}{0pt}}

\newcommand\eqalign[1]{%
	\vcenter{\normalbaselines\openup 1\jot\ialign{%
		\hfil$\displaystyle{##{}}$&
		$\displaystyle{{}##}$\hfil\cr
		#1\crcr
		}}}

\begin{document}

\thispagestyle{empty}

\begin{flushright}
UTTG-01-12 \\
ITEP-TH-61/11 \\
TAUP-2940-11
\end{flushright}

\vspace{10pt}
\begin{center}
{\Large \textbf{Baryonic Popcorn}}
\end{center}

\vspace{6pt}
\begin{center}
{\large{Vadim Kaplunovsky$^{a}$, Dmitry Melnikov$^{b,c}$ and Jacob Sonnenschein$^d$ }\\}
\vspace{25pt}
\textit{\small $^a$ Physics Theory Group, University of Texas,\\ 1 University Station, C1608, Austin, TX 78712, USA}\\ \vspace{6pt}
\textit{\small $^b$  International Institute of Physics, Federal University of Rio Grande do Norte, \\Av. Odilon Gomes de Lima 1722, Capim Macio, Natal-RN  59078-400, Brazil}\\ \vspace{6pt}
\textit{\small $^c$  Institute for Theoretical and Experimental Physics, \\B.~Cheremushkinskaya 25, Moscow 117218, Russia}\\ \vspace{6pt}
\textit{\small $^d$ The Raymond and Beverly Sackler School of Physics and Astronomy, \\ Tel Aviv University, Ramat Aviv 69978, Israel}\\ \vspace{6pt}
\end{center}

\begin{abstract}

In the large $N_c$ limit cold dense nuclear matter must be in a lattice phase. This applies also to holographic models of hadron physics. In a class of such models, like the generalized Sakai-Sugimoto model, baryons take the form of instantons of the effective flavor gauge theory that resides on probe flavor branes. In this paper we study the phase structure of baryonic crystals by analyzing discrete periodic configurations of such instantons. We find that instanton configurations exhibit a series of ``popcorn'' transitions upon increasing the density. Through these transitions normal (3D) lattices expand into the transverse dimension, eventually becoming a higher dimensional (4D) multi-layer lattice at large densities.

We consider 3D lattices of zero size instantons as well as 1D periodic chains of finite size instantons, which serve as toy models of the full holographic systems. In particular, for the finite-size case we determine solutions of the corresponding ADHM equations for both a straight chain and for a 2D zigzag configuration where instantons pop up into the holographic dimension. At low density the system takes the form of an ``abelian anti-ferromagnetic" straight periodic chain. Above a critical density there is a second order phase transition into a zigzag structure. An even higher density yields a rich phase space characterized by the formation of multi-layer zigzag structures. The finite size of the lattices in the transverse dimension is a signal of an emerging Fermi sea of quarks. We thus propose that the popcorn transitions indicate the onset of the ``quarkyonic'' phase of the cold dense nuclear matter.

\end{abstract}

\newpage

\tableofcontents

\section{Introduction}

The phase diagram of QCD at finite temperature and chemical potential has a very rich and non-trivial structure. It is believed to possess a number of interesting phases and transitions between them (for reviews see~\cite{Rajagopal:2000wf,Schafer:2005ff,Alford:2006wn,Stephanov:2007fk,McLerran:2007qj}.) At high temperature QCD must exist in a deconfined phase with restored chiral symmetry, while at small temperature and high density it is believed to be in a color superconducting phase. These two regimes attract a great deal of attention because they are ``accessible" experimentally in the heavy ion collisions and interior of compact stars.\footnote{Here the quotes stress our poor control and understanding of the experimental data.} Nevertheless the most challenging and perhaps interesting and rich physics resides outside these limits.

Despite decades of study of QCD and nuclear physics, the current theoretical  understanding of the QCD phase diagram is  still very limited. Due to strong coupling the direct approach to QCD is only possible via lattice methods. Strictly speaking the latter are only applicable in the zero density case, since finite density triggers the notorious sign problem. Fortunately some technologies were developed recently in the lattice community, which allowed to extrapolate the zero density results somewhat off the zero density regime, though not too far. It turns out, however, that  there is another loose end one can follow to untie the tangle: the large density case QCD is expected to become perturbative due to the presence of a large scale -- chemical potential. The analysis of this phase can then follow the one of weakly coupled BCS superconductors. Accidentally this two theoretical handles we have to study the phase space are applicable precisely in the accessible experimental regimes. Such a situation looks in fact unfortunate because it leaves us no efficient tools to understand the intermediate regime, for which we have to resort to all kinds of bottom-up or effective approaches.

In this work we will be interested in the properties of nuclear matter, which lies precisely in the intermediate part of the phase diagram. Therefore we have to choose an effective tool to study it. One of the relatively new effective and efficient tools was brought up by the ideas of holographic correspondence, or simply holography, which received a precise formulation with the advent of AdS/CFT~\cite{AdS/CFT}. Holography provided a new perspective for studying and understanding gauge theories in the regime of strong coupling. Among its many interesting developments it contains encapsulated toolkits to study theories which are cousins of QCD and hadron theories.

In this paper we have chosen to work with a generalization of Sakai-Sugimoto model~\cite{SakaiSugimoto2004}, which is so far the best-known holographic model of hadron physics. It is engineered from the following blocks. To construct the pure gauge sector one uses Witten's model of confining non-supersymmetric gauge theories~\cite{WittensModel}. Namely one works with a near-horizon limit of the theory on the D4-branes compactified on the circle in type IIA theory. The flavor sector is added following the ideas of Karch and Katz~\cite{KK}, by introducing additional D8-branes in the geometry of the D4. Notice that flavor branes are typically added assuming quenched approximation $N_f\ll N_c$, to neglect the backreaction on the background geometry. Sakai and Sugimoto were first to notice that D8 and anti-D8-branes smoothly connected in a U shape manner, model the phenomenon of chiral symmetry breaking in holography. Following a pioneering work by Kruczenski \emph{et al.}~\cite{Kruczenski:2003uq}, mesons can be analyzed as fluctuations of the fields in the effective low-energy theory of the flavor branes. Various achievements of holography applied to meson physics are reviewed in~\cite{Erdmenger:2007cm}. In turn baryons are introduced as other D4 branes wrapped on compact cycles according to the recipe suggested by Witten~\cite{WittenBaryons}. Since the work of Hata \emph{et al.}~\cite{Hata:2007mb} baryons in Sakai-Sugimoto model were studied in many works. See~\cite{Erdmenger:2007cm} and a more recent review~\cite{Kim:2011ey} for a (non-exhaustive) list of references.

The study of QCD phase diagram has recently drawn a revived attention mainly due to experiments on heavy-ion collisions and progress in theoretical understanding of QCD phases.  A number of papers have appeared that studied the phase diagram through holography including~\cite{Horigome:2006xu,Yamada:2007,Bergman,Rozali:2007rx,FiniteDensity}. In holography though for the sake of feasibility one is bound to study a specific limit of gauge theories, namely large number of colors $N_c$ and large 't Hooft coupling $\lambda$. In this limit the phase diagram gets important modifications. For the nuclear matter phase, which is of  main interest for us, this means two major changes. First, in the large $N_c$ limit the phase of  nuclear matter is a crystal, as opposed to ordinary nuclear matter. This happens because the potential energy of baryon interaction scales as $N_c$, while the kinetic energy is only $N_c^{-1}$. Second, in the large $\lambda$ limit baryon mass $m_B$ scales as $\lambda$, but baryon interactions only as $\lambda^0$. As a result baryon density grows very fast as a function of the chemical potential above the critical value $\mu_c\simeq m_B$. At densities such that baryons start to overlap, quarks no longer know which baryon they belong to and the matter essentially becomes a quark matter. Therefore crystalline nuclear matter occupies only a narrow region of the phase diagram plotted in units of $\mu_c$ and it disappears in the precise $\lambda\to\infty$ limit. In the case of Sakai-Sugimoto model one can find a $\lambda\to\infty$ diagram in the work of Bergman, Lifschytz and Lippert~\cite{Bergman}.

We are going to zoom on the phase diagram of Bergman \emph{et al.} around $\mu=\mu_c$ and $T=0$ in order to study crystals of holographic baryons. A general idea of construction of baryonic lattices in Sakai-Sugimoto model was already outlined in the work of Rho, Sin and Zahed~\cite{Rho:2009ym} in relation to skyrmion lattice, which we discuss below. We are going to put the ideas of Rho \emph{et al.} on a solid ground by working out several examples of lattices explicitly. In Sakai-Sugimoto model baryons can be identified with instantons of  $SU(N_f)$ gauge fields that live on the world volume of  flavor branes. Therefore a description of a baryon lattice requires a construction of the corresponding instanton solution of the gauge field equations of motion. Notice that these are curved space instantons, so the full solution for them is not known. Instead an approximation is derived employing first few orders in $1/\lambda$ expansion, following~\cite{Hata:2007mb}.

In the first order in $1/\lambda$ expansion the curvature can be neglected. The baryon is a flat space-time BPST instanton. Such a solution is characterized by moduli, such as  the instanton positions, radii and orientation. To compute interaction energy of instantons one has to go to the next order in perturbation theory. When interactions are turned on the moduli are partially stabilized. In particular the instanton acquires an equilibrium size. One can then quantize the resulting Hamiltonian to derive the baryon spectrum. Thus to find the ground state of a baryon system one has to find the relevant instanton solution, compute corrections to the BPS energy due to curvature and self-interactions on the BPS solution itself and minimize the net energy with respect to the moduli.

We will follow this general strategy to address the following question. The energy of a single baryon is minimized, when the baryon is sitting in the lowest point of the gravity potential (bottom of the U shape profile). When a finite density baryon configuration is considered the repulsive interaction between baryons should in principle be able to push them out from the bottom of the gravity well. In other words, suppose we consider a 3D lattice of instantons and squeeze it to an arbitrarily large density. Will the instantons stay bound to 3D or will they pop out from their 3D alignment?

There are several pieces of motivation for posing this question. First, this question appears to be natural from the point of view of holography. If the above expectations are confirmed then the next question is of course the interpretation of this result. On one hand the answer to these questions is partially known thanks to the work of Rozali, Shieh, Van Raamsdonk and Wu~\cite{Rozali:2007rx}. They have shown that a uniform distribution of instantons in 3D will necessarily incur a mesoscopic instanton density in holographic dimension as well. Finite instanton distribution must then be interpreted as a Fermi sea (of quarks, since the baryons in this picture are classical, neither fermions, nor bosons) and its edge as a Fermi surface. On the other hand the case of a uniform density is slightly different. A uniform density implicitly assumes the situation of overlapping baryons, when the quarks can no longer distinguish the baryons and start behaving in a qualitatively new way. Hence it will be interesting to understand physics in the non-uniform case as well. It turns out that the latter case has a rich structure.

Second, it is interesting to probe the holographic model of the nuclear matter phase against the predictions of other effective models. In particular, there are claims, most notably from the analysis of baryons lattices in Skyrme model~\cite{Skyrme,Klebanov}, that chiral symmetry restoring phase transition may take place in the nuclear matter phase at zero temperature and finite density~\cite{McLerran:2007qj}. In the skyrmion description this transition is also accompanied by the change of symmetry from skyrmion to half-skyrmion~\cite{Goldhaber:1987pb,KuglerShtrikman}. Since the Skyrme model is naturally embedded in holography, it is important to understand how those results are matched by holographic models and ultimately verify the statement about the chiral symmetry restoration (\emph{e.g.} the work of Rho \emph{et al.}~\cite{Rho:2009ym}). If the restoring transition is to take place at all this should definitely happen in the nuclear matter phase.

In this paper we will mostly address the first piece of the above motivation, while restricting to a summary of our expectations for the second. We consider two toy-models. In the first toy-model baryons are instantons in the limit of zero size, that is point charges in 5D. Since curvature of the fifth dimension binds the charges to the bottom of the flavor brane geometry, in the low density phase they form a 3D lattice. However if the high enough pressure is applied the 3D lattice becomes unstable and the baryons start occupying higher positions in the transverse fifth dimension. More precisely the lattice splits into several sub-lattices separated along the transverse dimension. The splitting continues as pressure is further increased.

In the second toy model we consider a 1D chain of baryons, described as finite size instantons in the generalized Sakai-Sugimoto model. To render the chain of repelling instantons stable in 3D we turn on curvature in all the transverse directions, including the spatial ones. The outcome of the analysis  is the following.  For large instanton spacing the periodic arrangement of instantons takes the form of a one dimensional straight chain with an abelian phase difference between neighboring instantons of $\phi=\pi$.   For a separation distance, below a critical value given in (\ref{Dcrit}) the straight chain becomes an unstable configuration and instead the chain takes the form of a zigzag structure (see figure~\ref{figZigzag}).  The zigzag amplitude $\epsilon$ as a function of $d$ is drawn in figure~\ref{SmallZigzag}.We further find that assuming an abelian phase between adjacent centers, the preferred value of the phase is $\phi=\pi$ for the spacing close to critical. We also consider the case of general non-abelian orientation angles and  find out that phases with abelian orientations are preferable. We show  that for small amplitude of the zigzag the neighboring instantons remain antiparallel as in the $\epsilon=0$ case. For larger zigzag amplitude, the relative orientation of the instantons changes from $\phi=\pi$ to $\phi\simeq 117^\circ$ in a first order phase transition. At even larger densities the  orientation changes smoothly to an asymptotical value $\pi/2$. For $\epsilon\gg d$ the instantons in each of the two layers of the zigzag become closer to each other than the original nearest neighbors and prefer to orient themselves in an antiferromagnetic way, $\phi=\pi$. Figure~\ref{figZigzagPD} summarizes the phase diagram of the chain up to 2 layers.

Notice that the approach used in this work is in fact independent of the concrete details of the holographic model. Although we start our discussion from Sakai-Sugimoto model and derive our effective action from it, the analysis is restricted only to the first non-trivial order in the $1/\lambda$ expansion. Up to this order all the dependence on a particular model can be absorbed in the definition of the coefficients that determine the characteristic scale, such as strong coupling scale $\Lambda$ (equivalently the Kaluza-Klein scale $M_{KK}$) or 't~Hooft coupling. Therefore the analysis itself is the same for all holographic models with large $\lambda$, \emph{e.g.} the flavored~\cite{DKS} Klebanov-Strassler geometry~\cite{KS}.

The paper is organized as follows. In section~\ref{secReview} we review the state of the art of the QCD phase diagram at finite temperature and chemical potential. Section~\ref{secRealQCD} describes realistic QCD with 3 colors and 2-3 (massless) flavors. In section~\ref{secLargeNQCD} we summarize possible modifications of the phase diagram when the number of colors $N_c$ is taken to be large. We also briefly mention there the consequences of large 't~Hooft coupling~$\lambda$. In section~\ref{secSkyrmions} we discuss the insight on the phase diagram obtained from the Skyrme model.

In section~\ref{secHBaryons} we review the story of baryons in holography. For a prototype holographic model of QCD we consider a generalization of the one proposed by Sakai and Sugimoto model. Section~\ref{secSakaiSugimoto} is dedicated to a review of the relevant facts about  the generalization of Sakai-Sugimoto model. The baryons are introduced in section~\ref{secSSBaryons} as instanons of the effective theory on the flavor D-branes. Here we summarize the general strategy to be used in our analysis of multi-baryon systems. Section~\ref{secOtherModels} contains a brief discussion of other holographic models of baryons, which also admit instanton description of baryons.

Section~\ref{secPopcorn} starts with  a review of previous results on the phase diagram of holographic QCD. In section~\ref{secOutline} we specify the part of the phase diagram that we are going to explore and summarize our expectations. We illustrate our expectations through a simple toy model of lattices of point charges, considered in the following section~\ref{secPointCharges}. In particular in section~\ref{sec1dPCharges} we consider a warm-up example of a 1D chain of point charges, while section~\ref{sec3dLattice} generalizes the result to 3D cubic lattice of point charges.

The effects of finite size are analyzed in section~\ref{secChain}. We start in section~{\ref{secStraightChain}} from a review of the BPST instanton solution describing an infinite straight chain with a constant orientation twist between neighboring pairs of instantons. Later we compute $1/\lambda$ corrections to the energy of the BPS configuration, which are generated by the non-BPS interaction of baryons. Assuming anti-ferromagnetic orientation of instantons in section~\ref{secZigzag} we show that for large density the straight chain is unstable against splitting along the holographic direction (zigzag deformation) in a second order phase transition. The analysis is generalized in section~\ref{secGeneralTwist} to the case of arbitrary twist between neighboring instantons. Here we discover another (first order) transition at larger densities.

In section~\ref{secSummary} we summarize our results and their interpretation and propose several open problems. We supplement the main text by several appendices. Appendix~\ref{secSLDensity} provides the details of calculations of the instanton self-interaction energy in the limits of large and small densities. Appendix~\ref{appMinEnergy} is dedicated to a proof of the fact that the straight chain always prefers anti-ferromagnetic orientation of instantons. Somewhat unexpected instability of the zigzag towards formation of the straight chain at high density is demonstrated in appendix~\ref{appHighDensity}. Finally in appendix~\ref{appGenTwist} we provide the details of the derivation of general twist zigzag instanton solution.


\section{Baryons at high density}
\label{secReview}

\subsection{Phase diagram of QCD}
\label{secRealQCD}

\begin{figure}[t]
\begin{center}
\vspace{3ex}
\includegraphics[width= 100mm]{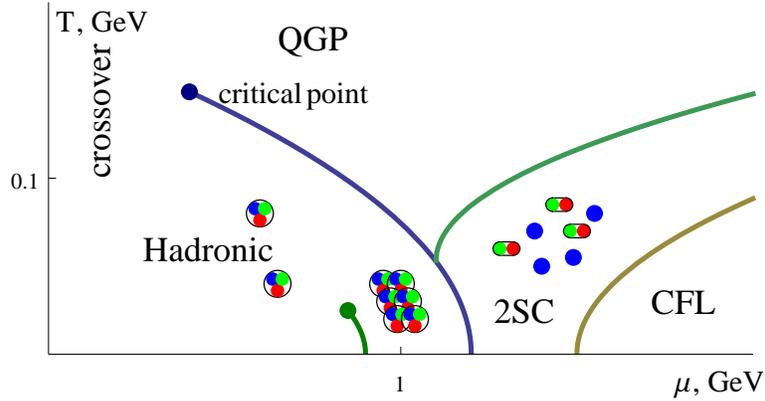}
\end{center}
\caption{Phase diagram of QCD (contemporary view~\cite{Rajagopal:2000wf,Alford:2006wn,Stephanov:2007fk}).}
\label{figQCDPhases}
\end{figure}

The precise shape of the QCD phase diagram is not firmly established. The current state of the art diagram (figure~\ref{figQCDPhases}) is a collection of lattice results, perturbative and non-perturbative estimates as well as educated guesses at various corners~\cite{Rajagopal:2000wf,Schafer:2005ff,Alford:2006wn,Stephanov:2007fk}. In fact not only the boundaries of the many phases of QCD cannot be fixed, but even the existence of some of the phases may be uncertain due to a crucial dependence on the true values of the parameters, e.g. quark masses, number of flavors \emph{etc.}

The low temperature and low chemical potential corner of the diagram is the hadronic phase. In this phase quarks are confined into hadrons and chiral symmetry is broken. When the temperature is increased a deconfinement transition occurs accompanied by the restoration of chiral symmetry. Let us review here both the case of massless and massive quarks.

\begin{figure}[b]
\begin{minipage}[b]{0.5\linewidth}
\begin{center}

\includegraphics[width=7.8cm]{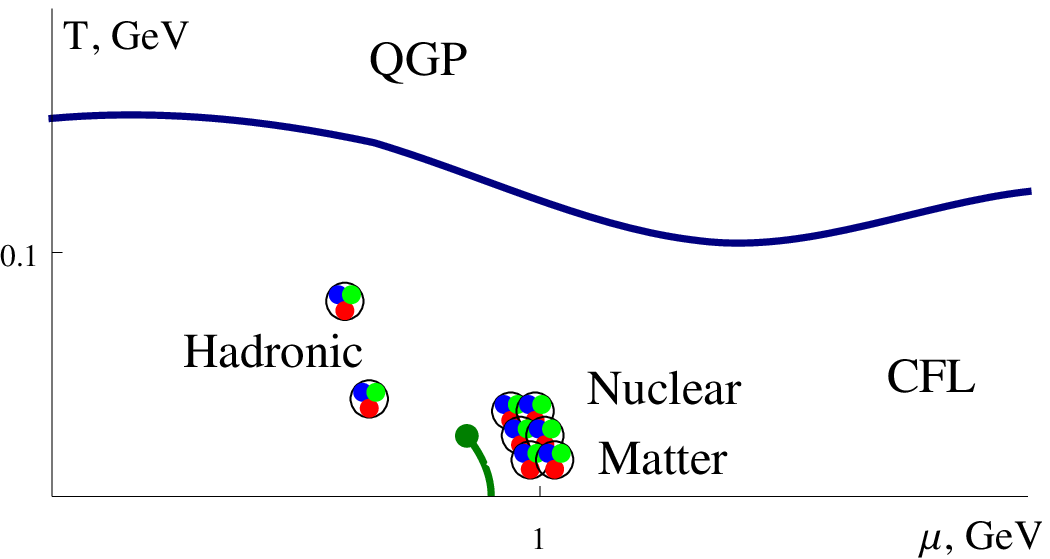}

(a)
\end{center}
\end{minipage}
\begin{minipage}[b]{0.5\linewidth}
\begin{center}

\includegraphics[width=8.cm]{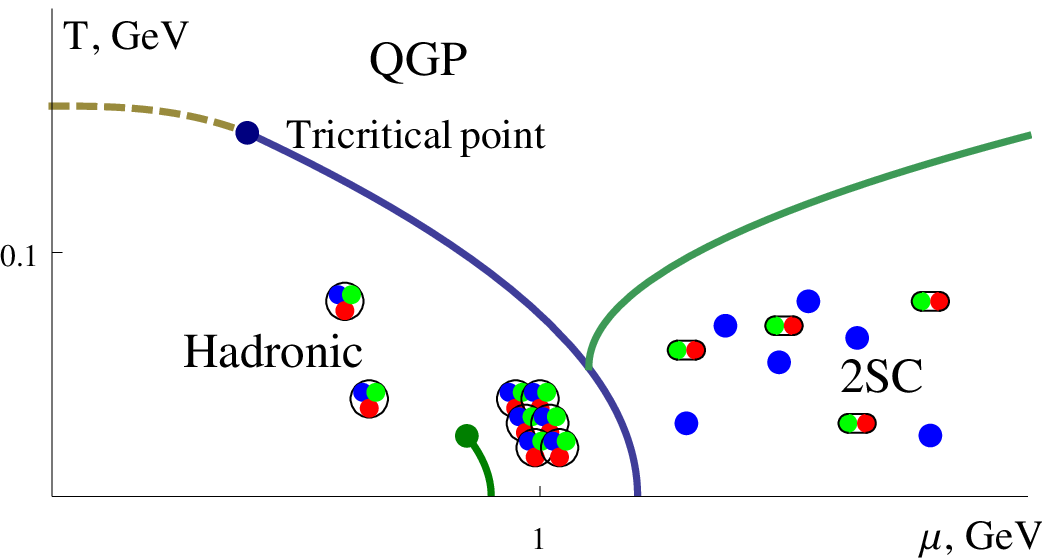}

(b)
\end{center}
\end{minipage}
\vspace{-0.6cm}
\caption{\small  The QCD phase diagram for three (a) and two (b) massless flavors~\cite{Rajagopal:2000wf,Schafer:2005ff,Stephanov:2007fk}.}
\label{figMasslessQCD}
\end{figure}

In the approximation of massless quarks  chiral symmetry is broken spontaneously by the quark condensate. For high temperature thermal fluctuations destroy the condensate and a symmetry restoring phase transition occurs. Lattice simulations in general support the conjecture that at zero chemical potential this phase transition is second order for two and first order for three massless flavors. On the other hand at finite (large) chemical potential many model calculations suggest that the phase transition to the restored symmetry phase becomes first order for both $N_f=2$ and $N_f=3$. Therefore for the case of two massless quarks a change from second order to first order phase transition is expected along the boundary of the hadronic and chiral symmetry restored (deconfined) phases. The point where the change occurs is called tri-critical (figure~\ref{figMasslessQCD}).

In the case of massive quarks, $m_u\simeq m_d\ll m_s$, the symmetry is broken explicitly. As a result the second order transition from a confined to a deconfined phase becomes a crossover, since in the absence of exact symmetry there is no critical behavior. Study of QCD at the crossover is extremely complicated, because neither the conventional strong coupling nor weak coupling descriptions work in this regime. The phase of QCD at the crossover is usually referred as to strongly coupled quark-gluon plasma (sQGP). The crossover changes to a first order transition at a (chiral) critical point similar to the critical point of water (figure~\ref{figQCDPhases}).

At small temperature and finite chemical potential physics of QCD is even more interesting. The natural energy scale for the interaction of the elementary degrees of freedom (quarks) is the Fermi momentum, which is related to the chemical potential as $\mu^2= k_F^2+ m_B^2$. Therefore for large enough chemical potential the physics is expected to be weakly coupled. The naive perturbation theory does not work though, because the perturbative ground state (free quark Fermi spheres) is unstable due to attraction of the quarks with equal and opposite momentum at the Fermi surface. This problem can be overcome in a fashion similar to the BCS theory of superconductivity. In the true ground state the quarks form Cooper pairs (condensate) while the colored excitations acquire a gap.

At asymptotically large chemical potential, one can ignore quark masses (with all 3 quark flavors participating). In this case the preferred quark condensate is the one that preserves maximal symmetry:
\be
\langle \epsilon^{\alpha\beta} q^{ia}_{\alpha L}(p) q^{jb}_{\beta L}(-p) \rangle= - \langle \epsilon^{\dot{\alpha}\dot{\beta}} q^{ia}_{\dot{\alpha} R}(p) q^{jb}_{\dot{\beta} R}(-p) \rangle = \Delta(p^2)\epsilon^{ijA}\epsilon^{abA}\,,
\ee
where Greek letters are reserved for spinor indices, $a$, $b$ and $i$, $j$ are flavor and color indices respectively. Such a condensate breaks both color and flavor symmetries, but leaves the diagonal subgroup of their product unbroken. The symmetry breaking proceeds as follows
$$
SU(3)_c\times SU(3)_L\times SU(3)_R \to SU(3)_{CFL}\,.
$$
The color and flavor transformations are locked together, hence the name of the phase: color-flavor-locking (CFL). There are in fact two condensates for the left and right fields separately. One of them locks $SU(3)_L$, while the other one $SU(3)_R$. So indirectly the chiral symmetry is broken.

Breaking of the color symmetry results in  masses for all gluons. This is analogous to the Meissner effect for ordinary superconductors. Therefore the CFL phase is superconducting. Since in the CFL phase all the symmetries are broken, while in the QGP they are unbroken, there should be at least one phase transition relating these phases. The phase diagram of QCD with three massless quark should thus look as shown in figure~\ref{figMasslessQCD}(a). It is not clear however if there is a phase transition between CFL phase and the nuclear matter phase. Since the former phase is deconfined while the latter is confined first order transition can be expected. However this phase transition will not restore the chiral symmetry, since it is broken on both sides.  The separation of deconfinement and chiral symmetry restoration can in principle occur as we discuss in more detail below.

For intermediate values of the chemical potential the difference in the quark masses cannot be ignored. A realistic approximation would be to choose $0=m_u=m_d\ll m_s$. First consider infinite $m_s$. In this case a condensate of quarks with two out of three colors is formed, e.g.
\be
\langle \epsilon_{ij3}\epsilon^{\alpha\beta} q^{i}_\alpha(p) q^j_\beta(-p) \rangle\,.
\ee
The condensate picks a direction in the color space. $SU(3)_c$ group breaks down to the $SU(2)_c$ and five gluons get mass. Gaps are formed for excitations of colors orthogonal to the condensate direction. The potentially light degrees of freedom are quarks of the condensate color and gluons of the unbroken subgroup. This phase is known as 2SC. In the case of two massless flavors this is the preferable phase at small temperature and large chemical potential as shown in figure~\ref{figMasslessQCD}(b).

The condensate is invariant under $SU(2)_L\times SU(2)_R$, which means the 2SC phase has unbroken chiral symmetry. Therefore a phase transition must separate this phase from the hadronic phase. This transition must be first order, since it involves a competition between the color and chiral condensates. Since there are no broken global symmetries the 2SC phase is not a superfluid.

For finite $m_s$ the CFL will be a dominating phase at large chemical potential. However the CFL phase requires that condensates that involve the strange quark and a light quark must produce a gap which is larger than roughly $m_s^2/2\mu$, which is the difference of the $s$-quark and light quark Fermi momenta. For not-so-large chemical potential the phase is 2SC as shown in figure~\ref{figQCDPhases}. Since the latter has restored chiral symmetry there is a phase transition in between. It cannot be second order, because this will require condensates $\langle us\rangle$, $\langle ds\rangle$ to be infinitesimal at the phase transition on the CFL side.

\begin{figure}[htb]
\begin{minipage}[b]{0.5\linewidth}
\begin{center}

\includegraphics[width=7.8cm]{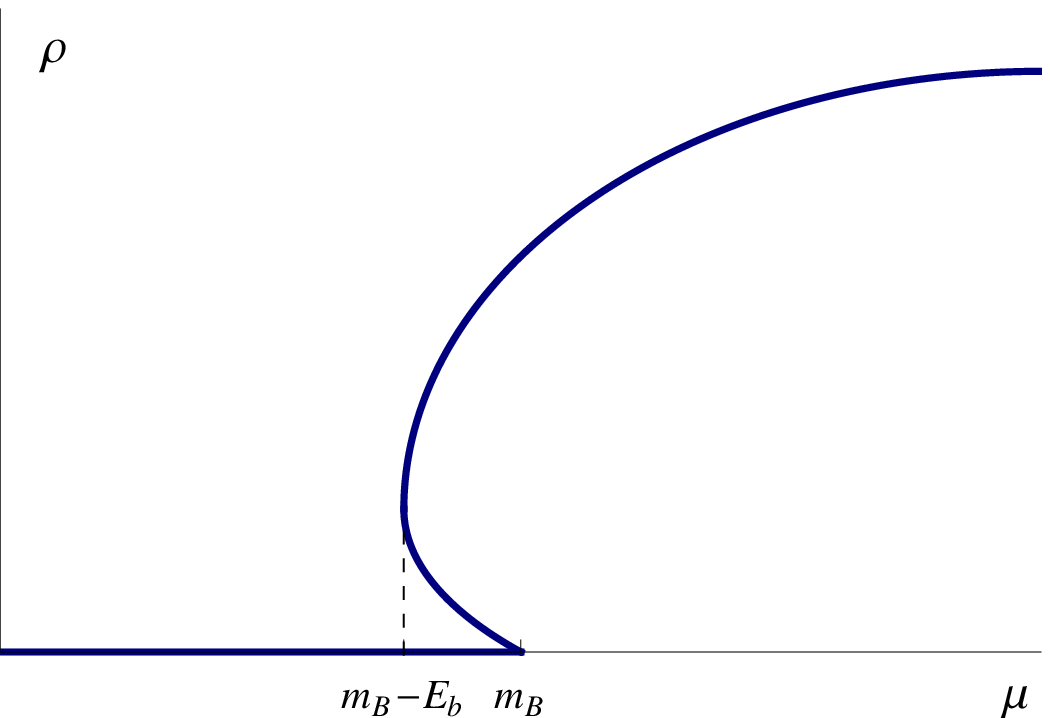}

(a)
\end{center}
\end{minipage}
\begin{minipage}[b]{0.5\linewidth}
\begin{center}

\includegraphics[width=8.cm]{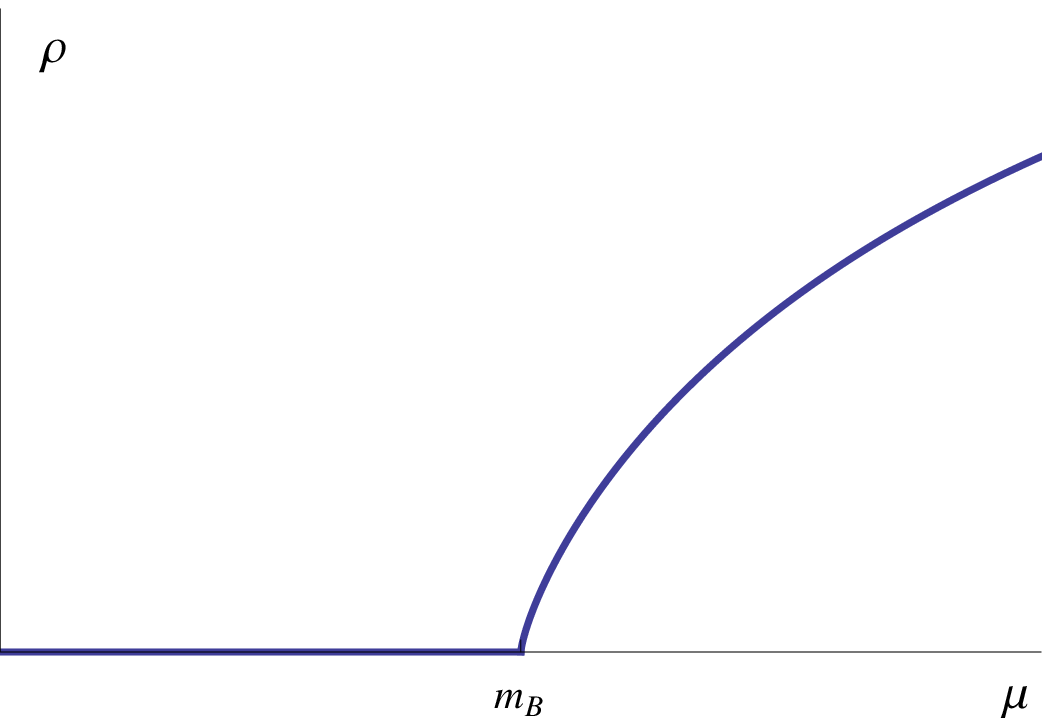}

(b)
\end{center}
\end{minipage}
\vspace{-0.6cm}
\caption{\small Zero temperature dependence of the baryon density on chemical potential in the case of attractive (a) and repulsive (b) equilibrium force between baryons. In the attractive case the transition to non-zero density is a sharp first order transition at chemical potential $m_B-E_b$. In the repulsive case it is a smooth second order transition.}
\label{figAttractiveRepulsive}
\end{figure}

Finally we will discuss one more phase of QCD, which is of most interest for this paper, namely the nuclear matter phase. For a chemical potential of order $\mu\sim\text{1GeV}$ and slightly below the baryons form a dilute gas: since it is hard to excite baryons, at zero temperature there is vacuum (a state with no baryons), while at finite temperature one expects a small population of baryons in the thermal ensemble. The baryons attract each other with a characteristic binding energy $E_b$, therefore it should cost nothing to produce them when $\mu$ reaches the value $\sim m_B - E_b$, where $m_B$ is the mass of the lightest baryons. The zero temperature case is demonstrated by figure~\ref{figAttractiveRepulsive}(a): at small chemical potential the density is zero, while for $\mu=m_B-E_b$ there is a first order phase transition to the equilibrium density of nuclear matter. If the interaction between baryons had been  repulsive as in the cases discussed below the transition would have been second order as in figure~\ref{figAttractiveRepulsive}(b).

In reality this picture is altered by the presence of the Coulomb repulsion between baryons. Because the Coulomb force is long range it precludes the existence of the bulk nuclear matter. Therefore the behavior corresponding to figure~\ref{figAttractiveRepulsive}(a) is an idealization, which assumes that the bulk Coulomb force is screened by either finite density of electrons or by idealized light $s$-quark. The real-life QCD is expected to have a first order gas-liquid phase transition similar to the phase transition in water. At finite temperature this phase transition must end at a (hadronic) critical point, because the Lorentz symmetry will be broken on both sides of the phase transition. The end point of the transition should appear at temperatures of the order of baryonic binding energy $E_b$.

As follows from the properties of ordinary nuclear matter, it is in a liquid phase. For increased chemical potential one may expect the liquid of baryons to turn solid. However it is not clear if such a phase exist at all in real life QCD. It may be that the liquid phase goes directly to a dense quark matter phase through a deconfining and chiral symmetry restoring phase transition.

In this part we have reviewed the phase diagram of real-life QCD with $N_c=3$ and three light flavors of quarks. The result is summarized in figure~\ref{figQCDPhases}. Similar diagrams with more detailed explanation and relevant references can be found in the reviews~\cite{Rajagopal:2000wf,Stephanov:2007fk}.


\subsection{QCD at large $N_c$}
\label{secLargeNQCD}

\begin{figure}[htb]
\begin{center}
\vspace{3ex}
\includegraphics[width= 100mm]{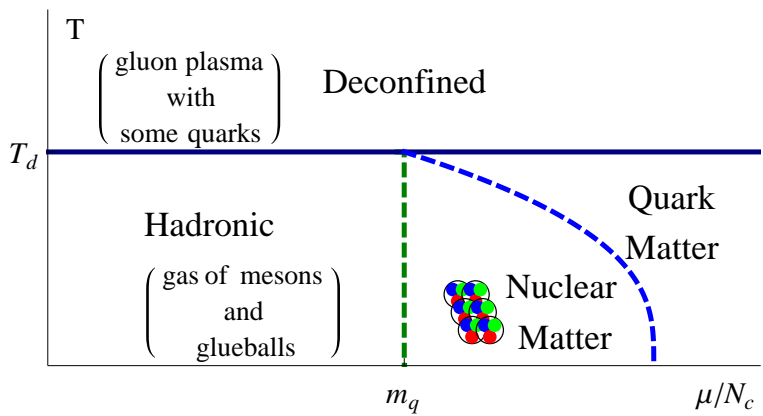}
\end{center}
\caption{Conjectured phase diagram for QCD at large $N_c$~\cite{McLerran:2007qj}}
\label{figLargeNQCD}
\end{figure}

Various properties of QCD are often studied qualitatively by taking the limit of large number of colors $N_c$~\cite{tHooftLargeN}. In this limit contribution of quark loops is suppressed and only planar gluon diagrams contribute. At zero temperature and chemical potential the large $N_c$ QCD is a theory of free mesons and glueballs, since there interactions are suppressed by powers of $N_c$, except for interactions with baryons. Masses of the mesons and glueballs are of order $\Lambda_{QCD}$, that is they scale as $N_c^0$. Baryons are special objects in the large $N_c$ limit~\cite{Witten:1979kh}. Because they are made out of $N_c$ quarks they are heavier than the mesons and glueballs with the mass of order $N_c\Lambda_{QCD}$. Their interaction energy is also non-vanishing and scales as $N_c$. Based on these properties of the large $N_c$ theory one can speculate on the structure of the $T-\mu$ phase diagram (see review~\cite{McLerran:2007qj} dedicated to this problem and references therein). One possibility is illustrated by figure~\ref{figLargeNQCD}. Below we will briefly explain this diagram.

At small temperature and chemical potential large $N_c$ QCD is a theory of mesons and glueballs. Since the baryons are so heavy ($\sim N_c$) their relative abundance is exponentially suppressed. When the temperature is increased a deconfining phase transition is expected. The temperature $T_d$ of this transition should be of the order $\Lambda_{QCD}$ and the transition is first order. It is implicitly assumed that $N_f\ll N_c$. In this regime the quark loops are suppressed as $N_f/N_c$ and the dynamics is completely captured by gluons, unless the chemical potential for quarks ($\mu_q=\mu/N_c$) scales as a power of $N_c$. Thus for $\mu_q$ of order $N_c^0$ the deconfining temperature is expected to be independent from $\mu_q$.

An interesting question is whether the deconfining phase transition coincides with the chiral symmetry restoration or not. Several field theory arguments, e.g.~\cite{Casher,McLerran:2007qj}, suggest that at $\mu_q=0$ there is a direct relation between confinement and chiral symmetry breaking. However at non-zero chemical potential this relation need not hold, and in fact there are various indications to the existence of confined, but chirally restored phase~\cite{McLerran:2007qj}. Moreover recent holographic models, which are large $N_c$ as well as large 't Hooft coupling $\lambda$ models, imply that the two transitions need not coincide even at zero chemical potential~\cite{Aharony:2006da}. At low chemical potential the latter models in fact exhibit phases that are deconfined, but without chiral symmetry.

For the values of the chemical potential $\mu$ above the mass of the lightest baryons ($\mu_q > m_B/N_c\equiv m_q$) and $T<T_d$ baryons persist in the nuclear matter phase with a non-zero density. Conversely to the ordinary one the large $N_c$ nuclear matter must be solid. To see this one can check the ratio of kinetic and potential energies in the large $N_c$ regime. Potential energy of the baryon-baryon interaction scales like $N_c$. Specifically,  large $N_c$ potential takes a form~\cite{Kaplan}
\begin{multline}
\label{Potential}
\Pi \ \sim \  N_c\times A_C(r) \ + \ N_c\times A_S(r) \left({\bf I}_1\cdot {\bf I}_2\right)\left({\bf J}_1\cdot {\bf J}_2\right) \ +
\\  + \  N_c\times A_T(r) \left({\bf I}_1\cdot {\bf I}_2\right)\left(3  \left({\bf n}\cdot {\bf J}_1\right) \left({\bf n}\cdot {\bf J}_2\right)- \left({\bf J}_1\cdot {\bf J}_2\right)\right) \ + \
 O\left(1/N_c \right),
\end{multline}
where $A_C$, $A_S$ and $A_T$ are the central, spin-spin and tensor potentials independent from $N_c$. Classically this potential tries to organize baryons in some kind of a crystal with interatomic separation independent from $N_c$. By analogy with the condensed matter systems the kinetic energy is the zero point motion energy in an appropriate potential well:
\be
\label{Kinetic}
{\rm K}\  \sim \ \frac{\pi}{2m_B d^2}\sim \frac{1}{N_c}\,\frac{1}{d^2\Lambda_{QCD}}\,,
\ee
where $d$ is the $N_c$-independent diameter of the potential well. Therefore the ratio
\be
\frac{\rm K}{\Pi}\ \sim \  \frac{1}{N_c^2}
\ee
is suppressed in the large $N_c$ regime and the cold dense nuclear matter is a crystal.

The type of the transition between the baryon vacuum and the nuclear matter will depend on whether the interaction between baryons is attractive or repulsive. Ordinary nucleons attract each other, but this attraction appears to be a result of a very subtle balance of various factors which can easily be violated by large $N_c$ effects (\emph{e.g.} a discussion in~\cite{Kaplunovsky:2010eh,Bonanno:2011yr}.) Thus both of the situations illustrated by figure~\ref{figAttractiveRepulsive} may be realized for large $N_c$. If the baryons still attract the density will behave according to the plot on figure~\ref{figAttractiveRepulsive}(a). There will be a first order phase transition from the vacuum to the nuclear matter phase at $\mu=m_B - E_b$.  Notice that we leave aside the question of what happens with the Coulomb interaction at large $N_c$ and its effect on nuclear matter. If the baryons repel the transition will be second order, figure~\ref{figAttractiveRepulsive}(b). This is the situation we will later assume in this paper.

The crystal phase of  cold high density baryons is mostly studied within the framework of Skyrme model. In the next section we will review the main results on Skyrme crystals, while here we continue the discussion of the phase diagram. Skyrme crystals exhibit a zero temperature finite density phase transition, which at least partially restores chiral symmetry. (This result is also supported by calculations in other qualitative models as well~\cite{DecChiSplitting}.) On the other hand it was mentioned before that the deconfining temperature $T_d$ should be independent of $\mu$ when the latter is not too large (quark contribution to the expectation value of the Polyakov loop is of order $1/N_c$, as compared to $N_c^{0}$ in the deconfined phase). Therefore the chiral phase transition and the deconfining phase transition are likely to separate in the dense baryonic phase.

The authors of~\cite{McLerran:2007qj} refer to the phase in figure~\ref{figLargeNQCD} below $T_d$ and to the right of $\mu_q=m_q$ as ``quarkyonic''. The reason for this is as follows. Similarly to finite $N_c$ QCD at asymptotically large $\mu_q$ a weakly coupled description is possible. One can compute the total pressure of finite density QCD perturbatively. The main contribution to the pressure is from quarks far from the Fermi surface that  scatter with the momentum exchange of order $\mu_q$. One obtains that the pressure scales as $N_c$ and this must be insensitive to non-perturbative corrections as one reduces the chemical potential down to $\mu_q\sim\Lambda_{QCD}$. However for such values of $\mu_q$ the phase should be that of dense baryons also with the pressure of order $N_c$. Effectively this looks like the phase is changing smoothly as one changes the chemical potential from $\mu_q\gg \Lambda_{QCD}$ down to $\mu_q \sim \Lambda_{QCD}$ and both descriptions in terms of quarks and baryons apply. More accurately, for large $\mu_q$ scattering of quarks far from the Fermi surface can be described by perturbation theory. Nevertheless for $\mu_q$ not too large baryons remain the lightest excitations near the Fermi surface: unlike for finite $N_c$ QCD, the quarks do not screen gluons until the chemical potential $\mu_q$ scales as a power of $N_c$. The ``quarkyonic'' phase is a phase with quark Fermi sea but with baryonic Fermi surface. When the chemical potential is increased from $\mu_q\sim \Lambda_{QCD}$ the initial baryonic Fermi see smoothly changes to a quark one with the baryonic Fermi surface of the width of order $\Lambda_{QCD}$.

When the chemical potential $\mu_q$ is as large as $N_c^{1/4}\Lambda_{QCD}$ the contribution of quarks to the pressure becomes comparable to that of gluons. Gluons start feeling quarks and the deconfining temperature $T_d$ should no longer be independent from $\mu_q$. At $\mu_q\sim N_c^{1/2}\Lambda_{QCD}$ the quark contribution completely dominates, the gluons are screened and the deconfining transition should end, either at $T_d=0$ or at some critical point.

At large $N_c$ for $\mu\gg \Lambda_{QCD}$ the phase of the quarkyonic matter is perhaps not a color superconductor. It was argued in~\cite{Rubakov} that color singlet chiral density wave (CDW) phase is dominating there. For large but finite $N_c$ it was proven that the CDW does not dominate until $N_c\sim 10^3$~\cite{SonShuster}.

In holographic models apart from the limit of large number of colors one also has to assume large values of the 't Hooft coupling constant $\lambda$. Since the main topic of this work is a holographic description of the nuclear matter phase let us discuss the implication of large $\lambda$ limit on the relevant part of the phase diagram. First the baryon mass  is proportional to both $N_c$ and $\lambda$. In turn the interaction of baryons is only of order $\lambda^0$. Therefore baryons are heavy, weakly interacting objects in the large $\lambda$ limit. Specifically the binding energy of baryons, if the net force between them is attractive, will be small anyway compared with the baryon mass.

\begin{figure}[htb]
\begin{minipage}[b]{0.5\linewidth}
\begin{center}

\includegraphics[width=7.8cm]{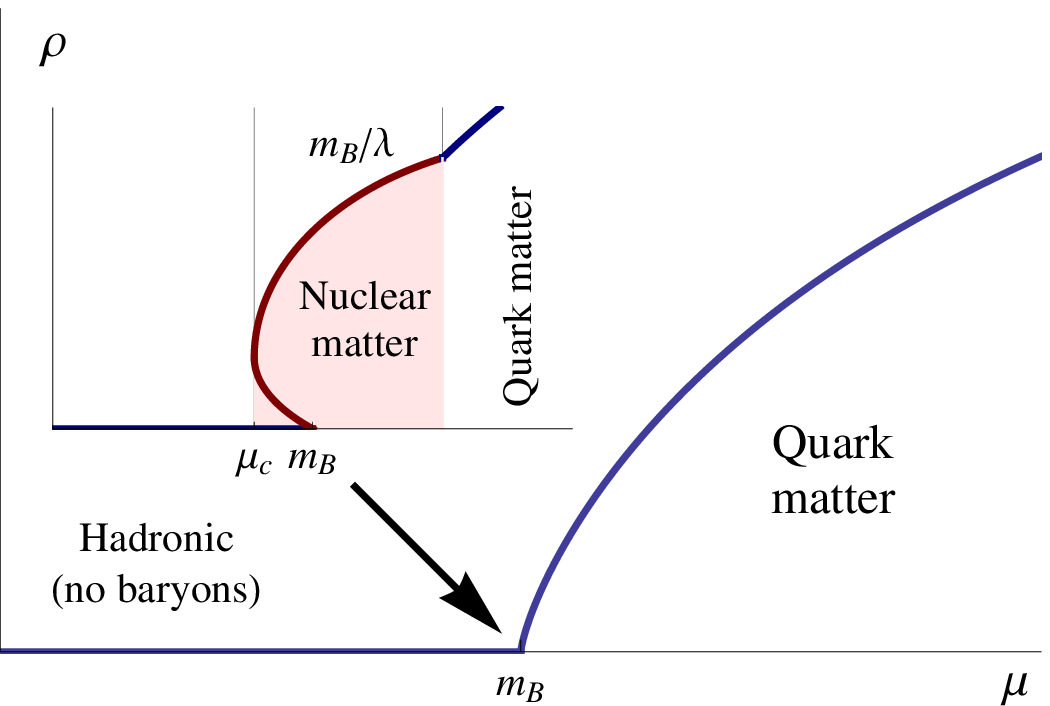}

(a)
\end{center}
\end{minipage}
\begin{minipage}[b]{0.5\linewidth}
\begin{center}

\includegraphics[width=7.8cm]{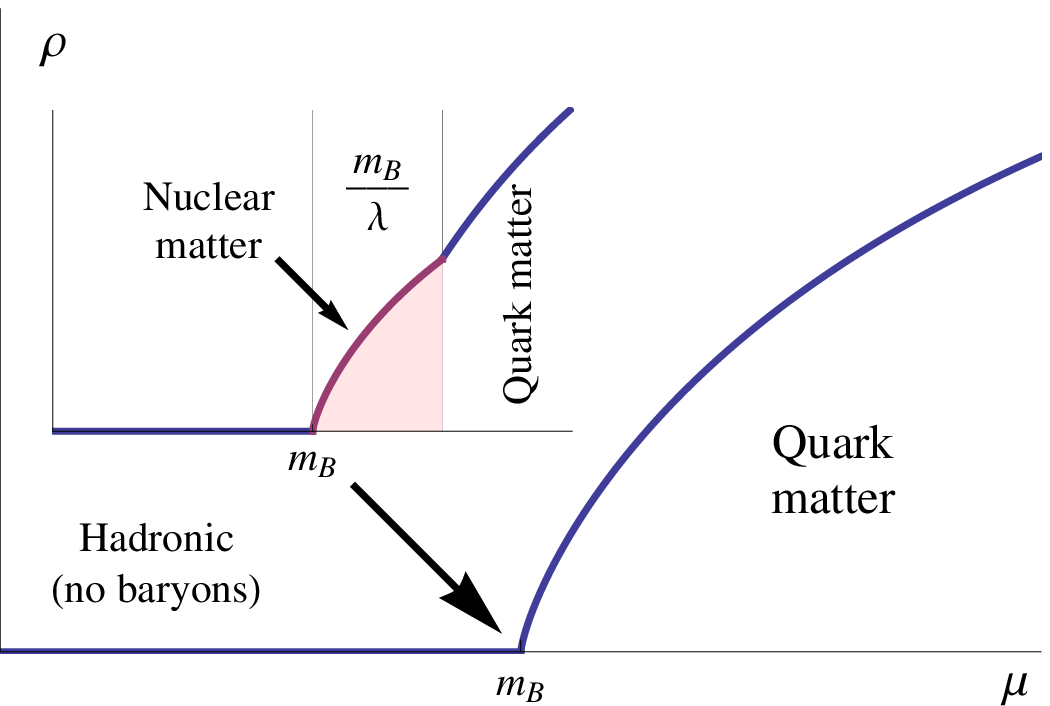}

(b)
\end{center}
\end{minipage}
\vspace{-0.6cm}
\caption{\small Density as a function of the chemical potential in the large $\lambda$ regime for attracting~(a) and repelling~(b) baryons. The nuclear matter phase is confined to a narrow window of the order $\Delta\mu\sim m_B/\lambda$. In the naive diagram the transition occurs directly from the no-baryon to a quark phase.}
\label{figWorkWindow}
\end{figure}

Because of the subleading scaling of the baryon interactions the characteristic scale of baryon physics will be proportional to $1/\sqrt{\lambda}$. A characteristic baryon size will be $\lambda^{-1/2}$ and the characteristic density of the $D$-dimensional baryon crystal $\rho\propto \lambda^{D/2}$. Such densities correspond to the values of the chemical potential $\mu\sim \mu_c +  O(\lambda^0)$, that is $\mu-\mu_c\ll \mu_c$. At larger values of the chemical potential, \emph{e.g.} $\mu - \mu_c\sim \mu_c$ the density will be too large for individual baryons to retain their identities. At such densities the nuclear matter will turn into some kind of a quark matter. The nuclear matter phase will thus be confined to a narrow window of values of the chemical potential. This is illustrated by figure~\ref{figWorkWindow}, where we show again both the case of attractive and repelling baryons.

\subsection{High density baryons in the Skyrme model}
\label{secSkyrmions}
In the 1980s the old idea of Skyrme that the baryons can be described as solitons of the chiral model~\cite{Skyrme} received a revived attention~\cite{Witten:1979kh}. These classical solutions, the skyrmions, satisfy the equations of motion derived from the Lagrangian proposed by Skyrme:
\be
\label{SkyrmeL}
\mathcal{L} = \frac{f_\pi^2}{16}\,\tr\partial_\mu U\partial^\mu U^\dagger + \frac{1}{32e^2}\,\tr\left[\partial_\mu U U^\dagger,\partial_\nu U U^\dagger\right]^2,
\ee
where $U(x)=\sigma + \vec{\pi}(x)\cdot \vec{\tau}$ is the $SU(2)$-valued pion field and $f_\pi$ is the pion decay constant. The second term in the Skyrme Lagrangian does not follow from chiral Lagrangian, but was added by Skyrme to stabilize the solutions. The parameter $e$ is typically fixed such that the mass of the single skyrmion matches the masses of $N$ and $\Delta$ nucleons. Skyrme model is effectively a model of large $N_c$ baryons.

Crystals of skyrmions were extensively studied in the 80's starting from the work of Klebanov~\cite{Klebanov}. To find a skyrmion solution corresponding to a lattice one imposes appropriate periodicity conditions on a single skyrmion in the unit cell of the lattice, consistent with the symmetries of the given lattice. The solution for various types of cubic lattices is typically obtained numerically. Analytical results are possible only within an approximation scheme, \emph{e.g.} replacing the unit cube by a 3-sphere~\cite{Kutschera:1984zm}, or employing the Atiyah-Manton ansatz~\cite{AtiyahManton}. (See also~\cite{Schroers:1993yk} for other approximation techniques.)

It was noticed that at higher densities skyrmion solutions exhibit an extended symmetry as compared with the symmetries allowed by the original lattice~\cite{Manton87,Wust}. This is most obvious in the $S^3$-approximation of the cubic cell. When the size of the spherically symmetric skyrmion is small compared with the volume of $S^3$ it  has only  an $SO(3)$ symmetry. However when its size becomes comparable to that of the sphere the symmetry extends to $SO(4)$. In the case of simple cubic lattice the additional symmetry corresponds to an exchange of the vertices of the elementary cells of the lattice with their centers. Instead of the cubic lattice it becomes more natural to consider unit cells of half the original volume consistent with the extended symmetry (Wigner-Seitz cells of the bcc lattice). The new symmetry imposes strong constraints on the skyrmion solution. As was demonstrated by Goldhaber and Manton~\cite{Goldhaber:1987pb} each elementary cell contains a half of the skyrmion charge and the value of the skyrmion field is different for the vertices of the original cube and its center (in particular it differs by the sign of $\sigma$). Thus instead of the original cubic lattice of skyrmions one obtains a bcc lattice of half skyrmions with different kind of skyrmions in the sites and in the centers of the cubic cell. It was also shown that, when averaged over the original unit cell, one obtains
\be
\label{SkyrmeAverage}
\langle\sigma\rangle = \langle\vec{\pi}\rangle = 0\,,
\ee
where brackets mean average over space, since the Skyrme model is classical. This serves as an indication of the chiral symmetry restoration at high density and zero temperature consistent with the phase diagram in figure~{\ref{figLargeNQCD}}.

It is not so easy to see chiral symmetry restoration in the case of cubic lattice, but it becomes obvious if one resorts to the spherical cell approximation~\cite{Jackson:1988bd}. When the skyrmion on $S^3$ is small it necessarily points in some direction in the group space. The solution is roughly a map $R^3\to S^3$ as in the case of the flat-space skyrmion, with $U\to 1$ at inifinity. However, when the size of the skyrmion is of order of the $S^3$ size the solution becomes  the identity map $S^3\to S^3$ of the spatial 3-sphere onto the group space. The integral of $U$ over the sphere vanishes in this case and specifically~(\ref{SkyrmeAverage}) holds. Further studies of excitations over the half-skyrmion lattice have shown the doubling of the spectrum, consistent with chiral symmetry restoration~\cite{ParityDoubling}.

It was shown by Kugler and Shtrikman~\cite{KuglerShtrikman} that the preferred configuration of skyrmions at small density is the fcc lattice, for which each skyrmion has 12 nearest neighbors. At higher densities this lattice becomes a simple cubic lattice of half skyrmions. The sc lattice of half-skyrmions can attain the smallest possible energy per skyrmion out of all configuration of skyrmions. This happens at the density corresponding to the lattice spacing $d=4.7$ in units of $1/e f_\pi$. This gives the energy $E=1.038$ in units of the topological lower bound on the skyrmion energy (to be compared with $E=1.23$ for a single skyrmion). As in the case of the Goldhaber-Manton transition the transition in the fcc skyrmion lattice also points to the chiral symmetry restoration.

It turns out that skyrmions can naturally be related to the baryons in holographic models. As will be discussed below baryons in holography are described by instanton solutions of the gauge fields on flavor branes. Remarkably the holographic description of baryons by instantons was anticipated by Atiyah and Manton~\cite{AtiyahManton}, who suggested the following approximation for skyrmions. The skyrmion field $U$ in 3D can be approximated by the holonomy of a 4D Euclidean instanton field. More precisely
\be
\label{AtiyahManton}
U \simeq P\exp \left(- \int A_4 \d x_4\right),
\ee
where the holonomy is evaluated for the additional 4th component of the gauge field along the additional direction. The instanton number in this case is mapped to the skyrmion number. It was later realized that such an approximation works extremely well. In particular, for a single skyrmion it predicts the energy $E=1.24$, while for the minimum energy of the Kugler-Shtrikman lattice of half-skyrmions the energy predicted by Atiyah-Manton ansatz is $E=1.04$.

For a while the coincidence of the results from skyrmion and instanton pictures remained somewhat mysterious. It was in fact holography which provided a resolution for this mystery~\cite{Sutcliffe}. In some holographic models baryons are described by instanton solutions of the non-abelian gauge fields that live on D-branes, e.g. Sakai-Sugimoto model~\cite{SakaiSugimoto2004}. For static baryons the instanton solution lives in 3 spatial and the holographic dimension ``$z$''. The pion-skyrmion field $U$ naturally appears in the holographic setting as the holonomy of the component of the non-abelian field along the holographic direction, \emph{cf.}~(\ref{AtiyahManton}),
\be
\label{HolographicSkyrmion}
U = P\exp\left(-\int A_z \d z\right).
\ee
The Skyrme Lagrangian~(\ref{SkyrmeL}) as seen by holography will be modified. An infinite number of terms coupling the pion field to the infinite tower of vector mesons will be added to it. This ``improved'' Skyrme model was derived by Nawa, Suganuma and Kojo in~\cite{Nawa} and further used to study various properties of baryons.

So why does instanton approximation describe skyrmion solutions so well? Holography provides formula~(\ref{HolographicSkyrmion}) for the skyrmion of the improved model. Unlike the approximation of Atiyah and Manton the instanon of equation~(\ref{HolographicSkyrmion}) lives in a curved space. The curved space is in fact responsible for the tower of vector mesons that couple to the skyrmion. In the limit of large 't Hooft coupling $\lambda$ the instanton size is much smaller than the curvature scale and one can think of it as of just a flat space instanton. Thus the formula of Atiyah and Manton gives an approximation to the improved Skyrme model and further to its truncation, the original Skyrme model. The fact that the approximation is so good can be attributed to the large value of 't Hooft coupling constant on the holographic side.

With the straightforward relation between skyrmions and holographic baryons a natural question to address is how the results on the skyrmion lattices can be translated to a holographic setup. It is also interesting if the chiral symmetry restoration at zero temperature and high density can be proved using holography. A qualitative proposal on mapping the skyrmion results to holography was made in the paper by Rho, Sin and Zahed~\cite{Rho:2009ym}. Later by studying toy models with 1D baryon lattices we will find some support to the ideas of that proposal. It still remains an open problem though to give an explicit description of skyrmion physics in the 3D case. Now let us move to a discussion of baryons in holographic models.


\section {Baryons in the Sakai-Sugimoto model}
\label{secHBaryons}

\subsection{The model}
\label{secSakaiSugimoto}

\begin{figure}[t]
\begin{center}
\vspace{3ex}
\includegraphics[width= 100mm]{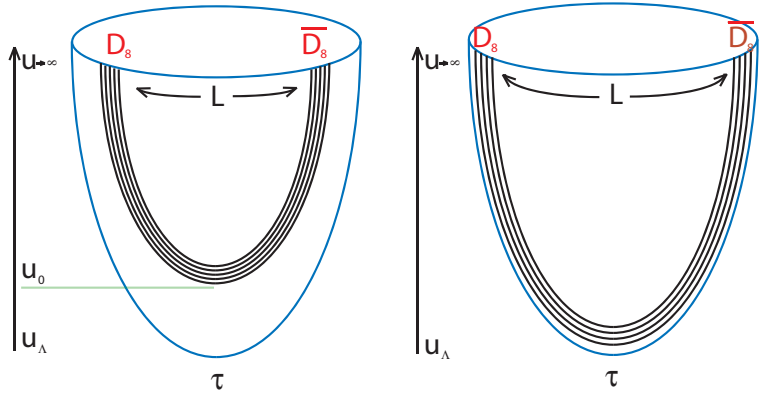}
\vspace*{-2cm}
\end{center}
\caption{The figure on the right is the generalized  non-antipodal configuration.
The figure on the right describes the limiting antipodal case $L=\pi R$, where
the branes connect at $u_0=u_{\Lambda}$.
\label{Ushape}}
\end{figure}

The holographic model proposed by Sakai and Sugimoto~\cite{SakaiSugimoto2004} describes large $N_c$, large $\lambda$ QCD in the quenched approximation $N_f\ll N_c$. Geometrically Sakai-Sugimoto model corresponds to an extension of the Witten's model~\cite{WittensModel} through introducing $N_f$ probe D8 and anti-D8 branes. In the Witten's model one considers $N_c$ type IIA D4-branes in the near-horizon limit. Four of the dimensions transverse to the D4 are compactified on the 4-sphere $S^4$. One dimension along the D4 is compactified on the circle $S^1$ with antiperiodic boundary conditions for fermions. The latter procedure breaks supersymmetry and introduces a (Kaluza-Klein) scale $M_{KK}$ into the theory, which is related to the radius $R$ of $S^1$. Correspondingly the geometry breaks into a warped product
$$AdS_6\times S^4\to M^{1,3}\times R^+\times S^1\times S^4.$$
For such a geometry to be smooth the radial direction $R^+$ should end at a finite distance from the origin, call it $u_\Lambda$ following the original papers. The value of $u_\Lambda$ is controlled by the $M_{KK}$ scale or $R$. The $R^+\times S^1$ part of the product thus looks like a cigar as shown in figure~\ref{Ushape}.

The geometry of the D4-brane is described by the type IIA supergravity solution for the metric, the 4-form and the dilaton
\begin{align}
\d s^2\ &
=\ \left( \frac{u}{R_{D4}}\right)^{3/2}
\Bigl[-\d t^2+\delta_{ij}\d x^i \d x^j+f(u)\d x_4^2\Bigr]\,
+\,\left( \frac{R_{D4}}{u}\right)^{3/2}
\left[\frac{\d u^2}{f(u)}+u^2\d\Omega_4^2\right],\nonumber\\
F_4\ &
=\ \frac{(2\pi)^3l_s^3 N_c}{V_4}\,{\rm vol}(S^4)\,,\qquad
e^{\phi}\ =\ g_s\left( \frac{u}{R_{D4}}\right)^{3/4},
\label{D4background}
\end{align}
where the curvature radius of $AdS_6$ and warping $f(u)$ is given by
\be
R_{D4}^3\ =\ \pi g_sN_cl_s^3\,,\qquad
f(u)\ =\ 1\,-\,\left( \frac{u_{\Lambda}}{u}\right)^3\,.
\ee
$V_4$ is the volume of the unit 4-sphere, $l_s=\sqrt{\alpha'}$ is the string length and $g_s$ is the string coupling. The radial coordinate $u$ runs from $u_\Lambda$ to infinity. In terms of the above metric the relation between $u_\Lambda$ and the circle radius $R$ reads
\be
\label{KKscale}
2\pi R\ =\ \frac{4\pi}{3}\left( \frac{R_{D4}^3}{u_{\Lambda}}\right)^{1/2}.
\ee

The theory on the world-volume of the D4-branes is a 5D Yang-Mills theory compactified on $S^1$. Antiperiodic boundary conditions break supersymmetry: the fermions receive mass of order $M_{KK}$. In the low energy limit the theory is a pure glue $SU(N_c)$ Yang-Mills with the strong coupling scale $\Lambda$ set by $M_{KK}$.

From the DBI action for the D4-branes one reads the coupling of the 5D Yang-Mills theory~\cite{Kruczenski:2003uq}
\be
g_5^2=(2\pi)^2 g_s l_s\,.
\ee
Below the compactification scale $M_{KK}$ the theory is effectively 4-dimensional with the coupling
\be
g_4^2=g_5^2/(2\pi R).
\ee
Note that the theory has a running dilaton so the effective coupling $g_{4}^2\propto e^{\phi}$ diverges in the UV limit, $u\to \infty$. Although the theory constructed so far is not UV-complete, one should not worry about this when the IR properties are studied. The 't Hooft coupling $\lambda$ is introduced as
\be
\label{'tHooftconst}
\lambda = g_4^2 N_c\,.
\ee

To introduce flavor degrees of freedom into the game one immerses $N_f$ D8 and anti-D8 flavor branes into the geometry of D4~\cite{SakaiSugimoto2004}. In such a scenario strings that connect D4 and D8 (anti-D8) carry quark (anti-quark) degrees of freedom. In general solutions for the geometry of D4-D8 systems are not known: one typically resorts to the probe approximation, in which the backreaction of the flavor branes on the geometry of D4 is ignored. The latter regime is achieved when $N_f\ll N_c$.

In the setup of the Sakai-Sugimoto model the D8-branes span the Minkowski directions and wrap the $S^4$ in the transverse directions. The remaining dimension spans a 1D curve on the cigar $R^{+}\times S^{1}$. At zero temperature stable embeddings correspond to the D8-anti-D8 connecting smoothly in a U-shape configuration (figure~\ref{Ushape}). This is a geometrical realization of the chiral symmetry breaking. By default there are two gauge groups $U(N_f)_L$ and $U(N_f)_R$ which live on the D8 and anti-D8. However when the D8 and anti-D8 reconnect only the diagonal subgroup of the product survives:
$$U(N_f)_L\times U(N_f)_R\to U(N_f).$$

The U-shape embeddings are parameterized by the asymptotic separation of the D8-anti-D8 branes on the circle -- $L$, or equivalently by $u_0$ -- the value of the $u$-coordinate of the embedding closest to the origin (figure~\ref{Ushape}). For $u_0=u_\Lambda$ the configuration is called antipodal since the D8 and anti-D8 remain at the opposite points on the circle $L=\pi R$ for all values of $u$. Note that in general there is no parameter in real QCD analogous to this new scale $L$ ($u_0$).

The background configuration of the Sakai-Sugimoto model is confining. The low-energy degrees of freedom are glueballs (closed strings), mesons (strings ending on both D8 and anti-D8) and baryons ($N_c$ strings ending at a point in the bulk). If one puts the Sakai-Sugimoto model at finite temperature -- there is a deconfining phase transition~\cite{Aharony:2006da}. At finite temperature the Euclidean time direction is also compactified and the background corresponds to a black hole. When the temperature is increased to the critical value $T=1/2\pi R$ the geometry changes: the radial direction $R^+$ and the Euclidean time now have the topology of a cigar, while $R^+\times S^1$ becomes a cylinder. Such a swap corresponds to a first order phase transition. There is a number of ways to see that this is the deconfining phase transition, e.g. one can look at the quark-antiquark potential or scaling of the free energy with the number of colors $N_c$ in the two phases.

Depending on the value of $L$ one can have a chiral symmetry restoration either together with deconfinement or at some large value of the temperature: two transitions do not necessarily coincide. Geometrically chiral symmetry restored phase corresponds to the D8 and anti-D8 separating from each other. In the large temperature phase the D8 branes live on the cylinder rather than on the cigar. Since there is a horizon (minimal value of the $u$-coordinate on the thermal cigar), there is a natural place for the D8 to end. There exist two classes of embeddings: the U-shape and parallel D8-anti-D8-branes ending under the horizon. At high enough temperature the second class becomes favorable.

The low-energy dynamics of the flavored degrees of freedom is described by the effective action comprised of the Dirac-Born-Infeld (DBI) action and the Chern-Simons (CS) term:
\be
S=S_{\rm DBI} + S_{\rm CS}\,.
\ee
The DBI term reads\footnote{We reserve the lowercase Latin indices for the world-volume of the D8-branes, Greek indices will parameterize coordinates in the Minkowski space and holographic direction, while the uppercase Latin indices will label four spatial indices in the latter 5D space-time.}
\be
\label{DBIaction}
S_{\rm DBI}\ =\ T_8\!\int\limits_{\text{D}8+\overline{\text{D}8}}\! \d^9x\, e^{-\phi}
{\rm str\,}\left(\sqrt{-\det(g_{mn} + 2\pi\alpha' \CF_{mn})}\right),
\ee
where $T_p = (2\pi)^{-p}l_s^{-p-1}$ is the D8-brane's tension, $g_{mn}$ is the nine-dimensional induced metric on the branes. $\mathcal{F}_{mn}$ is the $U(N_f)$ gauge field strength on the world-volume of the D8-branes,
\be
\CF = \d \CA + \CA\wedge \CA\,,
\ee
and ``${\rm str}$'' is the symmetrized trace over the flavor indices. The CS term comes from the coupling of the D8-branes to the bulk RR field,
\be
\label{CSterm0}
S_{\rm CS} = T_8\!\int\limits_{\text{D}8+\overline{\text{D}8}} \! C_3\wedge \tr e^{2\pi\alpha' \CF}\,, \qquad \text{where}\qquad F_4=\d C_3.
\ee

In terms of the effective action low energy degrees of freedom correspond to fluctuations of the embedding profile and of the gauge fields. In the weak field approximation (weak compared to the scale set by $\lambda$ or equivalently curvature) one can restrict to the leading orders in powers of $\CF_{mn}$ in the action:
\be
S_{\rm DBI} = S_0[g] + S_{\rm YM}+ O\left(\CF^4\right),
\ee
where $S_0$ is the action with the gauge fields set to zero, while $S_{\rm YM}$ corresponds to the first non-trivial order in gauge fields. The low-energy modes are independent from the coordinates on $S^{4}$: one can integrate out this directions and work with an effective 5D theory. Since the gauge fields along the $S^4$ directions will not play any role in our latter discussion and will be switched off, the $S_{\rm YM}$ becomes a 5D $U(N_f)$ Yang-Mills theory with a coupling depending on the holographic coordinate. In this respect a more natural way to parameterize the holographic direction is through a new coordinate, for which the action simply takes the form
\be
\label{actionz0}
S_{\rm YM}\ =\ -\frac{R_{D4}^3u_\Lambda}{48\pi^4g_s l_s^5M_{KK}}\,\! \int\!\d^4x\,\d z\
\frac{1}{2g^2_{\rm YM}(z)}\, \tr\CF_{\mu\nu}^2\,,
\ee
where the indices are raised with $\eta^{\mu\nu}$. The 5D Yang-Mills coupling constant is given by the equation
\be
\frac{1}{g_{\rm YM}^2(z)} = M_{KK}\frac{u(z)}{u_\Lambda}\,, \qquad \text{where} \qquad \frac{\d u}{\d z}=\sqrt{\frac{u^8 f(u)-u_0^8 f(u_0)}{u^5 R_{D4}^3}}\,.
\ee
The values of $z$ run from $-\infty$ to $+\infty$, with positive (negative) branch corresponding to the D8 (anti-D8).\footnote{Note that the coordinate $z$ here is not the same as in case of the original work of Sakai and Sugimoto~\cite{SakaiSugimoto2004}. It rather coincides with $w$ introduced in~\cite{Kaplunovsky:2010eh}. In terms of $z$ coordinate all components of the 5D metric are the same.} The coefficient in front of the action is often denoted as $\kappa$. Using~(\ref{KKscale}) and~(\ref{'tHooftconst}) it can be written in terms of the 't Hooft coupling and $N_c$
\be
\kappa = \frac{\lambda N_c}{216\pi^3}\,.
\ee
For our analysis below there is no need to know the explicit form of $u(z)$. In fact all we need is the expansion of $u$ for small $z$
\be
\label{Uexpansion}
\frac{u}{u_\Lambda} = \zeta \left(1 + \frac{8\zeta^3-5}{9\zeta^2}\,M_{KK}^2z^2 + O\left(M_{KK}^4z^4\right)\right), \qquad \zeta = \frac{u_0}{u_\Lambda}\,.
\ee

Finally we are interested in the case of two flavors $N_f=2$. It is convenient to decompose the $U(2)$ field $\CA$ into $SU(2)$ and $U(1)$ parts
\be
\label{ActoAs}
\CA\ =\ A_{SU(2)}\ +\ \tfrac12\,\hat A_{U(1)}\,.
\ee
The effective action of the $N_f=2$ Sakai-Sugimoto reads
\be
\label{ActionInstanton0}
S[A,\hat{A}]=S_{\rm YM} + S_{\rm CS}\,,
\ee
where Yang-Mills action~(\ref{actionz0}) takes the following form in terms of the $U(1)$ and $SU(2)$ fields:
\be
\label{actionz1}
S_{\rm YM}\ =\
-\kappa \int \d^4x\,\d z\ \frac{1}{2g_{\rm YM}^2(z)}\left(\tr F_{\mu\nu}^2+\frac12\,\hat{F}_{\mu\nu}^2\right),
\ee

The CS term~(\ref{CSterm0}) integrated over the $S^4$ reduces to
\be
\label{CSterm1}
S_{\rm CS}\ =\ {N_c \over 24\pi^2} \int\!
\tr\left (\mathcal{A}\mathcal{F}^2-{i \over 2}\mathcal{A}^3\mathcal{F}-{1 \over 10}\mathcal{A}^5\right ) .
\ee
For $N_f=2$ the only terms that remain are
\be
\label{CSterm2}
S_{\rm CS} = \frac{N_c}{16\pi^2}\int \hat{A}\wedge {\rm tr} F^2 \ + \ \frac{N_c}{96\pi^2}\int \hat{A}\wedge \hat{F}^2\,.
\ee

The above derivation ignores backreaction of the gauge fields on the geometry of embedding. In fact non-trivial values of the gauge fields on the D8-branes may change the embedding, \emph{e.g.}~\cite{Bergman}. In the weak field approximation the backreaction can be taken into account by adding scalar fields describing transverse fluctuations of the brane profile. The effective action for these scalar fields was derived in~\cite{Kaplunovsky:2010eh}. In particular it was shown that the leading order contribution comes from an isosinglet scalar field $\Phi$. Introduction of the scalar merely results in a renormalization of the effects induced by the abelian field $\hat{A}$. But before discussing these effects in detail let us remind how this effective action is linked to baryons.


\subsection{Baryons in the Sakai-Sugimoto model}
\label{secSSBaryons}

To introduce baryons in holography one has to have an object, where $N_c$ quark strings can end -- the baryon vertex (b.v.)~\cite{WittenBaryons}. One way to construct it is to consider D$p$-branes wrapped on compact $p$-cycles with $N_c$ units of flux. In the above setup this works as follows. There are $N_c$ units of the $F_4$-flux through $S^4$,
\be
\frac{1}{(2\pi)^3l_s^3}\int_{S^4} F_4 = N_c\,.
\ee
Introduce a D4-brane wrapped on $S^4$. The D4 is coupled to the RR field. As a result it picks $N_c$ units of charge from the $U(1)$ gauge field $B$ on its world-volume:
\be
T_4\int_{\rm D4} C_3\wedge e^{2\pi\alpha'\d B} = N_c \int B\,.
\ee
Due to the charge conservation the net charge in a closed space must vanish. To compensate for $N_c$ units of charge the same number of strings must end on the wrapped D4, since each of them carries precisely one unit of charge. This is an appropriate extension of a b.v. construction proposed by Witten for $AdS_5\times S^5$~\cite{WittenBaryons} and its generalization to confining backgrounds~\cite{Brandhuber:1998xy}.

In the non-supersymmetric background of Sakai and Sugimoto the strings pull the b.v. towards the D8-branes. For all values of $u_0$ ($L$, figure~\ref{Ushape}) the minimum energy configuration corresponds to the b.v. sitting precisely at $u_0$~\cite{Seki:2008mu}. In this case the b.v. ``dissolves'' in the D8-branes and can be described by an instanton -- $SU(N_f)$ flavor gauge field configuration in their world-volume~\cite{Dbrane-instanton}.

It is clear from the Chern-Simons action~(\ref{CSterm2}) that the $U(1)$ field $\hat{A}$ is sourced by the topological density of instantons. Indeed since we deal with the vector-like $U(1)$ of the product $U(N_f)_L\times U(N_f)_R$ spontaneously broken by the U-shape configuration, the bulk field $\hat{A}$ sources the baryon number current in the gauge theory on the boundary. According to the standard lore the subleading asymptotic in the $z\to\infty$ expansion of $\hat{A}$ corresponds to the vev of the dual operator, baryon charge density $\vev{\rho_B}$, and the leading one is the source, i.e. the baryon chemical potential $\mu_B$. Strings ending on the flavor branes (quarks) carry a unit charge under this $U(1)$. For a static $N$-instanton configuration of the $SU(2)$ fields one obtains from the first term in~(\ref{CSterm2}):
\be
\label{BaryonCharge}
 \frac{1}{2}\,N N_c\int \hat{A_0}\,.
\ee
Since the baryon number is $1/N_c$ times the $U(1)$ charge, for the particular normalization of the $U(1)$ generator in~(\ref{ActoAs}) it is a half of the instanton number $N$.

We will be interested in static baryon configurations, that is the instanton-like solutions for the non-abelian fields $A_\mu$ on the flavor branes. To find equilibrium configurations we must compute the free energy on a generic instanton solution and minimize it with respect to parameters, such as position, size and orientation. It turns out that the natural size for the instantons is set by the scale $1/M_{KK}\sqrt{\lambda}$. Although full instanton solutions for the curved space action~(\ref{ActionInstanton0}) are not known, the above scaling of the instanton radius allows to consistently expand the action in powers of $1/\lambda$ as first analyzed by Hata \emph{et al} in the work~\cite{Hata:2007mb}. The leading $O(\lambda)$ term in the full action~(\ref{ActionInstanton0}) is simply the flat-space 5D Yang-Mills action, $g_{\rm YM}(z)= \text{const}$. The Chern-Simons term is explicitly $O(\lambda^0)$, while $(M_{KK}z)^{2n}$ curvature corrections to $1/g_{\rm YM}^2$ will be of order $O(\lambda^{1-n})$. Indeed the $n^{\underline{\rm th}}$  moment of the instanton density should scale as $a^n$, where $a$ is the radius.

Thus the only $O(\lambda)$ term in the action~(\ref{ActionInstanton0}) is the $SU(N_f)$ Yang-Mills term with constant coupling. If the $O(1)$ terms in the action are ignored, static baryon is a BPS (BPST) instanton.\footnote{It can in fact be shown that the zero-size (anti-) self-dual gauge fields localized at $z=0$ minimize the full DBI action~\cite{SakaiSugimoto2004,Hata:2007mb,DKS,Kaplunovsky:2010eh}.} The instanton configuration is given by four $A_{M}$ components of the 5D field, while $A_0$ and $\hat{A}_\mu$ vanish. Parameters of the BPS instanton solution are free moduli. In other words the energy of a configuration does not depend on the values of the instanton parameters at this order. At the $O(1)$ order the interaction terms turn on and (partially) fix the moduli. In particular the instanton density sources a notrivial solution for $\hat{A}_0$, while $A_0$ and $\hat{A}_M$ can still be kept trivial. Notice that there is no need to compute corrections to the instanton solution $A_{M}$ at the next-to-leading order. Indeed corrections to the equations of motion due to $\delta A_{M}$ will be proportional to the variation of the action on the solution itself, which vanishes identically. Such corrections will only affect $O(\lambda^{-1})$ terms in the action. Therefore up to the $O(1)$ order the $A_M$ fields can be considered as self-dual.

The $O(1)$ equation for $\hat{A}_0$ reads
\be
\label{A0eq0}
\Box \hat{A}_0 = - \, \frac{N_c}{2\kappa\,\zeta M_{KK}}\, I(x)\,,
\ee
where the Laplacian is taken in the 4D space of $x^M$ and we have introduced the topological density
\be
I(x)= \frac{1}{32\pi^2}\,\epsilon_{MNPQ}\tr F^{MN}F^{PQ}\,.
\ee

As can be inferred from~\cite{Kaplunovsky:2010eh} the equation for the isosinglet scalar field $\Phi$, responsible for transverse fluctuations of the D8-branes, satisfies a similar equation to~(\ref{A0eq0}), except that the scalar couples to $\tr F_{MN}^2$ instead of the topological density. However according to the above remark at the $O(1)$ order in $1/\lambda$ the instanton solution is approximately self-dual. Therefore at this order $\Phi$ satisfies the same equation as $\hat{A}_0$ up to a numerical coefficient. Using~(\ref{A0eq0}) one can derive the following expression for the baryon free energy at the $O(1)$ order:
\begin{multline}
\label{EnergyInstanton0}
E[\text{moduli}] = \,  \kappa\,\zeta M_{KK}\int \d^3x\,\d z\ \left(1 + \frac{8\zeta^3-5}{9\zeta^2}\,M_{KK}^2z^2 + O\left(M_{KK}^4z^4\right)\right)\frac12\,\tr F_{MN}^2 \, +
\\ + \,\frac{N_c C}{4}\int \d^3x\,\d z\ \hat{A}_0\times I(x)  \
 + \ O(1/\lambda)\,,
\end{multline}
where the integrals are evaluated on the $O(\lambda)$ order flat space solutions, which define $\hat{A}_0$ through equation~(\ref{A0eq0}). Coefficient $C$ takes into account contribution of the scalar field. Since the scalar satisfies equation similar to~(\ref{A0eq0}) it merely renormalizes the instanton self interaction term
\be
C = 1 - \frac{1-\zeta^{-3}}{9}\,.
\ee
For the antipodal configuration of the D8-branes ($\zeta=1$) scalar fields have no effect on the profile and the scalar field decouples. Conversely the minimum value $C$ can take is 8/9, when $\zeta\to\infty$. In other words for any $\zeta$ the coefficient $C$ is positive. This means that in the Sakai-Sugimoto model the short distance interaction between baryons is always repulsive, never mind the value of $\zeta$~\cite{Kaplunovsky:2010eh}. At large distances the sign of the interaction is determined by the exponentially decaying Yukawa tails. Since the scalar meson mediating the attraction is always heavier than the vector in the Sakai-Sugimoto model, the net baryon force is repulsive also at large distances.\footnote{One may also recall that massless pions are also present in the Sakai-Sugimoto model. Therefore they should dominate the long-distance force. However in the bulk nuclear matter their contribution is expected to largely average to zero and isoscalar exchange is considered dominating. We thank the referee for bringing this issue to our attension.}

Expression~(\ref{EnergyInstanton0}) can be recast in a nicer form through a rescaling of parameters:
\be
\label{MlRescaling}
\frac{8\zeta^3-5}{9\zeta^2}\,M_{KK}^2 \to M^2\,, \qquad \frac{\lambda\zeta^2}{9\pi\sqrt{8\zeta^3-5}}\to \lambda \sqrt{C}\,, \qquad \hat{A}_0\to \frac{\hat{A}_0}{\sqrt{C}}\,.
\ee
Upon this rescaling coefficient $\sqrt{C}$ becomes an overall multiplicative factor in the expression for energy and will be dropped from here on. Equation~(\ref{EnergyInstanton0}) then takes form
\begin{align}
\label{ActionInstanton}
E[\text{moduli}] \ & = \ E_{\rm NA}+ E_{\rm C}=   \ \int \d^3x\,\d z\ \frac{8\pi^2}{g_{\rm YM}^2}\,I(x) \ + \ \,\frac{N_c}{4}\int \d^3x\,\d z\ \hat{A}_0\times I(x)  \ + \ O(\lambda^{-1})\,, \\
\label{5dcoupling}
\text{where}\quad \frac{8\pi^2}{g^2_{\rm YM}}\ & =\ N_c\lambda M\left(1+M^2z^2\right)\,\\
\label{A0eq2}
\text{and}\quad \square\hat A_0\ & =\ -\frac{4\pi^2}{\lambda M}\, I(x)\,.
\end{align}
Equation~(\ref{ActionInstanton}) will be the principle formula for our study of baryon systems below. We will refer to the first integral in~(\ref{ActionInstanton}) as to ``non-abelian'' term, while the second term will be called ``Coulomb'' or ``abelian''. After rescaling~(\ref{MlRescaling}) the new expression for the free energy is quite universal. The rescaling hides model-dependent coefficients in the definition of $\lambda$ and $M$. Therefore we expect~(\ref{ActionInstanton}) to be correct beyond the Sakai-Sugimoto model. Note also that this expression assume net repulsion between baryons (instantons). The adjustment for attracting baryons can be achieved \emph{e.g.} by changing the sign in~(\ref{A0eq2}).

Notice that the expression for the free energy depends on moduli of the instanton solution. This dependence shows up at $O(1)$ order due to the curvature correction and Coulomb term. These terms stabilize instanton moduli. For example in a single instanton case the energy will depend on the $z$-locus of the instanton and its size. Curvature terms will prevent instanton from being large and pin it to the bottom of the brane configuration $z=0$. On the other hand Coulomb self-interaction dislikes small charges and will stabilize instanton size at a finite value.

Apart from the two above parameters, the full moduli space of the flat-space static instanton also contains the coordinates in the remaining three spatial dimensions and an $SU(2)$ orientation up to an overall sign. For a single instanton these are the true moduli even when the NLO corrections to the energy are taken into account. Upon quantization the $SO(4)$ rotational symmetry of the orientation moduli space gives rise to the spin and isospin quantum numbers, which are locked together $I=J$. Fermionic baryons have half-integer spin (isospin). Spectrum of light baryons was studied in~\cite{Hata:2007mb} and~\cite{Seki:2008mu}.

In this work we will consider classical baryons with a fixed isospin orientation, which are neither fermions nor bosons. We will ignore the effects of quantum spin and isospin interactions. Let us use equation~(\ref{ActionInstanton}) to calculate the mass and the size of a single instanton. Evaluating~(\ref{ActionInstanton}) on the single instanton solution localized at $z=0$ one finds
\be
E[a]=N_c\lambda M \left(1 + \frac{a^2M^2}{2}\right)  + \frac{N_c}{5\lambda M a^2}\ + \ O(\lambda^{-1})\,,
\ee
where $a$ is the size of the instanton. Minimizing with respect to $a$ one finds the equilibrium radius of a standalone instanton:
\be
a_0\ =\ \frac{(2/5)^{1/4}}{M\sqrt{\lambda}}\,, \qquad \left(\text{or }\frac{9\pi^{1/2}}{M_{KK}\sqrt{\lambda}}\left(\frac{2C}{40\zeta^3-25}\right)^{1/4} \text{ for original }\lambda, M_{KK}\,.\right)
\label{OrigA0}
\ee
The mass is given by the non-abelian energy. Substituting $a_0$ for $a$ gives
\be
\label{mass}
m_B = N_c\lambda M \left(1 + \frac{1}{\sqrt{10}\lambda}\right)  + \ O(\lambda^{-1})\,, \qquad \left(\text{or }\frac{\lambda N_c\zeta M_{KK}}{27\pi\sqrt{C}}\left(1+\frac{9\pi\sqrt{C(8\zeta^3-5)}}{\zeta^2\sqrt{10}}\right) \,.\right)
\ee
As expected the mass of the baryon is proportional to $N_c\Lambda$: baryons are heavy in the large $N_c$ limit. In holography they are even heavier because of large $\lambda$.

Now it is clear that the curvature correction to the flat space Yang-Mills action as well as the Chern-Simons term give $O(\lambda^0)$ contribution. Indeed the $(zM)^{2n}$ corrections should be proportional to $\lambda a_0^{2n}\propto \lambda^{1-n}$, because $M$ effectively is the only scale in the problem. The instanton topological density sources the abelian field $\hat{A}_0$ in~(\ref{ActionInstanton0}). From the equation of motion for $\hat{A}_0$~(\ref{A0eq2}) it is obvious that the abelian field is subleading to the instanton solution. Therefore apart from the $O(\lambda)$ Yang-Mills term in~(\ref{actionz1}) we must only keep the flat space Maxwell action, which is $O(1)$, and drop the second term in the Chern-Simons action~(\ref{CSterm2}), which is only $O(\lambda^{-1})$.

An important comment is in order. The size of the real-life baryons is approximately $a_B\sim 4R_{\rm Yukawa}$, where $R_{\rm Yukawa}$ must correspond to $M_{KK}^{-1}$. In the large $\lambda$ limit the size acquires additional $1/\sqrt{\lambda}$ suppression~(\ref{OrigA0}). This signifies that in general it is not sufficient to just rely on the low energy DBI action~(\ref{DBIaction}). Stringy corrections including derivative terms may be equally important. Nevertheless we hope that the higher order corrections will not affect any of the conclusions below, but rather renormalize the exact values of physical quantities.

To summarize the strategy for studying static classical baryons is as follows. First one has to find a flat space instanton solution relevant for the problem. In general this is done with the help of ADHM construction~\cite{ADHM}. A particular formulation of the ADHM construction convenient for baryon study will be discussed in section~\ref{secChain}. Next one has to solve equation~(\ref{A0eq2}) for $\hat{A}_0$ and evaluate the energy~(\ref{ActionInstanton}) as a function of instanton moduli. The energy then needs to be minimized with respect to moduli to find stable configurations.

In the applications below we will be considering toy models of one dimensional baryon configurations. We will typically choose to label the direction along the configuration as $x_4$, while the transverse directions will be labeled $x_1$, $x_2$ and $x_3\equiv z$. To stabilize 1D configurations it will be useful to introduce an \emph{ad hoc} curvature in all dimensions transverse to $x_4$. In particular we will consider two situations: an isotropic transverse curvature, for which the curvature is the same in all transverse directions,
\be
\label{GRdependence}
\frac{8\pi^2}{g_{\rm YM}^2} \ = N_c\lambda M\left(1+ M^2 (x_1^2+x_2^2+x_3^2)\right);
\ee
and the anisotropic one
\be
\label{anisotropic0}
\frac{8\pi^2}{g_{\rm YM}^2} \ = N_c\lambda M\left(1 + M^2 x_3^2 + {M'}^2 (x_1^2+x_2^2)\right).
\ee
Notice that equations~(\ref{OrigA0}) and~(\ref{mass}) will be modified for the above choices. In particular, the equilibrium instanton size will be
\begin{align}
\label{OrigA}
\text{isotropic}\quad a_0\ & =\ \frac{(2/15)^{1/4}}{M\sqrt{\lambda}}\,,\\
\label{AnisRad}
\text{anisotropic}\quad a_0\ & =\ \frac{(1/5)^{1/4}}{\sqrt{\lambda MM'}}\,,
\end{align}
where in the latter case we have also assumed high anisotropy $M'\gg M$ .


\subsection{ Baryons as instantons in other holographic models}
\label{secOtherModels}

As was discussed in the previous section holographic nuclear physics based on the generalized Sakai Sugimoto model suffers from  several drawbacks. The most important ones are: (i) The size of the baryon scales as ${1}/{\sqrt{\lambda}}$~(\ref{OrigA0}) and hence in the large $\lambda$ limit stringy corrections cannot be faithfully neglected. (ii) Due to the fact that the lightest scalar particle that couples to the baryon is heavier than the lightest vector meson the interaction between two baryons is dominated in the far zone by repulsion. Thus no nuclei bound states of nucleons will be stable in this model.

This situation naturally raises the issue  of constructing other holographic laboratories for nucleons and  nuclear interactions. Such holographic models should be dual to a boundary field theory that admits confinement and chiral symmetry which is spontaneously broken. An additional constraint on the construction is that the corresponding baryonic vertices will be immersed in the flavor branes such that the baryons can be described as instantons of the flavor gauge symmetry. Geometrical realization of the spontaneous breaking of flavor chiral symmetry can be achieved, in a similar manner to the Sakai Sugimoto model, by incorporating a  U shape structure for a stack of $N_f$ branes and anti-branes (see figure~\ref{Ushape}). Indeed such a construction was implemented in other models and in particular in the  non-critical six dimensional holographic model~\cite{Kuperstein:2004yf} as well as in the model based on introducing D7 anti-D7-branes into the KS model which is described in~\cite{DKS} and~\cite{Dymarsky:2010ci}\footnote{We do not attempt to review here other holography inspired models, where baryon physics is intensively studied, \emph{e.g.}~\cite{Nawa,Pomarol}. In those bottom-up models there is no $\lambda$, it is rather explicitly or implicitly set to something of order one. Thus the problem of large $\lambda$ is replaced by a lack of microscopic or top-down justification of the models.}. Let us now describe some features of the models and of the corresponding instantons as baryons in those models.

\subsubsection {The six dimensional non-critical model}
The non-critical holographic model of~\cite{Seki:2008mu} is based on six dimensional background associated with large $N_c$ D4-branes. By compactifying one world-volume coordinate of the D4-branes, the background is rendered into a confining one. One then incorporates $N_f$ flavor D4 and anti-D4-branes which are perpendicular to the ``colored" D4 branes. It is well known that this model, similar to other non-critical models suffers from the problem that the curvature of the background cannot be made small by taking the large  $N_c$, and it is of order unity.

In the absence of compactified extra dimensions (in addition to the circle mentioned above), the baryonic vertex in this model is not a wrapped brane but rather simply a D0-brane. One can argue that similarly to~(\ref{BaryonCharge}) the action on the worldline of the D0 brane has a CS term that forces the baryonic vertex to have $N_c$ strings attached to it and connecting it to the flavor branes. Due to the fact that the model is at $\lambda \sim 1$, the problem of the stringy scale baryon radius does not occur for this model as was shown in~\cite{Seki:2008mu}. Non-critical models also correctly predict certain model-independent baryon properties of large $N_c$ QCD~\cite{Cherman:2009gb}.\footnote{In the earlier versions of this paper, citing~\cite{Cherman:2009gb}, we mentioned some issues with the derivation of the model-independent properties in critical string models. This issue was later resolveld in~\cite{Cherman:2011ve}. We thank the referee for communicating this result.}

The disadvantage of having 't Hooft parameter of order one is in fact advantageous also for the meson spectrum. Generically in critical holography the masses of the scalar and vector mesons scale like the scale of the system ($M\sim 1/R$ where $R$ is the radius of the compact direction) and the masses of the higher spin mesons, which are described by semi-classical stringy configurations scale as $\sqrt{T}\sim M\sqrt{\lambda}$. This implies that there is a huge gap between the low and high spin mesons. In the non-critical holographic model on the other hand since $\lambda\sim 1$ there is no such a gap. This feature may be an asset also in terms of the baryon interaction. If the mesons that couple to the baryon are stringy mesons that reside on a Regge trajectory, than indeed the lightest scalar meson is lighter than the lightest vector meson which will guarantee the domination of the attraction over the repulsion in the far zone.

\subsubsection{The DKS model}
This critical model is based on incorporating  D7 and anti-D7 flavor branes in the Klebanov-Strassler background~\cite{KS}. It was proposed by Dymarsky, Kuperstein and Sonnenschein (DKS) in~\cite{DKS} (see also an earlier realization~\cite{Kuperstein:KW} for the Klebanov-Witten background~\cite{KW}). The D7-branes admit a U-shape profile whereby the spontaneous breaking of the chiral flavor symmetry is geometrically realized.

{
\begin{figure}[htb]
\begin{minipage}[b]{0.5\linewidth}
\begin{center}

\includegraphics[width=6.cm]{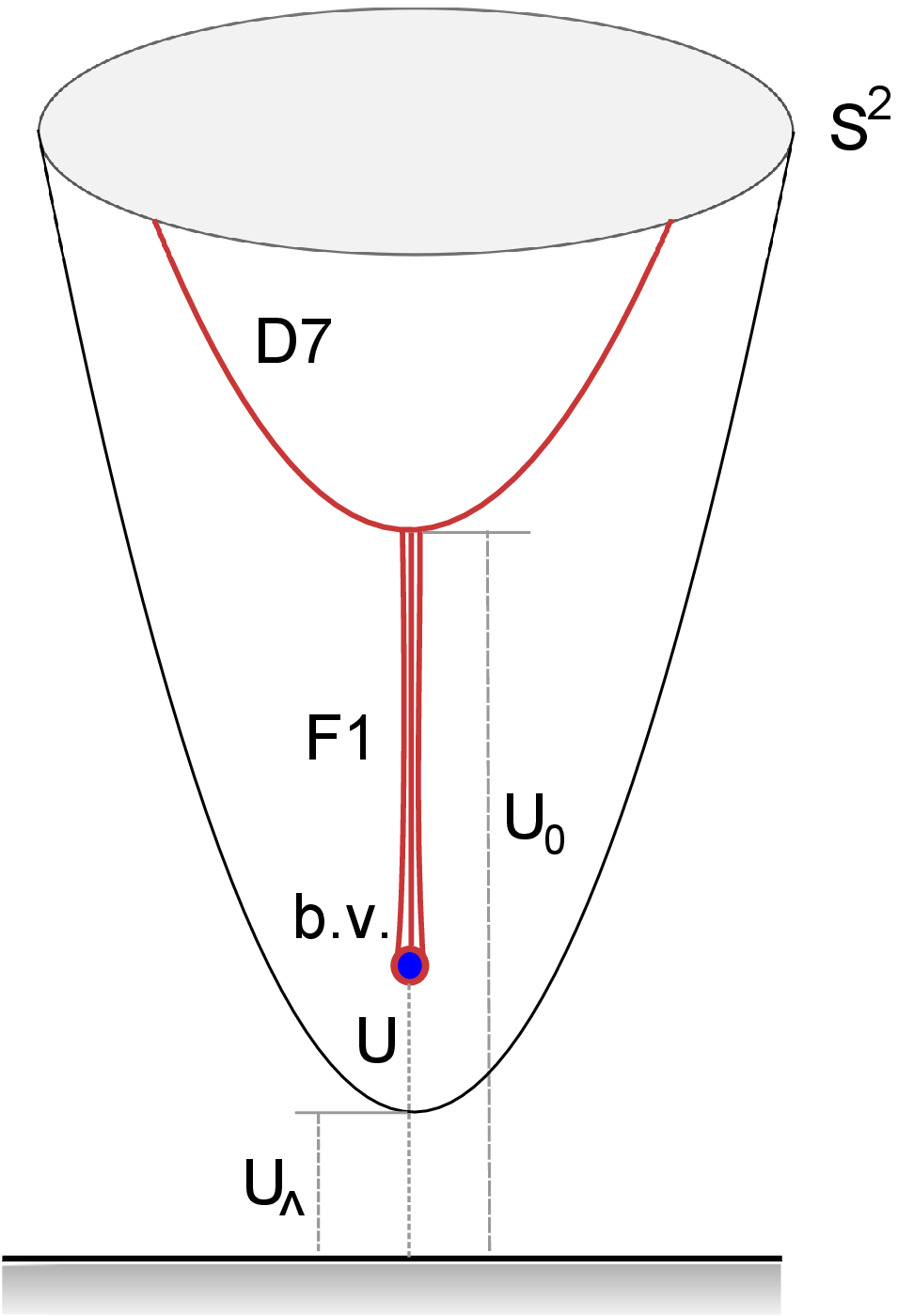}

(a)
\end{center}
\end{minipage}
\begin{minipage}[b]{0.5\linewidth}
\begin{center}

\includegraphics[width=6.cm]{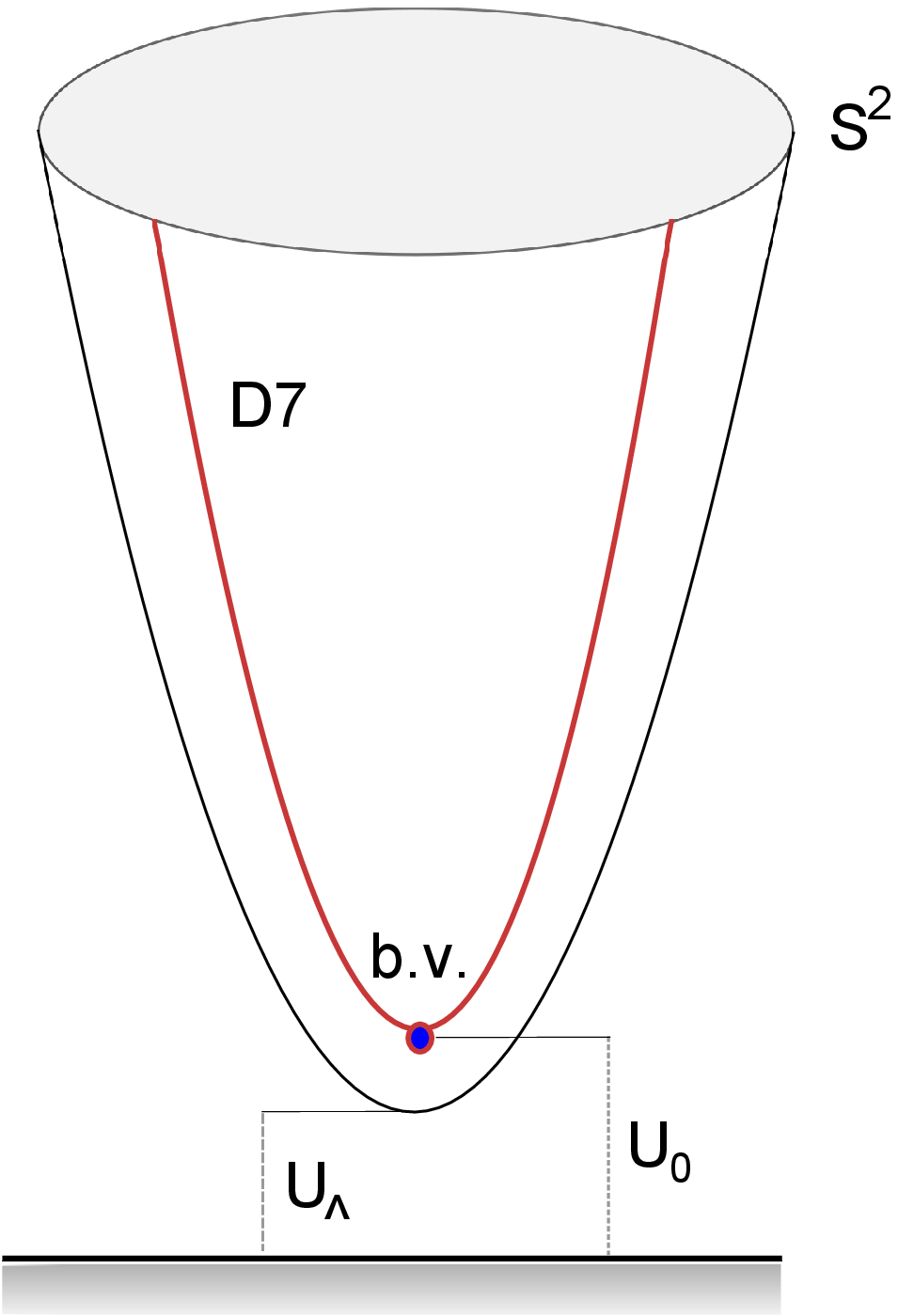}

(b)
\end{center}
\end{minipage}
\vspace{-0.6cm}
\caption{\small Baryon in the DKS model. For highly non-antipodal embedding $U_0\gg U_\Lambda$ baryon is represented by the baryon vertex and $N_c$ strings stretching towards the D7-branes~(a) For nearly-antipodal $U_0\simeq U_\Lambda$ configuration baryon vertex ``dissolves'' in the D7-branes and can be described as a world-volume instanton configuration of flavor fields~(b).}
\label{fig svi}
\end{figure}}

A baryon in this model  consists of a baryonic vertex in the form of a D3-brane wrapping the $S^3$ of the base of the conifold with $N_c$ strings connecting it to the flavor branes. Depending on the way the U shape flavor branes are embedded in the background, there are two possibilities for the location of the the baryonic vertex~\cite{Dymarsky:2010ci}. When the tip of the U-shape is very far from the tip of the conifold the baryonic vertex will approach the tip of the cone, namely, will be separated from the flavor branes, whereas when the tip of the U-shape is close to the tip of the conifold the baryonic vertex dissolves in the flavor branes (see figure~\ref{fig svi}). In this latter case one can describe  baryons in terms of flavor instantons in a similar manner that was described above for the generalized Sakai-Sugimoto model.

The leading order interaction between the baryons in the far zone where the separation distance is much larger than the inverse of the basic mass scale (the analog of $M_{KK}$) is determined via a competition between the attraction
and the repulsion due to the exchange of isoscalar scalar mesons $0^{++}$ and isoscalar vector mesons $1^{--}$ respectively. The unique property of the DKS model is that the lightest $0^{++}$ particle is in fact a
pseudo-Goldstone boson associated with the spontaneous breaking of scale invariance. In a certain range of parameters this meson  is parametrically lighter than any other massive state. As a result  unlike the Sakai-Sugimoto model, in the DKS model the baryons attract each other in the far zone. At short distances the potential admits a repulsive hardcore.

Thus we see that the two major problems of the Sakai Sugimoto model, namely the size of the baryon and its interactions, can be bypassed in other models. The former in the non-critical model and the latter in the DKS model. It is still to be seen if the non-critical model admits the desired property of the domination of the attraction over the repulsion. Alternatively, a new  consistent holographic model where the two problems are been solved simultaneously is obviously very welcome.

In the following sections we discuss the structure of the holographic nuclear matter based on determining the lowest energy configurations of the multi instanton solutions. It is important to emphasize that, though the analysis was based on the Sakai Sugimoto as a prototype model, it could be adopted also for the cases of the non-critical and the DKS model and in fact to any holographic model in which flavor instantons play the role of baryons.


\section{Baryonic popcorn}
\label{secPopcorn}

In this section we turn on finite density for holographic baryons. Due to repulsive interaction between baryons it is natural to expect that at large density the condition $z=0$ (location at the tip of the U shape) for the baryons may no longer minimize the energy. In other words one can expect that densely packed baryons will overcome the gravity force, pinning them down to the tip, and expand into the holographic dimension. Such effect was observed in~\cite{Rozali:2007rx}, where a uniform distribution of instantons was assumed. Formation of a macroscopic layer of baryons in the holographic dimension was identified with a developing Fermi sea of quarks. Uniform density of baryons may indeed be a good approximation at large enough densities, when matter is expected to be quarkyonic. Instead we consider discrete instanton solutions, which should be a better description for the nuclear matter phase in the vicinity of phase transition. It is in this regime of relatively small densities we expect to find the onset of the quarkyonic phase.


\subsection{Multi-baryon systems}
\label{secMultiBaryons}

Sakai-Sugimoto model at finite baryon density was originally considered in~\cite{Kim:2006gp}. For the analysis of properties of dense hadronic matter Kim, Sin and Zahed employed the fact that the asymptotic value of the time component $\hat{A}_0$ of the $U(1)$ gauge field on the world-volume of the D8-branes corresponds to the baryon chemical potential in the boundary gauge theory. Phase diagram of large $N_c$ QCD was later studied in~\cite{Horigome:2006xu,Yamada:2007,Bergman,Rozali:2007rx} using Sakai-Sugimoto model as a prototype. For other holographic models with finite baryon density see \emph{e.g.}~\cite{FiniteDensity}.

\begin{figure}[t]
\begin{center}
\vspace{3ex}
\includegraphics[width= 100mm]{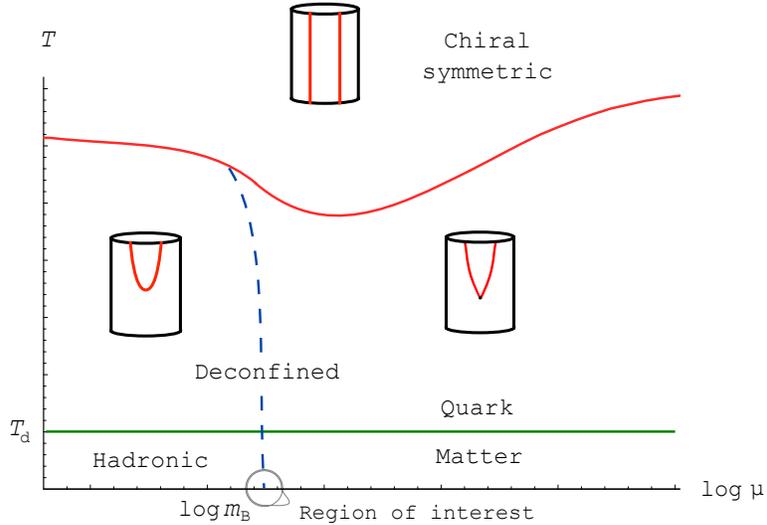}
\end{center}
\caption{Phase diagram of the holographic QCD as seen by~\cite{Bergman}. The figure demonstrates flavor brane configurations in different phases.}
\label{fighQCDPhase}
\end{figure}

Phase diagram of Sakai-Sugimoto model at finite temperature and density was summarized by Bergman \emph{et al.} in~\cite{Bergman}. In accordance with the review in section~\ref{secReview} phase diagram of the holographic analog of QCD has the following features. At low temperature and chemical potential the phase is confined with broken chiral symmetry. For the values of chemical potential up until the mass of lightest baryons $m_B$ the thermal density of baryons is exponentially suppressed due to their large mass (proportional to $\lambda N_c$). This is the hadronic phase with no baryons, but with a finite density of mesons and glueballs (lower left corner of the diagram in figure~\ref{fighQCDPhase}).

At high temperature holographic QCD is in a deconfined phase with restored chiral symmetry. Chiral symmetry restoration is realized through a separation of D8- and anti-D8-branes, which is favorable for large enough temperature and both  zero~\cite{Aharony:2006da} and finite chemical potential~\cite{Horigome:2006xu}. For moderate chemical potential critical temperature of chiral symmetry restoring transition is in general a function of temperature, while the critical temperature of the deconfining transition is temperature independent in both conventional large $N_c$ and holographic models. Therefore there might exist an intermediate phase, in which the matter is deconfined, but the symmetry is broken. This indeed happens for sufficiently small values of asymptotic separation $L$ between the D8-branes as in the case shown in figure~\ref{fighQCDPhase}. Both deconfining and chiral symmetry restoring phase transitions are first order. As noticed by the authors of~\cite{Bergman}, in the situation presented in figure~\ref{fighQCDPhase} the shape of the chiral symmetry restoring curve has a peculiar form: it first dips for $\mu<m_B$; it reaches a minimum for some $\mu>m_B$ and then raises again. In other words the chiral condensate first decreases when $\mu$ is increased and then increases again, which is perhaps a similar behavior to some limits of QCD, e.g. figure~\ref{figMasslessQCD}(a).

For the values of chemical potential $\mu\gtrsim m_B$ in the chiral symmetry broken phase there is a non-zero baryon density. As we will explain in the next section this phase should be some kind of quark matter, since the density is always large enough for baryons to overlap. If the baryon density is approximated by a uniform distribution as in~\cite{Bergman}, there is a direct transition from no-baryon hadronic matter to the quark matter. The transition would be second order with zero critical baryon density, as shown by figure~\ref{figAttractiveRepulsive}(b). This is consistent with the observation that the net baryon-baryon force is repulsive in this model. At zero temperature the critical chemical potential $\mu_c=m_B$, while if the temperature is raised the value of $\mu_c$ will slightly decrease.

In the finite baryon density phase the geometry of the D8 embedding is modified as described by~\cite{Bergman}: the brane profile changes from the U-shape to a V-shape as illustrated by figure~\ref{fighQCDPhase}. The new geometry has a cusp at the lowest point of the profile. The reason for this modification is the coupling of the baryons to scalar fields, mentioned in the previous section. Fluctuations of the scalar fields parameterize transverse displacements of the D8-brane profile. In the antipodal case however, only U-shape configuration is possible. Indeed in this case the coupling of the scalars to the baryons is forbidden by an additional symmetry and thus vanishes, no matter what the baryon density is.

Studying finite density baryon systems along the lines sketched in the previous section is challenging. For generic flat space multi-instanton solution one has to resort to the ADHM construction and solve infinite dimensional (in principle) matrix equations, not to mention instanton solutions in curved space, which are not known so far. Bergman \emph{et al.} approximated instanton density by a uniform distribution in three spatial dimensions and a delta-function localization at the bottom of the profile $z=0$. However as we see in the analysis of the previous section, baryon positions are determined by the gravity potential, which prefers to have them at $z=0$, and the Coulomb-like repulsion of individual baryons. If the baryons are very densely packed, it is conceivable that the Coulomb repulsion will overcome the gravity force and push the baryon density out to finite values of $z$. In this respect a more appropriate approximation was considered by Rozali \emph{et al}. In the work~\cite{Rozali:2007rx} they also considered instanton density as uniform in spatial dimensions, but the distribution in the holographic dimension was determined dynamically. A non-trivial distribution of density was discovered, with a sharp edge at some finite value of $z$. It was shown that the expansion into the holographic direction goes along with the increase of chemical potential. Rozali \emph{et al} have interpreted the finite density in the holographic direction as the Fermi sea with the sharp edge -- Fermi surface. The Fermi sea must be a Fermi sea of quarks, because in this description baryons are neither fermions, nor bosons.

Although uniform density is an approximation it is likely an appropriate description of the quark phases, when the densities are high. One can expect that in such a case we deal with a weakly interacting quark fluid. For smaller densities, such that the size of the baryon $a$ is much smaller than the average inter-baryon distance, a better approximation may be required. Recall that in the large $N_c$ limit and chemical potential $\mu\sim m_B$ baryons should form a crystal, see explanation in section~\ref{secLargeNQCD}. Therefore a better description at small densities should be provided by periodic arrays (lattices) of instantons. We will provide such a description in terms of the generalized Sakai-Sugimoto model summarized in section~\ref{secHBaryons} and investigate the phase space in the intermediate regime of small densities. We will probe a small window of values of the chemical potential around $\mu=m_B$, where very interesting physics resides. In particular, in this window we can observe transitions to structures with finite width in the holographic dimension, which we interpret as an onset of the quarkyonic phase (the terminology was explained in section~\ref{secLargeNQCD}). In this respect we will supplement the phase diagram in figure~\ref{fighQCDPhase} by zooming on the values of $\mu\sim m_B$.

We also expect that by looking at baryon densities for which it make sense to talk about lattices  we should be able to establish connections to the results on finite density Skyrme model reviewed in section~\ref{secSkyrmions}. Recall that skyrmion lattices exhibit transitions to the new ones with half-skyrmion symmetries. Since there is a natural relation between holographic baryons and skyrmions, one can expect that the features of the skyrmion lattices should be inherited by the lattices of holographic baryons. So what are the skyrmion-half skyrmion transitions in terms of instantons? To answer this question in principle one can use the same approach as used below, but this will involve solving for a 3D instanton lattices, which is beyond the scope of the present work. A qualitative explanation was proposed by Rho, Sin and Zahed in~\cite{Rho:2009ym}. Let us briefly summarize the idea.

As will be clear from section~\ref{secChain} a periodic 1D chain of instantons on $R^4$ with the same orientation twist between all nearest neighbors, or equivalently a single instanton on $R^3\times S^1$ with a non-trivial holonomy around $S^1$ is similar to a pair of 3D BPS monopoles if the instanton separation (radius of $S^1$) is small enough~\cite{Lee:1997vp}. For the twist angle $\phi=\pi$ the monopoles have equal mass and opposite electric and magnetic charges. The separation between monopoles depends on the radius of $S^1$. When the radius is small (high density) the topological density is independent from the coordinate on $S^1$, but localized in two well separated points (monopole centers) in $R^3$. For the values of the $S^1$ radius comparable to the instanton size and larger the topological density is localized at one point and does not look as 3D solution any more.

The authors of~\cite{Rho:2009ym} have conjectured that the high density splitting of instantons into monopoles is akin to the skyrmion-half-skyrmion transition. This conjecture is qualitatively supported by numerical studies of skyrmion chains in~\cite{Harland}: at large densities the chain becomes thicker in the transverse directions exhibiting skyrmion constituents with fractional topological charge.  Some details however remain unclear. In particular for a 1D chain the instanton-monopole transition is a smooth crossover, while for skyrmions it is expected to be a second order transition. We will also demonstrate below that the monopole separation does not depend on the baryon density in the same way as for half-skyrmions. Perhaps these issues will be naturally resolved in the case of 3D instanton lattices, but the latter have yet to be analyzed.


\subsection{Lattices of holographic baryons}
\label{secOutline}

Let us now explain in more detail why we are interested in the small range of values of chemical potential around the critical value $\mu=m_B$ and what we expect to find there. As we briefly mentioned in the end of section~\ref{secLargeNQCD}, it is important that we are working in the regime of large $\lambda$. In this regime the mass of the baryon is $O(\lambda)$ while the baryon interactions are $O(1)$. One consequence of this fact is that the natural scale for the baryon physics is proportional to $1/\sqrt{\lambda}$ in units of $1/M_\text{KK}$. For example, such is the scaling of the baryon radius as we saw in section~\ref{secSSBaryons}. As we are going to see later $1/\sqrt{\lambda}$ is also a characteristic scale for the equilibrium lattice spacing in the holographic crystal. At this scale the baryon  lattice undergo a sequence of phase transitions.

\begin{figure}[t]
\begin{center}
\vspace{3ex}
\includegraphics[width= 100mm]{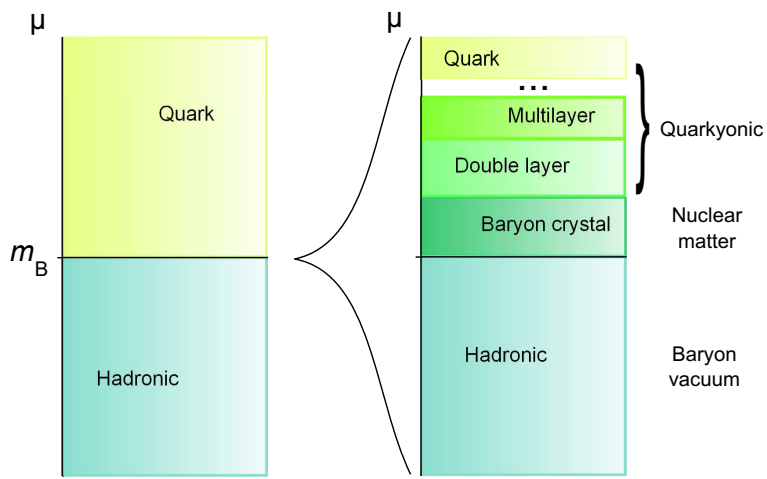}
\end{center}
\caption{When zooming on the small vicinity of the critical chemical potential the phase diagram on figure~\ref{fighQCDPhase} exhibits a richer structure. After the chemical potential reaches critical value $\mu_c$ baryons are created and form a crystal. The crystal becomes multi-layered in the holographic dimension, when the chemical potential is increased, where layer production must correspond to the emergence of the quark Fermi sea of the quarkyonic phase conjectured in~\cite{McLerran:2007qj}.}
\label{figzooming2}
\end{figure}

In the meantime baryon density pertinent to the scale $1/\sqrt{\lambda}$ correspond to values of the chemical potential in the small vicinity of $\mu_c$:
\be
\mu = \mu_c + O(\lambda^0)\,.
\ee
To probe the values of the chemical potential at finite distance away from $\mu_c$ one has to consider lattices with spacing much smaller than $1/\sqrt{\lambda}$, that is smaller than the size of the baryon. Accordingly the baryons will overlap so that the constituent quarks will not know which baryon they belong to. So this phase must be some kind of a weakly interacting quark fluid. In such a case the exact instanton description of the baryons becomes tricky, while uniform density may be a decent approximation. In summary  nuclear matter phase is contained in an asymptotically small window of the chemical potential $\Delta\mu\sim \mu_c/\lambda$, \emph{cf.} figure~\ref{figWorkWindow}. As seen by the phase diagram on figure~\ref{fighQCDPhase} this phase immediately changes to a quark matter phase once the chemical potential is increased beyond $\mu_c$. Nevertheless nuclear matter phase is quite interesting and have a rich structure as demonstrated by figure~\ref{figzooming2}.

At arbitrary low densities we have a normal 3D crystal sitting at the bottom of the potential well in the holographic dimension. When the density is increased the baryons will start feeling tight in 3D due to Coulomb repulsion. It is thus conceivable that at some large density the baryons will pop out from the 3D alignment into the holographic dimension, forming a 4D structure. Analogous effect has already been observed by Rozali \emph{et al.}~\cite{Rozali:2007rx} in the approximation of the uniform baryon density. In the latter study the 4D thickness of the baryon substance was increasing with the chemical potential and was interpreted as a development of the quark Fermi sea. The uniform density of baryons however is not an appropriate description for the nuclear matter phase. Instead we need to consider honest lattices of baryons described by periodic instanton solutions.

Working with instantons we will follow the procedure outlined in section~\ref{secSSBaryons}. We will consider periodic arrays of instantons for which equilibrium density is related to the spacing $d$ between instantons. We will describe the baryon lattice by constructing an appropriate instanton solution in flat space, which corresponds to the zeroth order approximation of non-interacting baryons. The zeroth order solution will then be used to compute corrections to the free energy due to interactions between instantons. Flat space solutions depend on various moduli including instanton positions, size and orientation. Instanton interactions will lift the degeneracy. By varying the free energy with respect to moduli we will find their equilibrium values. This way we will see that at large enough density the 4D lattice will replace the 3D one in a series of phase transitions.

Although our ultimate goal is to study realistic 3D instanton lattices, in this paper only a 1D chain will receive the full treatment. For the 3D lattices we will restrict ourselves to the point charge approximation. As we will see this is enough to demonstrate the main feature of the holographic nuclear matter: at some critical density the 3D baryons will pop out to the holographic dimension. More precisely the lattice will split into layers separated along this dimension. Finite instanton size will have the following consequences. First of all finite size will reduce the Coulomb interaction of closely packed instantons: the interaction will be partially screened. Second, phase space of lattices of finite size instantons have a much reacher structure due to orientation and size dependence of interactions, while the zero-size instantons do not care about their orientation.

In general we expect the following to happen with the 3D or 1D lattice of instantons when we vary the density (figure~\ref{figzooming2}). At small densities the gravity force (curvature) in the transverse dimensions manages to keep instantons aligned in the space dimensions despite their electrostatic repulsion. At a critical density instanton repulsion will push the instantons out to a direction transverse to the lattice (holographic dimension).  The 1D lattice will split  forming a zigzag-like configuration, when every pair of nearest neighbors shift in the opposite direction, while the 3D lattice will analogously split into two layers. The transition will be second order, with the order parameter being the amplitude of the displacement of nearest neighbors. The displacement will grow with density further separating layers in the transverse dimension. For non-infinitesimal separation between the layers one can expect other kind of phase transitions between different types of lattices related to the change of the dimensionality of the lattices from 3D (1D) to 4D (2D).

For larger densities (layer separation) the nearest neighbors in each sublattice will become close. One can then expect a new transition to a multi-layer lattice (not necessarily to a 3-layer). The transition will most likely be first order. Therefore while the density is increased the lattice undergoes a sequence of phase transitions changing the lattice structure and multiplying layers in the transverse dimension. When the lattice becomes mesoscopic in this direction the instantons will start overlapping in the 3D projection. Here the story should match the unform density analysis of~\cite{Bergman,Rozali:2007rx}. The baryons will melt into a quark liquid. The finite width of the liquid in the holographic dimension will be analogous to the Fermi sea of quarks. Therefore we conclude that beyond the first transition in the sequence the phase is resembling the quarkyonic one conjectured by McLerran and Pisarski in~\cite{McLerran:2007qj}: a Fermi sea with both quark and baryonic manifestation. The first transition just marks the onset of the quarkyonic phase.

In the remainder of this paper we will consider two toy-models. The first one is a lattice of point charges to demonstrate the first transition from a single-layer 3D lattice to a multi-layer lattice in 4D. The second toy-model is a 1D chain of instantons. To keep such a chain stable we introduce an \emph{ad hoc} curvature in the transverse spatial directions~(\ref{GRdependence}) so that the chain can only zigzag into the holographic direction. This also allows to keep the size of the instanton small $a\ll d$ in a controllable way. The 1D chain will be shown to exhibit the zigzag transition as well as a first order transition to another lattice structure.


\section{First toy model. 3D lattice of point charges}
\label{secPointCharges}

In this section we will consider an approximation of instantons by point charges. This will allow us to demonstrate the instability of 3D lattices against splitting in the holographic dimension. Point charges give a good qualitative explanation of what happens with baryon crystals. Taking into account full instanton solutions leads to a partial screening of effective interaction between individual instantons and to a richer phase space as we will see in the following section, where a 1D chain of finite size instantons will be analyzed. For a warm-up and for the sake of comparison with the full 1D instanton solution in this section we first consider the case of a 1D lattice of point charges and afterwards proceed  to a discussion of 3D lattices.


\subsection{Warm-up exercise}
\label{sec1dPCharges}

Consider a 1D lattice (periodic chain) of point charges. We can think of point charges as of a zero-size limit of instantons. To cook-up a 1D lattice in the holographic setup one can turn on potential in all transverse directions. Here we will consider a case of anisotropic curvature, \emph{i.e.} use the coupling~(\ref{anisotropic0}) in the expression  for the energy~(\ref{ActionInstanton}). In such a case dislocations in the holographic direction $z$ are preferable as long as $M'\gg M$. Recall that we label the direction along the chain as $x_4$ and the transverse spatial coordinates as $x_1$ and $x_2$, while $x_3\equiv z$ is the holographic coordinate . The density in this case is $1/d$, where $d$ is the lattice spacing.

\begin{figure}[t]
\begin{center}
\vspace{3ex}
\includegraphics[width= 150mm]{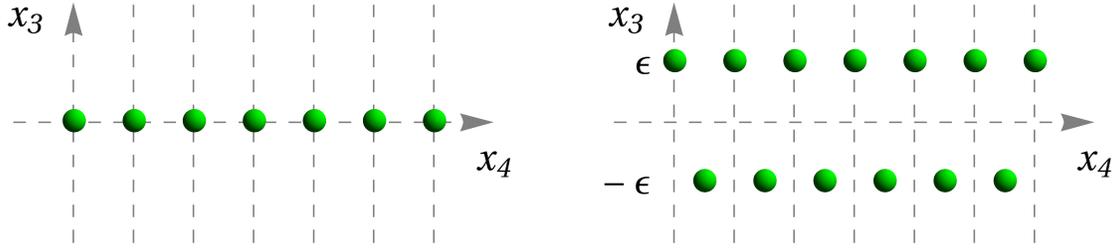}
\caption{Splitting of the 1D chain into the zigzag configuration. At low densities the chain of baryons is straight. When the density is increased the chain splits, forming a zigzag configuration. In the case of anisotropic curvature~(\ref{anisotropic0}) the deformation of the chain occurs along the holographic direction.}
\label{figZigzag}
\end{center}
\vspace{-0.5cm}
\end{figure}

Due to the gravity force the chain must be straight at arbitrary low densities. When the density is increased the instantons will be driven out of the 1D alignment as long as resisting to the Coulomb repulsion will cost more energy than climbing up the gravitational well. Apparently each pair of nearest neighbors will prefer to displace in the opposite directions away from the chain. As a result the 1D lattice will split into two sublattices separated in the holographic dimension forming a ``zigzag'' as in figure~\ref{figZigzag}. From the point of view of the D8-branes neighboring instantons move to the opposite branches of the U-shape configuration.

Let us now study this transition quantitatively. We replace the instanton density $I(x)$ by the sum of delta-functions
\be
I({\bf x}) = \sum\limits_{n=-\infty}^{\infty}\delta^{(4)}({\bf x}-n {\bf d})\,,
\ee
where ${\bf d}$ is a 4-vector generating translations from one site of the chain to a neighboring one, here chosen to be along $x_4$. For the straight chain the non-abelian part of the energy, calculated per instanton, gives
\begin{multline}
\label{nonAbelianEnergy}
E_{\rm NA}= N_c\lambda M\int\limits_0^d \d x_3\int \d^3 x \ I(x)\left(1 + M'^2(x_1^2+x_2^2)+ M^2 x_3^2\right) =
\\ = N_c\lambda M\left(1+{M'}^2(x_1^2+x_2^2)^2 + M^2 x_3^2\right).
\end{multline}
The energy is minimized for the choice $x_1=x_2=x_3=0$. The solution to equation~(\ref{A0eq2}) in the case of the point charges is the 4D Coulomb potential. The abelian energy per instanton is then given by the sum
\be
\label{AbelianEnergy}
E_{\rm C} =\frac{N_c}{4\lambda M}\sum\limits_{n\neq 0}\frac{1}{(nd)^2}=\frac{N_c}{\lambda M}\,\frac{\pi^2}{12 d^2}\,.
\ee

Let us investigate stability of the charges in the chain against the transition to the zigzag. The condition $x_1=x_2=0$ holds since we have frozen these directions by requiring $M'\gg M$. In the zigzag phase every even charge is displaced by amount $\epsilon$ in one direction along $x_3$, while every odd one is displaced by the same amount in the opposite direction. As a result the average non-abelian energy (per instanton) will be shifted $x_3^2\to \epsilon^2$:
\be
\label{NAshift}
E_{\rm NA}\to E_{\rm NA} + N_c\lambda M^3\epsilon^2\,.
\ee
It is also straightforward to evaluate the Coulomb energy per instanton in this case
\be
\label{AbelianEnergy2}
E_{\rm C} = \frac{N_c}{4\lambda M}\left(\sum\limits_{{\rm even}~n\neq 0 }\frac{1}{(nd)^2}+\sum\limits_{{\rm odd}~n}\frac{1}{(nd)^2+(2\epsilon)^2}\right) = \frac{N_c}{\lambda M}\left(\frac{\pi^2}{48 d^2}+\frac{\pi}{16\epsilon d}\,\tanh\frac{\pi\epsilon}{d}\right).
\ee

\begin{figure}[t]
\begin{center}
\vspace{3ex}
\includegraphics[width= 100mm]{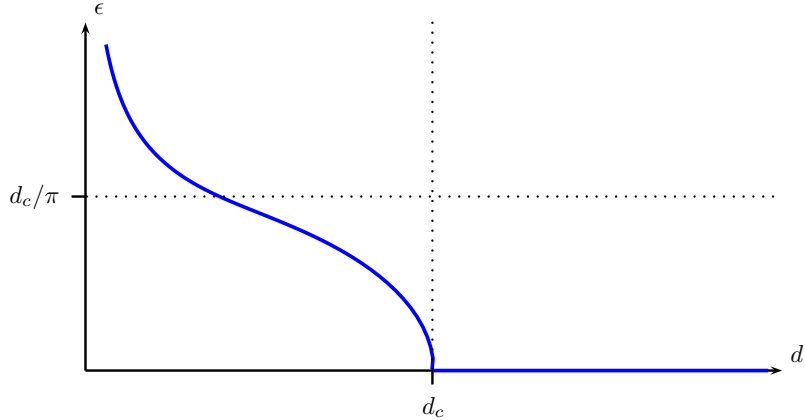}
\end{center}
\caption{Zigzag amplitude $\epsilon$ as a function of the lattice spacing $d$ for point-like instantons.}
\label{figzigzagpar0}
\end{figure}

The expression for the total energy can be expanded at small $\epsilon$:
\be
\label{1dChargesEn}
E=E_0 + N_c\lambda M^3\epsilon^2 + \frac{N_c}{\lambda M}\left(-\frac{\pi^4\epsilon^2}{48 d^4} + \frac{\pi^6\epsilon^4}{120 d^6}+ O(\epsilon^6)\right).
\ee
The sign of the $\epsilon^2$-term depends on the density (lattice spacing). It is positive for small densities and negative for large. Therefore there is a second order phase transition, in which $\epsilon$ acquires a non-trivial vev, \emph{i.e.} the straight chain transforms to a zigzag. The critical density corresponds to the point, where $\epsilon^2$-term changes sign:
\be
\label{PointChargeDc}
d=d_c \equiv \frac{\pi}{2\cdot 3^{1/4}M\sqrt{\lambda}}\,.
\ee
For the spacing slightly smaller than $d_c$ the order (zigzag) parameter has the mean field behavior:
\be
\label{Trans1DCharge}
\langle\epsilon\rangle\simeq \pm \frac{\sqrt{5}}{\pi}\,\sqrt{d_c(d_c-d)}\,.
\ee
More generally $\epsilon$ is given by the solution to
\be
\label{Esolution}
\frac{(\pi\epsilon/d)^3\cosh^2(\pi\epsilon/d)}{\sinh(2\pi\epsilon/d)\,-\, (2\pi\epsilon/d)}\
=\ \frac{3}{4}\left(\frac{d_{c}}{d}\right)^4.
\ee
Graphical solution to this equation is presented in figure~\ref{figzigzagpar0}. The zigzag amplitude grows with density. Notice that the units for the critical spacing $d_c$ are $1/(M\sqrt{\lambda})$ -- the same as for the equilibrium size of instantons derived in section~\ref{secSSBaryons}.

One can further study the fate of the zigzag in this toy-model. Once the zigzag amplitude is large enough one can think of the zigzag as of a 2D lattice with 2 layers in the holographic dimension. For larger densities the number of layers will grow, although the sequence of transitions is not necessarily trivial. More precisely if we study a sequence of transitions up to 3-layer lattices we will find that the straight chain turns first to the zigzag and then to 3 layers. However if 4-layer lattice configurations are also included in consideration the sequence becomes as follows
\be
\label{1-4sequence}
1\to 2\to 4\to 3\to 4\,,
\ee
where numbers indicate the number of layers in the lattice and the corresponding configurations are summarized in table~\ref{tabPCSequence}.

\begin{table}
\begin{center}
 \begin{tabular}[htb]{c|c|m{0.4\linewidth}|m{0.4\linewidth}}
  n  & $\rho_c$ &  \begin{center}Configuration\end{center} &  \begin{center}{Free energy}\end{center} \\
\hline
 \Trule\Brule 1 &  --- & \parbox[c]{\linewidth}{\includegraphics[width=\linewidth]{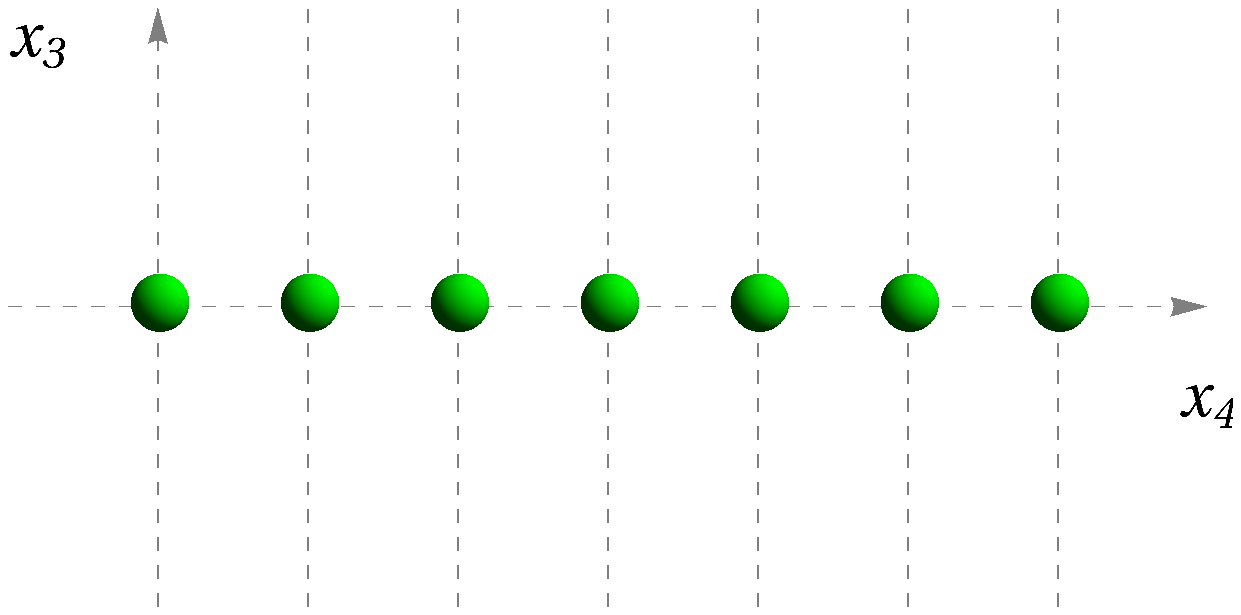}}
 &  \be\nonumber \frac{N_c}{\lambda M}\,\frac{\pi^2\rho^2}{12} \ee \\
\hline
 \Trule\Brule 2 & $\rho_c$  & \parbox[c]{\linewidth}{\includegraphics[width=\linewidth]{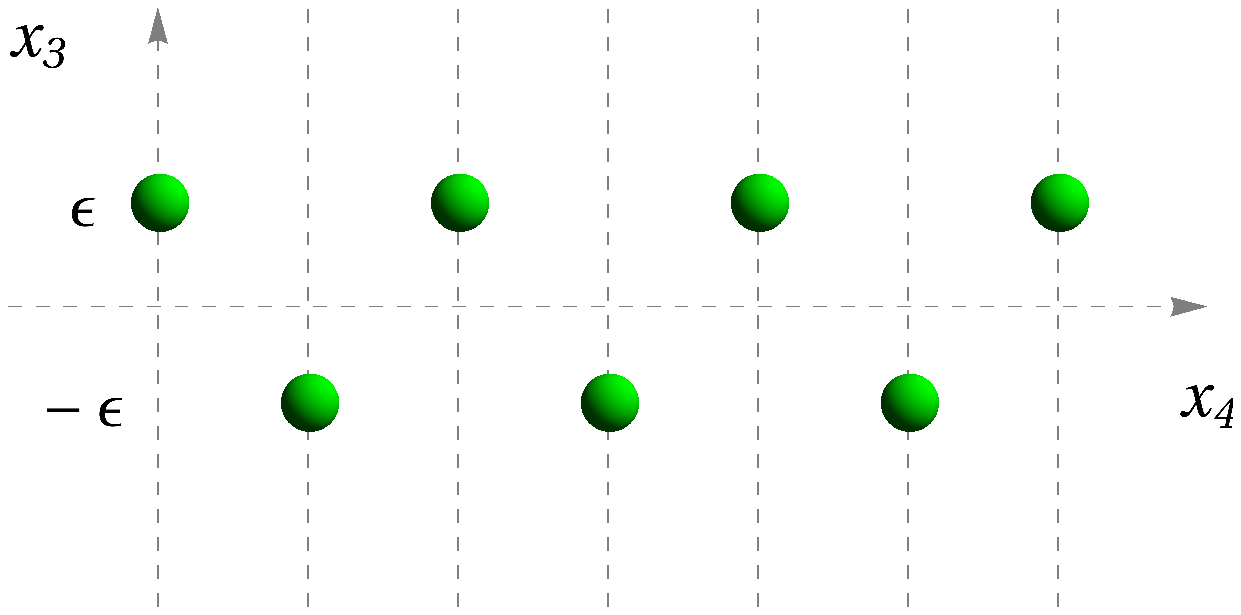}}
   &  \begin{multline}\nonumber
  N_c\lambda M^3\epsilon^2 + \frac{N_c}{16\lambda M}\left(\frac{\pi^2\rho^2}{3}+\frac{\pi\rho}{\epsilon }\,\tanh{\pi\epsilon\rho}\right)
	\end{multline} \\
\hline
 \Trule\Brule 4 & $2.26\rho_c$ & \parbox[c]{\linewidth}{\includegraphics[width=\linewidth]{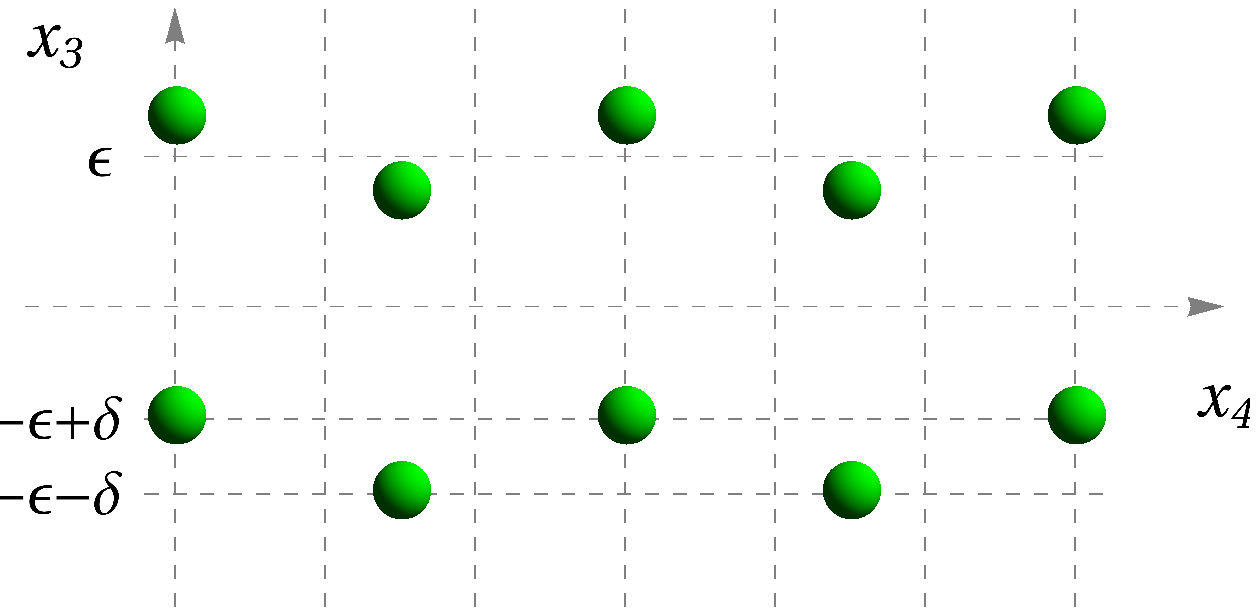}}
 &  \begin{multline*}
     N_c\lambda M^3 \left(\epsilon^2+\delta^2\right) +  \frac{N_c}{32\lambda M}\left(\frac{\pi^2\rho^2}{6} \ +  \right.
\\  + \frac{\pi  \rho }{ \epsilon }\, \coth\frac{\pi  \epsilon  \rho }{2} \ +\  \frac{\pi\rho\tanh{\pi(\epsilon+\delta)\rho}/{2}}{2 (\epsilon+\delta)}\  +
\\ \left.  + \frac{ \pi  \rho }{  \delta  }\, \tanh\frac{\pi  \delta  \rho }{2} \ +  \ \frac{\pi\rho\tanh{\pi(\epsilon-\delta)\rho}/{2}}{2 (\epsilon-\delta)} \right)
    \end{multline*}  \\
\hline
 \Trule\Brule 3 & $2.4\rho_c$  & \parbox[c]{\linewidth}{\includegraphics[width=\linewidth]{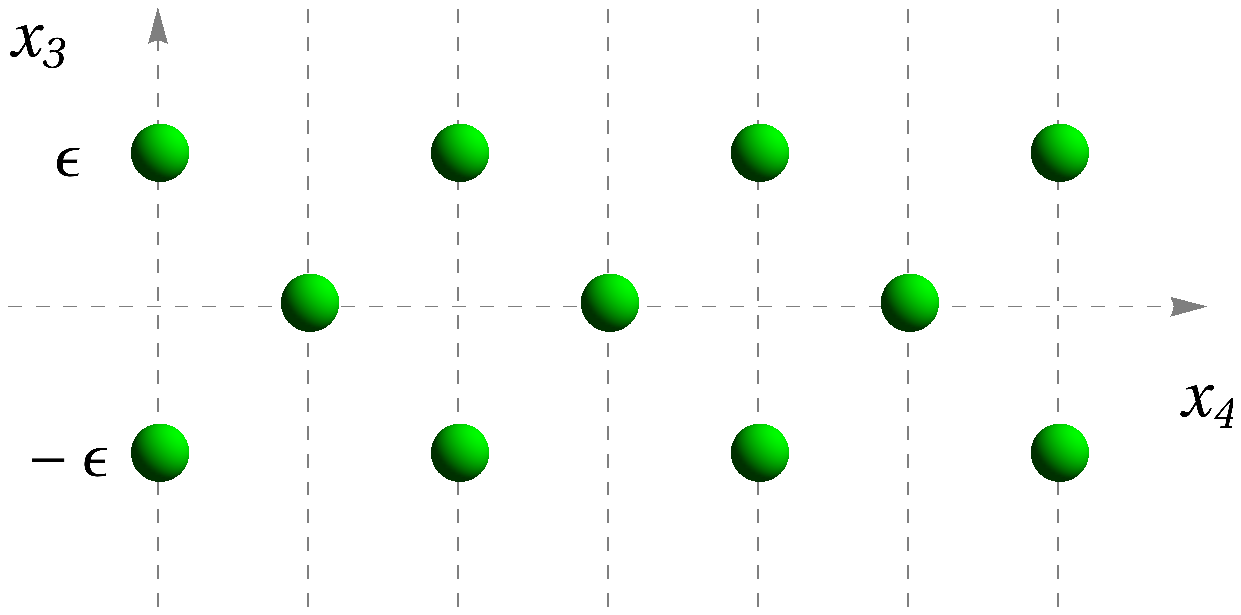}}
&   \begin{multline}\nonumber
\frac{2}{3}\,N_c\lambda M^3\epsilon^2 +  \frac{N_c}{36\lambda M}\,\left(\frac{\pi^2\rho^2}{3} \ + \right.
\\ + \ \left. \frac{\pi\rho}{\epsilon}\,\coth\frac{2\pi\epsilon\rho}{3} + \frac{4\pi\rho}{\epsilon}\,\tanh\frac{\pi\epsilon\rho}{3}\right)
	\end{multline} \\
\hline
 \Trule\Brule 4 & $3.08\rho_c$ &  \parbox[c]{\linewidth}{\includegraphics[width=\linewidth]{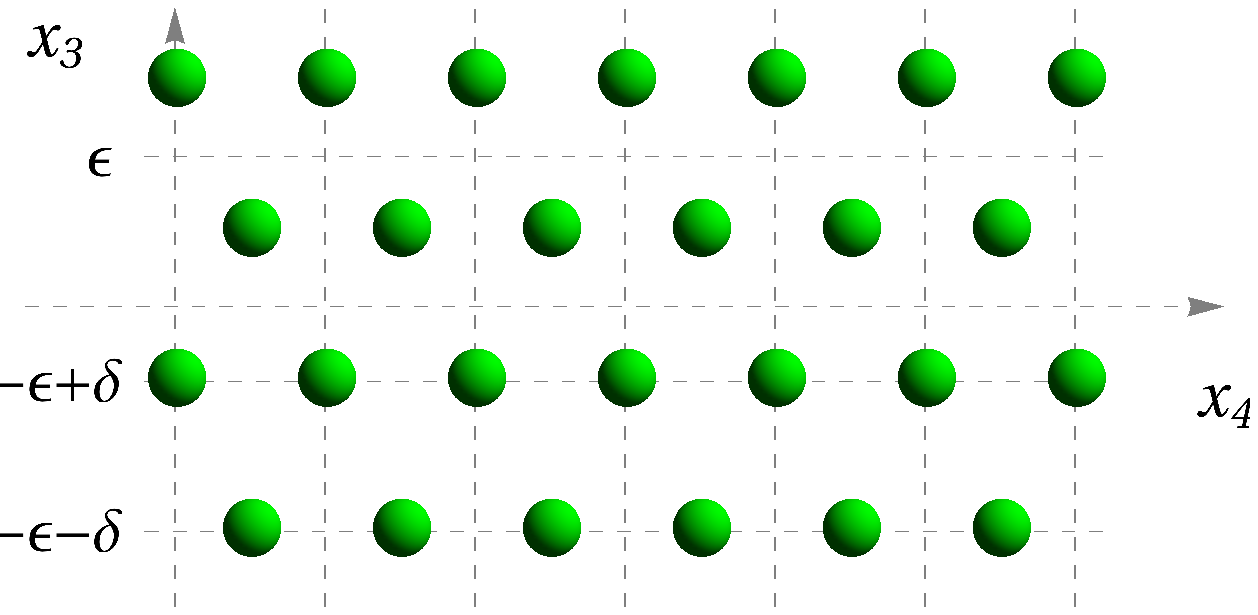}}
&  \begin{center} same as in the 3$^{\frac{\text{rd}}{}}$ row \end{center} \\
\hline

 \end{tabular}
\end{center}
\caption{Various $n$-layer configurations that appear in the sequence of transitions in the 1D lattice of point-like instantons. The 2$^\frac{\rm nd}{}$ column and the 4$^\frac{\rm th}{}$ column contain the critical density $\rho_c$ and the free energy (per instanton) respectively  for a configuration shown in the 3$^\frac{\rm rd}{}$ column. Critical density is measured in units of the density of the zigzag transition given by~(\ref{PointChargeDc}). Notice that lattices in the 3$^\frac{\rm rd}{}$ and 5$^\frac{\rm th}{}$ rows are the same, albeit at different densities. Only configurations with no more than 4 layers are analyzed.}
\label{tabPCSequence}
\end{table}

Table~\ref{tabPCSequence} demonstrates configurations that appear in the sequence of transitions of the 1D chain in the case one restricts to the lattices with no more than 4 layers. It shows a formula for the free energy (per charge) and a critical density at which the transition to this configuration occurs. The critical density is shown in units of the density of the zigzag transition, which from~(\ref{PointChargeDc}) can be found to be $\rho_c\simeq 0.84$ in units of $M\sqrt{\lambda}$.  The apparent 3-layer configuration is shown in the fourth row of table~\ref{tabPCSequence}. If we just have restricted our attention to configurations with at most 3 layers, then after the zigzag transition we would have a first order transition to the 3-layer lattice at $\rho\simeq 2.3\rho_c$. In the equation for the free energy of 3 layers we denote $\epsilon$  the distance between nearest layers as shown in the corresponding diagram. It turns out however that the transition to 4 layers occurs at smaller densities, $\rho=2.26\rho_c$. Interestingly  the free energy of the 4-layer lattice is smaller than that of the 3-layer one for $2.26\rho_c\geq \rho\geq 2.4\rho_c$ and for $\rho>3.08\rho_c$, while in between 3 layers win. In both cases the 4-layer configuration is the same, as one can see from the third and the fourth rows of table~\ref{tabPCSequence}, although in the former case it looks more like a deformation of 2-layer lattice, rather than a 4-layer one. That is why in the formula for the free energy  we choose $\epsilon$ to be a distance from the central axis to the imaginary 2 layers, while the parameter $\delta$ measure the amount of deformation, or the ``zigzag parameter'', of imaginary layers.  All the transitions in table~\ref{tabPCSequence} are first order, except for the zigzag transition, which is second order.

\begin{figure}[t]
\begin{minipage}{0.45\linewidth}
\begin{center}
\includegraphics[height=70mm]{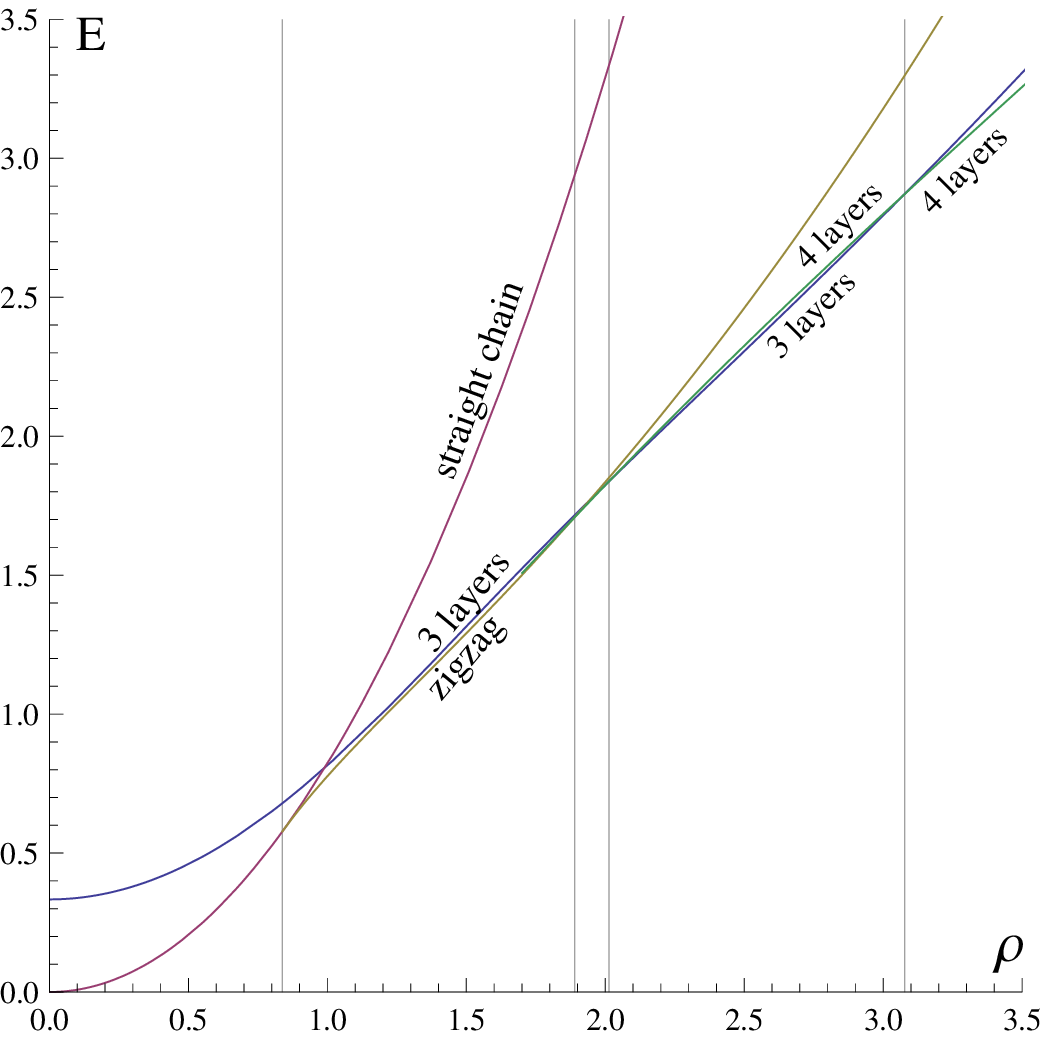}

(a)
\end{center}
\end{minipage}
\hfill
\begin{minipage}{0.45\linewidth}
\begin{center}
\includegraphics[height=70mm]{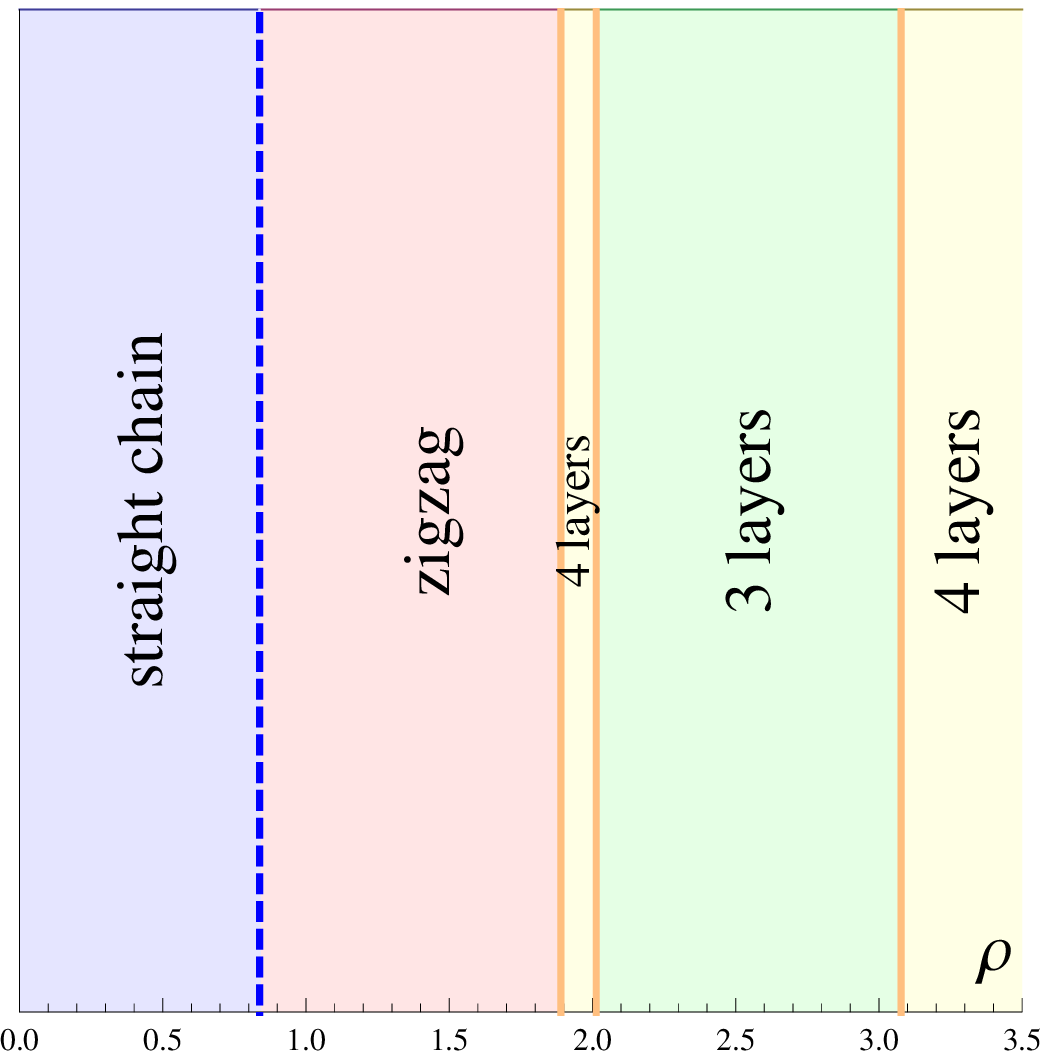}

(b)
\end{center}
\end{minipage}
\caption{(a) Free energy (per instanton) of various 1D point charge configurations as functions of the density in units of $M\sqrt{\lambda}$. (b) Zero temperature phase diagram of point-like-instanton chain. Dashed line indicates the second order phase transition, while the solid lines are first order transitions.}
\label{figChainPhase}
\vspace{-0.5cm}
\end{figure}

In figure~\ref{figChainPhase} we plot the free energies of various configurations as functions of the density and summarize the phase diagram of the 1D chain of point-like instantons (with the restriction of 4 layers). As can be inferred from the left part of the figure the energies of various configurations are often very close. For example, for $\rho_c<\rho<2.4\rho_c$, the energies of the zigzag, 3-layer (and also 2-layer) configurations are impressively close to each other. Although we have not studied the transitions beyond 4 layers it is clear that the sequence of transitions will not be trivial. Specifically it looks likely that the transition to 5 layers will occur before the second transition to 4-layers. Complexity will grow fast with the number of layers and present a challenge for numerical analysis.

This otherwise trivial 1D exercise demonstrates main features of the crystal of holographic baryons. When squeezed enough the baryons cannot seat in the regular space dimensions and pop into the holographic one. The phase space of the crystal has a very rich structure as it involves various crystal configurations in the additional dimension. In section~\ref{secChain} we promote the chain of point charges to the full instanton solution. Interference between finite size instantons will effectively reduce their repulsion and therefore increase the critical density. The phase space structure will also depend on the mutual orientation of instantons. But first, let us slightly elaborate on the point charge case and consider a 3D lattice.

\subsection{3D lattices}
\label{sec3dLattice}

Let us start our discussion from the simple cubic (sc) lattice. Similarly to the 1D lattice we expect that at large density Coulomb repulsion will expel instantons from the 3D alignment out to the holographic dimension. The lattice will become 4D with multiple number of 3D layers. The most probable first transition must be the one, in which any pair of nearest neighbors displace in the opposite directions along the holographic dimension, in analogy with the zigzag transition. For the sc lattice this transition corresponds to every even site moving one way and every odd one in the opposite. In this second order phase transition the original sc lattice will split into two fcc layers.

The above transition can be studied quantitatively. The analysis of stability proceeds similarly to the 1D case. The only difference is that in 2 and 3 dimensions the Coulomb energy $E_{\rm C}$ per instanton will diverge. In the case of a 3D lattice it is linearly divergent:
\be
\sum\limits_{(n_1,n_2,n_3)\neq 0}\frac{1}{d^2\left(n_1^2+n_2^2+n_3^2\right)}\propto L\,,\qquad L\to \infty\,,
\ee
where $L$ is the linear size of the system. However for the question of stability one does not need to compute the sum itself, but rather its variation in case of the lattice expansion to the 4th dimension. In other words we are not interested in the infinite constant $E_0$, \emph{cf.}~(\ref{1dChargesEn}), but rather in the finite $\epsilon^2$ and $\epsilon^4$ terms. It is easy to show that to find the energy shift for a split configuration one needs to compute the following two quantities
\be
\Delta \mu^2 = \sum\limits_{\text{odd}}\frac{1}{\left(n_1^2+n_2^2+n_3^2\right)^2}\,, \qquad \ell = \sum\limits_{\text{odd}}\frac{1}{\left(n_1^2+n_2^2+n_3^2\right)^3}\,,
\ee
where the sums are taken over odd sites: $n_1+n_2+n_3$ -- odd. Numerical evaluation of the sums gives values $\Delta\mu^2\simeq 10.0$, $\ell\simeq 6.60$. The regularized expression for the energy per instanton of the split lattice is
\be
\label{3dChargesEn}
E-E_0 = N_c\lambda M^3\epsilon^2 + \frac{N_c}{\lambda M}\left(-\frac{\Delta\mu^2\epsilon^2}{d^4} + \frac{4\ell\epsilon^4}{d^6}+ O(\epsilon^6)\right).
\ee
The critical density corresponds to the lattice spacing
\be
\label{scCritSpacing}
d_c= \frac{1}{M}\sqrt{\frac{\Delta\mu}{\lambda}} \simeq\ 0.69\,\frac{1}{M\sqrt{\lambda}}\,,
\ee
while the generalization of equation~(\ref{Trans1DCharge}) for the absolute value of the order parameter reads
\be
\label{Trans3DCharge}
\langle\epsilon\rangle\simeq \pm \frac{\Delta\mu}{\sqrt{\ell}}\,\sqrt{d_c(d_c-d)}\,.
\ee

Thus for a sc lattice with the lattice spacing smaller than~(\ref{scCritSpacing}) it is favorable to split. In reality however the sc phase of repelling instantons is not the most favorable configuration at a given density. The minimum energy configuration must correspond to close packing. Close packing gives the largest interatomic distance between the nearest neighbors for a given density. In the 3D case this is achieved in the fcc lattice.

Stability analysis of the fcc lattice is trickier because for the fcc lattice there is no natural way to split into two sublattices. One can expect two things to happen when the density is increased. First there can be a first order transition to a multi-layer 4D lattice. Second the splitting into two sublattices can occur through a breaking of the cubic symmetries (half of nearest neighbors will go one way and another half the other way). The latter will be restored as soon as the separation between the sublattices will be large enough. Let us also mention the bcc lattice. For the latter there is a natural splitting to two sc sublattices.

We expect the finite density phase space of 3D lattices to be qualitatively similar to that of 1D lattices. When the lattices are squeezed the instantons will pile-up in the holographic dimension producing more layers and rearranging the lattice structure. In the meantime the former phase space will be richer as more lattice configurations are possible. For small thickness of a lattice in holographic dimension its structure will be defined by the 3D intuition, but when the number of layers is sufficiently large the lattice will start exhibiting its 4D nature.

In this section we have shown that lattices of point-like instantons (baryons) undergo popcorn transitions to configurations with finite thickness in the holographic dimension. In other words the boundary theory lattices become bulk lattices. The characteristic scale for such transitions is provided by the same combination $1/(\sqrt{\lambda}M)$ as in the case of the baryon size. In fact the same scale will remain for the transitions in the case of finite size instantons as we are going to see in the next section. This is an indication that the transitions occur when the baryons start overlapping. We will continue this discussion in the concluding section~\ref{secSummary} after the analysis of the effects of finite instanton size.

\section{  1D chain of instantons -- a second  toy model}
\label{secChain}

In this section we will analyze a model of a 1D crystal (chain) of baryons. In contrast with the previous section the full instanton solution will be used for the chain. Full solution introduces additional parameters, such as instanton size and orientation. We start from a straight chain: this solution was originally found by Kraan and van Baal~\cite{Kraan:1998pm}; and compute corrections to the energy of the flat space instanton. The corrections stabilize the size of the instantons and their orientation angle. In particular we show that the preferable configuration at low densities is ``antiferromagnetic''. In the case of a generic instanton configuration the Coulomb energy integral cannot be computed analytically. Instead of evaluating it numerically we restrict our analysis to the case of small instantons, where analytical derivation is possible.

Next we derive the zigzag solution to the ADHM equations. We show that at large densities the zigzag configuration is preferable over the straight chain.\footnote{This statement is of course true only for the densities, which are not too high. At higher densities the zigzag will yield to  multi-layer structures as in the case of point charges. Remarkably even the straight chain is more favorable than the zigzag at asymptotically high densities as we show in appendix~\ref{appHighDensity}.} At small values of the zigzag amplitude the twist angle remains $\phi=\pi$. By analyzing more general twist configuration (with non-abelian twist) we show that at larger zigzag amplitude the twist angle becomes $\phi=\pi/2$.


\subsection{Straight chain}
\label{secStraightChain}

\subsubsection{Straight periodic twisted chain of instantons}
\label{secStraightChainSol}

A solution for an infinite periodic chain of $SU(2)$ instantons was first  obtained by Harrington and Shepard in~\cite{Harrington:1978ve} in the context of finite temperature field theory. This solution corresponds to parallel-oriented instantons. It was later noticed by Rossi~\cite{Rossi} that through a gauge transformation it can be transformed into a single 3D BPS monopole solution. In particular the instanton density does not depend on the coordinate along the chain. This instanton-monopole equivalence was generalized by Lee an Yi through a study of moduli space of calorons -- instantons on $R^3\times S^1$~\cite{Lee:1997vp}. More generally a $SU(N)$ caloron solution with a non-trivial holonomy around the  $S^1$ is equivalent to $N$ BPS monopoles. Net electric and magnetic charges of the monopoles vanish and the net topological charge (mass) is that of the instanton. The holonomy of an instanton on $R^3\times S^1$ has an interpretation of a constant relative orientation twist between the instantons in the infinite chain in $R^4$. An explicit solution for a twisted instanton chain was found by Kraan and van Baal in~\cite{Kraan:1998pm}. Let us briefly review this result.

To find a solution for parallel instantons it is enough to write down the 't Hooft ansatz. In the case of instantons of variable orientation one has to resort to the details of the ADHM construction~\cite{ADHM}. ADHM data are matrices, which contain the information about the locations of the instanton centers, their radii and orientations. For an infinite chain the matrices are infinite dimensional. More specifically, $Sp(k)$ $N$-instanton solution is encoded by two quaternionic matrices: an $N\times N$ symmetric matrix $X$ and a $k\times N$ vector $Y$, which satisfy the constraint
\be
\label{ADHMeqs}
X^\dagger X + Y^\dagger Y \qquad \text{is real symmetric.}
\ee

In the case of $Sp(1)\simeq SU(2)$ the solution can be reformulated in terms of four real symmetric matrices $\Gamma^\mu$ and real vectors $Y^\mu$:
\be
\label{GammaDef}
X=\Gamma^\mu\tau^\mu\equiv \Gamma^4+ i\tau^j \Gamma^j\,, \qquad Y=Y^\mu\tau^\mu\equiv Y^4 + i\tau^jY^j\,,
\ee
where $\tau^j$, $j=1,2,3$ are Pauli matrices. The real matrices satisfy the constraint
\be
\left([\Gamma^\mu,\Gamma^\nu]\,+\,Y^\mu\otimes Y^\nu\,-\,Y^\nu\otimes Y^\mu\right)\
=\ \frac12\,\epsilon^{\mu\nu\kappa\lambda}
\left([\Gamma^\kappa,\Gamma^\lambda]\,+\,Y^\kappa\otimes Y^\lambda\,-\,Y^\lambda\otimes Y^\kappa\right),
\label{GammaRel}
\ee
where $\bigl(Y^\mu\otimes Y^\nu\bigr)_{mn}=Y^\mu_m\times Y^\nu_n$. Physically, the diagonal matrix elements $\Gamma^\mu_{nn}$ are the 4D coordinates of the instanton centers, $Y_{n}^\mu$ combine the radii and the $SU(2)$ orientations of the instantons, while the off-diagonal matrix elements $\Gamma^\mu_{m\neq n}$ are determined by the condition (\ref{GammaRel}). (Modulo a common $O(N)$ symmetry of all the $\Gamma^\mu_{mn}$ and $y^\mu_n$.)

For the purposes of this paper there is no need to find the expression for the gauge potential itself. What we only need to know is just the expression for the instanton density. In terms of the matrices $\Gamma^\mu$ and $y^\mu$ it can be constructed as follows. Provided that~(\ref{GammaRel}) is satisfied define a real symmetric $N\times N$ matrix
\be
L(x)\ =\ (x^\mu\mathbbm{1}-\Gamma^\mu)(x^\mu\mathbbm{1}-\Gamma^\mu)\ +\ Y^\mu\otimes Y^\mu .
\label{Ldef}
\ee
The instanton density is then given by~\cite{Osborn}
\be
I(x)\ =\ -\frac{1}{16\pi^2}\,\square\square\log\det(L(x))\,.
\label{Idensity}
\ee

We are interested in constructing a solution, which corresponds to an array of equal size instantons arranged in a 1D periodic lattice. Periodicity means discrete translational symmetry ${\bf S}:x^\mu\to x^\mu+d^\mu$ of the ADHM solution; in the language of the $\Gamma^\mu_{mn}$ and $Y^\mu_m$, this symmetry acts as
\begin{align}
\Gamma^\mu\ &\to\ S^{-1}\Gamma^\mu S\ =\ \Gamma^\mu\ +\ d^\mu\mathbbm{1}\qquad
\comment{to keep the $x^\mu\mathbbm{1}-\Gamma^\mu$ invariant}
\label{SymAct1}
\\
(Y^\mu_n\tau^\mu)\ &\to\,\sum_m G(Y^\mu_m\tau^\mu)S_{mn}\ =\ (Y^\mu_n\tau^\mu)\,,
\label{SymAct}
\end{align}
for some $O(N=\infty)$ matrix $S_{mn}$ and $SU(2)$ matrix $G$.

Physically, $G$ rotates the orientation of an instanton relative to its immediate neighbor. The more distant neighbors are related by $G^{n-m}$ rotations, which generate a $U(1)$ subgroup of the $SU(2)$. Without loss of generality, we take $G=\exp\bigl(i\phi\tau_3/2\bigr)$ for some ``twist'' angle $\phi$ between 0 and $2\pi$. We also take the direction of the instanton chain to be the $x^4$ while the transverse directions are $x^1$, $x^2$, and $x^3=z$; in terms of equation~(\ref{SymAct1}) this means $d^\mu=(0,0,0,d)$. Finally, we take $S_{m,n}=\delta_{m,n+1}$ (shifts from the $n{\rm th\over}$ instanton to the $(n+1){\rm st\over}$) and $Y_0^\mu=(0,0,0,a)$ (where $a$ is the radius of the $0{\rm th\over}$ instanton).  Consequently, the translational symmetry~(\ref{SymAct1})--(\ref{SymAct}) requires
\begin{align}
i\tau^\mu Y^\mu_n\ &=\ a\,\exp\left(in\tfrac\phi2\tau_3\right)
\quad\Longleftrightarrow\quad
Y^\mu_n\ =\ \bigl(0,0,a\sin(n\phi/2),a\cos(n\phi/2)\bigr)\\
{\rm and}\quad\Gamma^\mu_{mn}\ &
=\ d\,\delta^{\mu4}\times n\,\delta_{mn}\ +\ \hat\Gamma^\mu(m-n)\,,
\end{align}
where the
$\hat\Gamma^\mu(m-n)$ do not have separate dependence on $m$ and $n$ but only on $m-n$. Combining these symmetry conditions with the ADHM constraint (\ref{GammaRel}), we get
\begin{align}
Y^\mu_m\otimes Y^\mu_n\ &=\ a^2\,\cos\left[(m-n)\phi/2\right],\\
\Gamma^4_{mn}\ &=\ dn\,\delta_{mn}\,,\\
\Gamma^1_{mn}\ &=\ \Gamma^2_{mn}\ =\ 0,\\
\Gamma^3_{mn}\ &=\ \frac{a^2}{d}\times\frac{\sin\left[(m-n)\phi/2\right]}{m-n}\quad
{\rm for}\ m\neq n,\ {\rm but\ 0\ for}\ m=n.
\end{align}

To calculate the instanton density of the periodic chain we need the determinant of the infinite matrix $L$~(\ref{Ldef}). This determinant is badly divergent, but we may obtain it up to an overall infinite-but-constant factor from the derivatives
\be
\partial_\mu\log\det(L(x))\ =\ 2\tr\left((x^\mu\mathbbm{1}-\Gamma^\mu)L^{-1}(x)\right).
\label{Derivatives}
\ee
For 3 of these derivatives ($\mu=1,2,3)$ the trace converges while for $\mu=4$ the trace diverges but can be regularized using symmetry $x_4\to-x_4$, $n\to-n$.  To evaluate those traces, it is natural to use Fourier transform from infinite matrices to linear operators acting on periodic functions of $\theta~(\rm mod~2\pi)$\footnote{This operation is also known as Nahm transform~\cite{Nahm}.}. Consequently, $L$ becomes
\begin{align}
L\ &=\ x_1^2\ +\ x_2^2\ +\ \bigl(x_3-\Gamma^3(\theta)\bigr)^2\
+\ \left(x_4+id\frac\partial{\partial\theta}\right)^2\ +\ T(\theta),\\
{\rm where}\ T(\theta)\ &=\ \pi a^2\delta\big(\theta-\frac{\phi}{2}\big)\ +\ \pi a^2\delta\big(\theta+\frac{\phi}{2}\big),
\label{Tantiferromagnetic}\\
{\rm and}\ \Gamma^3(\theta)\ &=\ \frac{\pi a^2}{d}\times\begin{cases}
	1-\frac\phi{2\pi} & {\rm for}\  0 <\theta<\frac{\phi}{2} \ \rm{and} \ 2\pi-\frac{\phi}{2}< \theta < 2\pi,\\
	\quad-\frac\phi{2\pi}& {\rm for}\ \frac{\phi}{2}< \theta < 2\pi - \frac{\phi}{2},
	\end{cases}
\end{align}
and $L^{-1}$ becomes the Green's function of this operator. Calculating this Green's function one obtains the following expression upon integration of the traces~(\ref{Derivatives}):
\begin{align}
\det(L)\ =\ &
\left(\cosh\frac{\phi r_1 }{d}\,+\,\frac{\pi a^2}{dr_1}\,\sinh\frac{\phi r_1 }{d}\right)
\left(\cosh\frac{(2\pi-\phi) r_2 }{d}\,+\,\frac{\pi a^2}{dr_2}\,\sinh\frac{(2\pi-\phi) r_2 }{d}\right)\nonumber \\
&+\ \frac{r_1^2+r_2^2-(\pi a^2/d)^2}{2r_1r_2}\,\sinh\frac{\phi r_1 }{d}\,\sinh\frac{(2\pi-\phi)r_2}{d}\nonumber\\
&-\ \cos \frac{2 \pi x_4}{d}\,,
\label{BigFla}
\end{align}
where
\begin{align}
r_1^2\ &=\ x_1^2\ +\ x_2^2\ +\ \left(x_3+\frac{a^2(\phi-2\pi)}{2d}\right)^2,\nonumber\\
r_2^2\ &=\ x_1^2\ +\ x_2^2\ +\ \left(x_3+\frac{a^2\phi}{2d}\right)^2.
\label{Radii}
\end{align}
This is precisely the result obtain by Kraan and van Baal in~\cite{Kraan:1998pm}.


\subsubsection{Total energy of the straight baryon chain}
\label{secStraightChainEn}

Let us now apply the above instanton solution to the holographic model of baryons. The goal is to find the energy of the corresponding multi-baryon configuration. As outlined in section~\ref{secHBaryons} to find the energy up to $O(\lambda^0)$ order we need to plug the $O(\lambda)$ (flat space) instanton solution of interest into equation~(\ref{ActionInstanton}) and then minimize the energy with respect to the moduli. To stabilize the chain in 1D we introduce curvature in all directions transverse to the chain.

The expression for the instanton density follows from~(\ref{Idensity}) after plugging solution~(\ref{BigFla}). The non-abelian part of the energy~(\ref{ActionInstanton}) reduces to a calculation of moments of the instanton density. Although the instanton density following from~(\ref{BigFla}) is way too complicated to print, the
integrals
\be
\vev{x_i^2}\ \equiv \int_0^d\!\!\d x_4\int\!\d^3x\, x_i^2\times I(x)
\label{vevx2}
\ee
can be calculated rather easily by integrating by parts. This gives us
\be
\vev{x_1^2}\ =\ \vev{x_2}^2\ =\ \frac{a^2}{2}
\label{Size12}
\ee
--- exactly as for a single stand-alone instanton of radius $a$ --- but
\be
\vev{x_3^2}\ =\ \frac{a^2}{2}\ +\ \frac{a^4}{4d^2}\times\phi(2\pi-\phi)\,,
\label{Size3}
\ee
where the second term is due to interference between the instantons. Curiously, the interference term vanishes for $\phi=0$, $i\,.e.$ for instantons with the same $SU(2)$ orientations.

For a gauge coupling rising isotropically in all directions transverse to the instanton chain, as in the case described by equation~(\ref{GRdependence}), and instantons of small size $a\ll M^{-1}$, equations~(\ref{Size12}) and~(\ref{Size3}) give us the non-abelian energy (per instanton) of the Kraan--van Baal chain as
\begin{align}
E_{\rm NA}\ 
&=\ N_c\lambda M\left(1\,+\,\frac32M^2 a^2\,
+\,\frac{M^2a^4}{4d^2}\times\phi(2\pi-\phi)+\ O(M^4a^4)\right).
\label{NAenergy}
\end{align}

To compute the Coulomb energy one first has to find a solution to~(\ref{A0eq2}). In light of equation~(\ref{Idensity})
\be
\label{Asoln}
\hat A_0(x)\ =\ \frac{1}{4\lambda M}\,\square\log\det(L(x))\ +\ \rm const.
\ee
Consequently, the abelian Coulomb energy per instanton is
\begin{align}
E_{\rm C}\ &
=\ \frac{N_c\lambda M}{16\pi^2}\int_0^d\!\!\d x_4\int\!\d^3x\,(\partial_\mu\hat A_0)^2\nonumber\\
&=\ \frac{N_c}{256\pi^2 \lambda M}\int_0^d\!\!\d x_4\int\!\d^3x\,
(\partial_\mu\square\log\det(L))^2.
\label{Cenergy}
\end{align}
For generic lattice spacings $d$, this integral is too hard to take analytically. But it becomes much simpler in the $d\gg a$ limit of well-separated instantons, and also in the opposite $d\ll a$ limit, where the instantons merge into a continuous line. Here we present the results for the energy in these two limits and refer to appendix~\ref{secSLDensity} for more details.

For large lattice spacing $d\gg a$ the Coulomb energy~(\ref{Cenergy}) evaluates to
\be
\label{ECstraight}
E_{\rm C}\ \approx\ \frac{N_c}{\lambda M}\left[
\frac{1}{5a^2}\ +\ \frac{4\pi^2+3(\pi-\phi)^2}{30d^2}\ +\ O(a^2/d^4)\right].
\ee
Combining it with the non-abelian energy (\ref{NAenergy}) and minimizing with respect to the instanton radius~$a$ and twist angle~$\phi$, we find the minimum at
\be
\label{phiastraight}
\phi[{\rm@min}]\ =\ \pi,\qquad  a[{\rm@min}]\ =\ a_0\ -\ \frac{\pi^2 a_0^3}{12d^2}\
+\ O(a_0^4/d^2)\,,
\ee
where $a_0$ is the equilibrium radius of a standalone instanton~(\ref{OrigA}).

In the opposite limit $d\ll a$ of densely packed overlapping instantons, the instanton density becomes independent of the $x_4$ coordinate, while in the 3 transverse dimensions it becomes concentrated at two widely separated points
\be
\vec X^{(1)}\ =\ \Bigl(0,0,\frac{a^2}{2d}(2\pi-\phi)\Bigr)\qquad
{\rm and}\qquad\vec X^{(2)}\ =\ \Bigl(0,0,\frac{a^2}{2d}(-\phi)\Bigr);
\ee
note that the distance $b=\pi a^2/d$ between these points is much larger than the naive instanton radius $a$. Effectively the solution looks like a pair of monopoles on $R^3$, in accordance with the analysis of~\cite{Lee:1997vp}. Monopole centers sit at $\vec X^{(1)}$ and $\vec X^{(2)}$. The net topological charge of the monopoles is equal to one --- instanton charge per unit cell, while the distribution of the charge between them depends on the value of $\phi$. Indeed for $d\ll a$  the 3D monopole density takes the form (see details in appendix~\ref{secSLDensity})
\be
\label{3DMonDensity}
I_{3D}(\vec X)\ =\ d\times I_{4D}(\vec X)\
\approx\ \frac{\phi}{2\pi}\times\delta^{(3)}(\vec X-\vec X^{(1)})\
+\ \frac{2\pi-\phi}{2\pi}\times\delta^{(3)}(\vec X-\vec X^{(2)})\
+\ O\bigl(e^{-2\phi r/d}\bigr).
\ee
This density holds   for well separated monopoles such that  their topological charges are $\phi/2\pi$ and $(2\pi-\phi)/2\pi$ respectively.  One can also show that the monopoles have opposite electric and magnetic charges. In~\cite{Lee:1997vp} the monopoles are obtained from an instanton solution on $R^3\times S^1$ --- caloron. By comparing~(\ref{3DMonDensity}) to the monopole charge densities in~\cite{Lee:1997vp} one concludes that the twist angle $\phi$ must be identified with the holonomy of the caloron solution around $S^1$.

The Coulomb energy of the large density solution can be expressed as
\be
E_{\rm C}\ =\frac{N_c}{8\pi\lambda M d}\left(
	\frac{\phi^2}{2\rho_1}\,+\,\frac{(2\pi-\phi)^2}{2\rho_2}\,
	+\,\frac{\phi(2\pi-\phi)}{b}\right),
\label{GenCoul}
\ee
where $\rho_1$ and $\rho_2$ are effective radii of the monopole density concentrations at $X^{(1)}$ and $X^{(2)}$. Formal expressions for them are given by equations (\ref{rho1app}) and~(\ref{rho2Aapp}) in the appendix~\ref{secSLDensity}. Plugging these charge radii into equation~(\ref{GenCoul}), we write the Coulomb energy
of large instantons as
\be
E_{\rm C}\ =\ \frac{N_c}{\lambda M}\left(
	\frac{C_1(\pi^2+3(\pi-\phi)^2)}{32d^2}\
	+\ \frac{(C_2+2)\pi^2+(C_2-2)(\pi-\phi)^2}{16\pi^2 a^2}\
	+\ O(d^2/a^4)\right).
\label{EClarge}
\ee
Here $C_1\approx\ 1.174$ and $C_2\approx\ 1.761$ are constants defined by some dimensionless integrals~(\ref{C1defapp}) and~(\ref{C2defapp}).

Combining this Coulomb energy with the non-abelian energy (\ref{NAenergy}) and minimizing with respect to the instanton radius~$a$ and twist angle~$\phi$, we find the minimum at
\be
\phi[{\rm @min}]\ =\ \pi,\qquad
a [{\rm @min}]\ \approx\ \left(\frac{\sqrt{C_2+2}}{2\sqrt{2}\pi}\times
	\frac{d}{\lambda M^2}\right)^{1/3}\
\approx\ 0.842\, a_0^{2/3}d^{1/3}.
\label{MinLarge}
\ee

\begin{figure}[t]
\begin{center}
\vspace{3ex}
\includegraphics[width= 150mm]{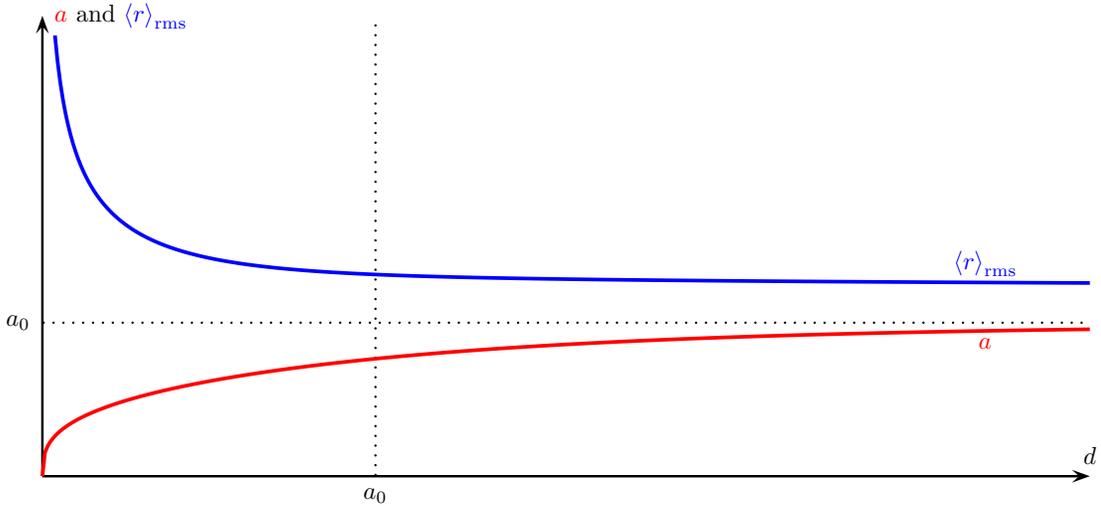}
\caption{Plots of the equilibrium (red) and root-mean-square (blue) instanton radii. Note that only the asymptotic regions $d\gg a_0$ and $d\ll a_0$ of this graph are based on real calculations, but the crossover region $d\sim a_0$ is only an interpolation.}
\label{plotaRrms}
\end{center}
\vspace{-0.5cm}
\end{figure}

Note that when the lattice spacing $d$ becomes much smaller than the equilibrium radius $a_0$ of a standalone instanton, the formal radius parameter $a$ of the instantons in the chain decreases as $a\propto a_0^{2/3}d^{1/3}$. However, the actual concentration of the instanton density shrinks to two separate 3D zones of a much smaller size $O(d)\ll a$ but separated by a much larger distance $b=\pi a^2/d$ that increases for small $d$ as
\be
b\ \approx\ \frac{2.23\,a_0^{4/3}}{d^{1/3}}\,.
\ee
In terms of the root-mean-square (rms) density radius
\be
\vev{r}_{\rm rms}\ \equiv\ \left[\int\!\!d^3x(x_1^2+x_2^2+x_3^2)I_{3D}(x)\right]^{1/2}\
=\ a\times\left[\frac32\,+\,\frac{\pi^2 a^2}{4d^2}\right]^{1/2}
\label{RMS}
\ee
we have something like the blue line in figure~\ref{plotaRrms}.  Here $I_{3D}(x) = d\times I_{4D}(x)$ and exact expressions for the moments are used, \emph{cf.} equations~(\ref{Size12}) and~(\ref{Size3}). Equation~(\ref{RMS}) --- and hence the red line on the graph --- presume $\phi=\pi$ for all lattice spacings $d$. For large and small $d$ we have seen that this is indeed the equilibrium (\emph{i.e.} the lowest-energy) value of the twist angle. For the intermediate $d\sim a_0$ an argument is given in the appendix~\ref{appMinEnergy} that $\phi[{\rm @min}]=\pi$, independent of lattice spacing.

Let us briefly comment on a possible relation between these results and skyrmion-half-skyrmion phase transition discussed in section~\ref{secSkyrmions}. Rho \emph{et al.}~\cite{Rho:2009ym} have suggested that the splitting of each instanton into a pair of monopoles, which occurs in the large instanton regime $d\lesssim a$, is equivalent to the latter transition. Indeed, we have proven that the preferable twist angle is $\phi=\pi$ and therefore each monopole carries a half of the instanton charge.\footnote{The reader should not be confused by the fact that the splitting in our model occurs along the holographic direction. This can be cured by a different choice of the twist direction. All twist directions are degenerate in the case of isotropic gauge coupling~(\ref{GRdependence}). In a 3D lattice the twist direction and thus the instanton splitting, if it occurs, will necessarily be along the spatial directions.} On the other hand the splitting is merely an alternative description of the instanton solution related to the fact that for $a\gg d$ the instanton density is independent from $x_4$. It is a smooth crossover rather than a phase transition, and we do not expect chiral symmetry restoration in the sense of section~\ref{secSkyrmions}. This is perhaps a drawback of the 1D model, which will not be present in 3D lattices. If we look at the parameter $b$ of the monopole separation, it smoothly interpolates between a $a_0^2/d$ behavior for $a\ll d$ and $a_0^{4/3}/d^{1/3}$ for $a\gg d$, that is it smoothly grows with density. For the 3D lattice case the smooth behavior would break the discrete symmetries. It is then conceivable that in the latter case the crossover will be replaced by a phase transition.


\subsection{Abelian zigzag. Exact solution}
\label{secZigzag}

Our main interest in the instanton chain is not in its equilibrium configuration but rather its stability (or instability) against transverse motion (in $x_{1,2,3}$ directions, where $x_3\equiv z$) of the instantons that would get them out of the linear alignment. If such motion can decrease the chain's net energy, than the chain is unstable.

In the analysis of stability it will be useful to break the symmetry in the directions transverse to the chain and introduce anisotropic gauge coupling~(\ref{anisotropic0}). The main motivation for introducing the anisotropy is to make the chain more unstable in the holographic $x_3$ rather than transverse spatial $x_1$, $x_2$ directions. For this apparently we must impose $M<M'$.  The anisotropy also breaks the degeneracy between different directions $\vec n$ of the $SU(2)$ twist $\exp\bigl(i\phi(\vec n\cdot\vec\tau)/2\bigr)$ between adjacent instantons. The lowest-energy direction of the twist is now $\tau_3$ -- which is precisely what we have used in our formulae in the previous section. Specifically, while the Coulomb energy does not depend on the twist direction, the non-abelian energy is minimized when the largest $\vev{x^2}$ of the instantons is oriented in the lowest-potential direction $x_3$. According to equations~(\ref{Size12})--(\ref{Size3}), the instantons are larger in the direction of the twist than in other directions, hence the non-abelian energy
\be
E_{\rm NA}\ =\ N_c\lambda M\left(1\ +\ a^2 M^{\prime2}\ +\ \frac12\, a^2 M^2
+\ \frac{a^4\phi(2\pi-\phi)}{4d^2}\times
\left(M^{\prime2}(n_1^2+n_2^2)\,+\,M^2n_3^3\right)\
+\ O(a^4M^4)\right)
\ee
is minimized for $\vec n=(0,0,\pm1),\ i.\,e.\ SU(2)$ twist in the $\tau_3$ direction.

Also, $M'>M$ reduces the equilibrium size of standalone or far-apart instantons from
(\ref{OrigA}) to
\be
a'_0\ =\ \frac{(1/5)^{1/4}}{(\lambda^2M^2(M^{\prime2}+{1\over2}M^2)^{1/4}}\
\approx\ \frac{(1/5)^{1/4}}{\sqrt{\lambda MM'}}\quad({\rm for}\ M'\gg M).
\label{InstSize}
\ee
Consequently the anisotropy, apart from stabilizing the chain in the $x_1$, $x_2$ directions,  provides an additional control over the equilibrium size of instantons. This will allow for analytical analysis of phase transitions below.

Before we start the stability analysis we must describe the possible displacement in proper ADHM terms of $\Gamma^\mu_{mn}$ matrices and $Y^\mu_n$ vectors. Moving the instantons' centers in $x_{1,2,3}$ directions without changing their $x_4$ locations along the chain or any radii or $SU(2)$ orientations means keeping
\be
Y_n^\mu\ =\ (0,0,a\sin(n\phi/2),a\cos(n\phi/2)),\qquad
\Gamma^4_{mn}\ =\ d\,n\times \delta_{mn}\,,
\ee
exactly as for the straight chain but changing the $\Gamma^{i=1,2,3}_{mn}$ matrices
\be
\Gamma^3_{mn}\ \to\ \Gamma^3_{mn}[\mbox{straight chain}]\
+\ \delta\Gamma^3_{mn}\,,\qquad
\Gamma^1_{mn}\ \to\ \delta\Gamma^1_{mn}\,,\qquad
\Gamma^2_{mn}\ \to\ \delta\Gamma^2_{mn}\,,
\ee
in a manner that preserves the self-duality equations~(\ref{GammaRel}). The solution corresponding to the displacement in the $x_3$ direction reads
\be
\delta\Gamma^{1,2}_{mn}\ \equiv\ 0,\qquad
\delta\Gamma^3_{mn}\ =\ \delta_{mn}\times\delta X^3[n]\,.
\label{SD3mat}
\ee

The gauge coupling keeps the instanton centers lined up along the $x_4$ axis for low instanton density. At high density, such alignment becomes unstable because the abelian Coulomb repulsion between the instantons squeeze them out in the directions transverse to the chain. Since the repulsion is strongest between the nearest neighbors, the leading instability should have adjacent instantons moved in opposite ways forming a zigzag pattern (figure~\ref{figZigzag}),
\be
\delta X^3[n]\ =\ \epsilon\times(-1)^n.
\ee

After defining the ADHM matrices we make a Fourier transform to map the infinite-dimensional matrices to differential operators on a circle. In the following it will be natural to combine wave functions $\psi(\theta)$ and $\psi(\theta\pm\pi)$ on the circle into a two-component wave function. Two-component functions provide a natural description for a two-layered chain. For some particular cases of the twist angle $\phi=0$ or $\pi$ all expressions take particularly compact form. More generally for $n$ layers $n$-component functions should be used instead. Here we choose two-component functions with the following boundary conditions
\be
\label{TwoComponent}
\Psi(\theta)\ =\,\begin{pmatrix} \psi(\theta)+\psi(\theta +\pi)\\ \psi(\theta)-\psi(\theta+\pi)\end{pmatrix},\quad
-\frac{\pi}{2}\le\theta\le\frac{\pi}{2},\quad \Psi(\pi/2)\,=\,\Sigma_3\Psi(-\pi/2),
\ee
where $\Sigma_1,\Sigma_2,\Sigma_3$ are Pauli matrices acting on two components. In the space of such two-component wave functions, the $\Gamma^3$ operator becomes
\be
\label{Gamma3beta0}
\Gamma^3\ = \ \epsilon\times\Sigma_3 +\frac{\pi a^2}{2d}\Theta\bigl(-\frac{\phi}{2}<\theta<\frac{\phi}{2}\bigr)\times\Sigma_1 + \frac{\pi a^2}{2d}\left(\frac{\phi}{\pi}-\Theta\bigl(-\frac{\phi}{2}<\theta<\frac{\phi}{2}\bigr)\right)\times\mathbbm{1}\,,
\ee
where $\Theta$ is the step-function, \emph{i.e.} $\Theta=1$ if $- \phi/2<\theta<\phi/2$ and $\Theta=0$ otherwise; and we restrict $\phi$ to the values $0\leq\phi\leq\pi$. For $\phi=\pi$ the latter expression takes the form
\be
\Gamma^3\ =\ \frac{\pi a^2}{2d}\times\Sigma_1\ +\ \epsilon\times\Sigma_3\,.
\ee

Matrix $T\equiv Y^\mu\otimes Y^\mu$ in the two-component notation becomes
\be
\label{Tbeta0}
T(\theta)= \frac{\pi a^2}{2}\left(\mathbbm{1}+\Sigma_1\right)\left(\delta\bigl(\theta-\frac{\phi}{2}\bigr) + \delta\bigl(\theta+\frac{\phi}{2}\bigr)\right), \qquad 0\leq \phi<\pi\,.
\ee
For $\phi=\pi$ the latter can just be written as $T=\pi a^2\delta(\theta\pm \pi/2)\times\mathbbm{1}$, \emph{cf.} equation~(\ref{Tantiferromagnetic}).

Next we plug the data above in the formula for the operator $L$~(\ref{Ldef}). To calculate the determinant of this operator, one needs to calculate its Green's function $L^{-1}(\theta,\theta_0)$ and consequently the traces~(\ref{Derivatives}). As in the example of the straight chain the naive determinant is divergent and needs to be regularized, which is done by extracting an infinite but constant prefactor. One can reconstruct an analytic expression for $\det(L)$ with any value of $\phi$. However only for $\phi=0$ and $\pi$ the expression is compact enough to present in the paper. Accidentally $\phi=\pi$ is the minimal energy configuration of instantons for both the straight chain at low density and in a zigzag phase provided that the zigzag amplitude is not too large. A proof of this fact will be presented shortly. In the remainder of this section we discuss the case $\phi=\pi$. One finds that the regularized determinant reads
\begin{align}
\frac{\det(L)}{\rm const}\ =\ &
\left( \cosh\frac{\pi r_1}{d}\,+\,\frac{\pi a^2}{dr_1}\,\sinh\frac{\pi r_1}{d}\right)\times
\left( \cosh\frac{\pi r_2}{d}\,+\,\frac{\pi a^2}{dr_2}\,\sinh\frac{\pi r_2}{d}\right)\nonumber\\
&\qquad+\ \frac{r_1^2+r_2^2-(\pi a^2/d)^2}{2r_1 r_2}\,
	\sinh\frac{\pi r_1}{d}\,\sinh\frac{\pi r_2}{d}\
-\ \cos\frac{2\pi x_4}{d}\nonumber\\
&+\ \sin\nu\times\cos\frac{\pi x_4}{d}\times\left(
	\left( 2\cosh\frac{\pi r_1}{d}\,+\,\frac{\pi a^2}{dr_1}\,\sinh\frac{\pi r_1}{d}\right)\,
	-\,\left( 2\cosh\frac{\pi r_2}{d}\,+\,\frac{\pi a^2}{dr_2}\,\sinh\frac{\pi r_2}{d}\right)
	\right)\nonumber\\
&+\ \sin^2\nu\times\left(
	-1\,+\,\cosh\frac{\pi r_1}{d}\,\cosh\frac{\pi r_2}{d}\,
	-\,\frac{r_1^2+r_2^2}{2r_1r_2}\sinh\frac{\pi r_1}{d}\,\sinh\frac{\pi r_2}{d}
	\right),
\label{DetLmod}
\end{align}
where
\begin{align}
r_{1,2}^2\ &=\ x_1^2\ +\ x_2^2\ +\ (x_3\,\mp\,\tfrac12 b_\epsilon)^2,\\
\label{be}
b_\epsilon\ &=\ \sqrt{4\epsilon^2+(\pi a^2/d)^2},\\
\nu\ &=\ \arctan\frac{2\epsilon}{\pi a^2/d}\ =\ \arcsin\frac{2\epsilon}{b_\epsilon}\,.
\label{nu}
\end{align}

Naively, we would expect the zigzag deformation to have no effect on the width of the instanton chain in the $x_1$ and $x_2$ directions while the width squared in the $x_3$ direction should increase by $\epsilon^2$. And indeed, this is precisely what happens for any instanton radius and lattice spacing: evaluating (\ref{vevx2}) by integrating by parts, we obtain precisely
\be
\vev{x_1^2}\ =\ \vev{x_2^2}\ =\ \frac{a^2}{2},\qquad
\vev{x_3^2}\ =\ \frac{a^2}{2}\ +\ \frac{\pi^2 a^4}{4d^2}\ +\ \epsilon^2.
\ee
Consequently, the non-abelian energy of the zigzag is, {\it cf.} (\ref{NAshift}),
\begin{align}
E_{\rm NA}\ &
=\ N_c\lambda M\left( \bigl(\tfrac12 M^2+M^{\prime2}\bigr)\times a^2\
	+\ M^2\times\frac{\pi^2 a^4}{4d^2}\ +\ M^2\times\epsilon^2\right)\nonumber\\
&=\ E_{\rm NA}[\epsilon=0]\ +\ N_c\lambda M^3\times\epsilon^2.
\label{ENAmod}
\end{align}

\begin{figure}[t]
\begin{center}
\vspace{3ex}
\includegraphics[width= 100mm]{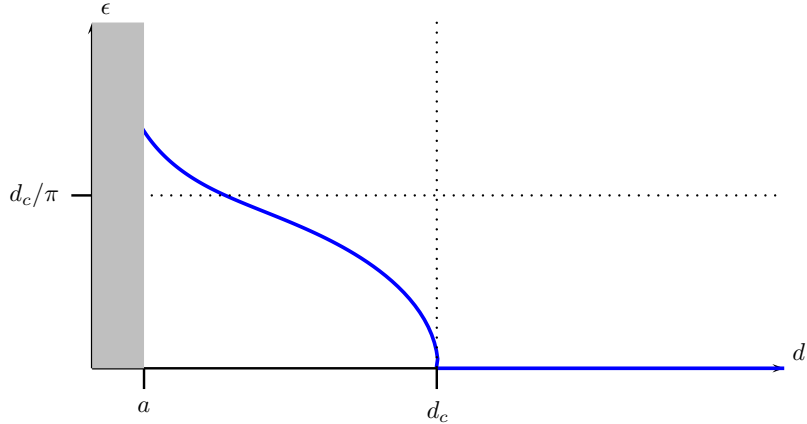}
\end{center}
\caption{The zigzag amplitude $\epsilon$ as a function of the lattice spacing $d$ for small instantons, $a\ll d$.}
\label{SmallZigzag}
\end{figure}

As to the Coulomb energy, we will restrict ourselves to the $d\gg a$ regime, where analytical calculation is possible.
For small, widely-separated instantons, we approximate
\begin{align}
\mbox{for}\ r_1\sim a\ll d &\ \mbox{but finite}\ \epsilon/d,\nonumber\\
\det(L)\ \approx\ &
\frac{\pi^2}{d^2}\left( (a^2+r_1^2+x_4^2)\,+\,\frac{\pi^2}{12d^2}\,(r_1^4+2a^2r_1^2-x_4^2)\,
	+\,O(a^6/d^4)\right)\times{}\nonumber\\
&\qquad\qquad\qquad\times\left( \cos\frac{\pi x_4}{d}\,+\,\cosh\frac{\pi r_2}{d}\,
	+\,\frac{\pi a^2}{2r_2d}\,\sinh\frac{\pi r_2}{d}\right)\nonumber\\
&+\ \frac{\pi^2 a^4}{8d^2\epsilon^2}\times\frac{\pi r_2}{d}\,\sinh\frac{\pi r_2}{d}
\label{modDsmall}
\end{align}
and consequently obtain
\be
E_{\rm C}\ =\ \frac{N_c}{\lambda M}\left[ \frac{1}{5a^2}\ +\ \frac{3\pi^2}{80d^2}\
	+\ \frac{\pi^2}{80d^2}\times\frac{\tanh(\pi\epsilon/d)}{\pi\epsilon/d}\
	+\ O(a^4/d^6)\right].
\label{ECsmallMod}
\ee
Remarkably, the $\epsilon$--dependent part of this Coulomb energy is precisely 5 times smaller than the energy of point charges in a similar zigzag formation~(\ref{1dChargesEn}). Mathematically, this fivefold reduction stems from the last term in equation~(\ref{modDsmall}), which accounts for the interference between the instantons. It is not clear why the interference between well-separated small instantons
has such a drastic effect on the Coulomb energy of the zigzag. Anyhow, the net energy cost of a small zigzag deformation $\epsilon\ll d$ is
\be
\Delta E_{\rm net}\ =\ \Delta E_{\rm NA}\ +\ \Delta E_{\rm C}\
=\ N_c\lambda M^3\times\epsilon^2\ +\ \frac{N_c}{\lambda M}\left[
	-\frac{\pi^4 \epsilon^2}{240 d^4}\
	+\ \frac{\pi^6\epsilon^4}{600 d^6}\
	+\ O(\epsilon^4/d^6)\right].
\label{ECsmallEdep}
\ee
This cost is positive --- and hence the straight chain is stable --- for all sufficiently large lattice spacings
\be
d\ >\ d_{c}\ \equiv\ \frac{\pi}{\root 4\of{240}}\times\frac{1}{M\sqrt{\lambda}}\,.
\label{Dcrit}
\ee
For smaller lattice spacings $d<d_{c}$, the energy function~(\ref{ECsmallEdep}) has a negative coefficient of $\epsilon^2$ but positive coefficient of $\epsilon^4$. Thus, for $d<d_c$ the straight chain becomes unstable and there is a second-order phase transition to a zigzag configuration. For lattice spacings just below critical, the zigzag parameter $\epsilon$ is
\be
\vev{\epsilon}\ \approx\ \pm\frac{\sqrt{5}}{\pi}\times\sqrt{d_c(d_c-d)}\,.
\label{SecondOrder}
\ee
For smaller lattice spacing (but larger than the instanton size), the zigzag parameter satisfies the same equation as in the point charge limit~(\ref{Esolution}) with the new $d_{\rm c}$. Notice again that the new critical spacing is precisely $\sqrt[4]{5}$ times smaller than the one in the point charge limit. Graphically, the zigzag parameter $\epsilon$ as a function of the lattice spacing behaves as demonstrated in figure~{\ref{SmallZigzag}}.
Note that this curve can be trusted only for lattice spacings larger than the instanton size since it is based on equation~(\ref{ECsmallMod}) for the Coulomb energy which presumes $d\gg a$. In units of the equilibrium instanton size~(\ref{InstSize}), the critical lattice spacing~(\ref{Dcrit}) is
\be
\frac{d_{\rm crit}}{a\approx a_0'}\
=\ \frac{\pi}{\root 4\of{48}}\times\left(\frac{M^{\prime2}+{1\over2}M^2}{M^2}\right)^{1/4}\
\approx\ 1.2\sqrt{M'\over M}\quad({\rm for}\ M'\gg M).
\ee
For a highly anisotropic gauge coupling with $M'\gg M$, the critical spacing is much larger than the instantons, which justifies our approximations. However, for near-isotropic couplings with $M'\approx M$, the critical spacing is only $d_{\rm crit}\approx 1.32 a_0$, and in this regime we cannot be sure our analysis of the zigzag instability is even qualitatively correct. That is, we are not sure that  there is a transition from a straight instanton chain to a zigzag
for $M'\approx M$, never mind any details of such transitions. On the other hand, for the highly-anisotropic $M'\gg M$ setups, we can trust that the transition does happen at the critical density~(\ref{Dcrit}), and that the zigzag amplitude below the transition behaves according to the curve~(figure~\ref{SmallZigzag}), at least for the lattice spacings not too much smaller than the critical.


\subsection{General twist. Approximate solution}
\label{secGeneralTwist}

The zigzag instanton solution considered in the previous section  was derived assuming $\phi=\pi$. When the density is increased and the zigzag amplitude grows large enough one expects a series of transitions similar to the ones discussed in the case of point charges (table~\ref{tabPCSequence} and figure~\ref{figChainPhase}). However the phase space of the instanton chain must be richer: \emph{e.g.} one can expect transitions driven by a change of their mutual orientation. Indeed, for the zigzag amplitude $\epsilon$ comparable to the chain spacing $d$, it is natural to expect that the orientation of instantons will respect the essentially 2D character of the lattice. In particular twists between adjacent instantons may appear along multiple directions in the $SU(2)$ space rather than a single $U(1)$ subgroup -- a non-abelian zigzag. In this section we consider a more general ansatz for the instanton twists and show that while for small amplitudes the $\phi=\pi$ twist is indeed preferred, for large ones it asymptotes to $\phi=\pi/2$, that is an anti-ferromagnetic order inside each of the two zigzag layers. Surprisingly non-abelian twist configurations are always disfavored, for any values of density in the $d\gg a$, $M'\gg M$ approximation.

Expressions for instanton solutions obtained for generic twist are quite monstrous. Even though their derivation is straightforward in principle the results can hardly be presented in a compact form. On the other hand for the sake of keeping our computations analytical we have to resort to the $d\gg a$ expansion anyway, when computing the Coulomb energy. Therefore in this section we will employ this expansion already at the level of the instanton solution.


\subsubsection{General observations}
\label{sssecNAgeneral}

Let us summarize some observations about the instanton size and energy, based on the examples of single and multi-instanton solutions considered above. A single stand-alone instanton of radius $a$ centered at $x_{1,2,3}=0$ has non-abelian energy (of the $SU(2)$ magnetic fields)
\be
E_{\rm NA}\ =\ N_c\lambda M\times\Bigl(1\,+\,(M^{\prime2}+\tfrac12\, M^2)\times a^2\Bigr)\,,
\ee
while the $U(1)$ electric fields (due to Chern-Simons coupling to the instanton density) have 5D Coulomb energy
\be
E_{\rm C}\ =\ {N_c\over\lambda M}\times{1\over 5a^2}\,.
\ee
Minimizing the net energy $E_{\rm NA}+E_{\rm C}$ with respect to the instanton radius, we find
\begin{align}
a\ &
=\ {1\over\sqrt{\lambda MM'}}\times\root 4\of{2M^{\prime2}\over5(2M^{\prime2}+M^2)}\
\xrightarrow[M'\gg M]\ {5^{-1/4}\over\sqrt{\lambda MM'}}
\label{StandAloneA}\\
{\rm and}\quad \frac{E_{\rm net}}{\lambda N_c}\ &
=\ M\ +\ \frac{1}{\lambda}\sqrt{\frac45}\,\sqrt{M^{\prime2}+\frac12\, M^2}\
\xrightarrow[M'\gg M]\ M\ + \ \sqrt{\frac45}\, \frac{M'}{\lambda}\,.
\label{StandAloneE}
\end{align}

For an infinite chain (straight or zigzag shaped) of instantons with lattice spacing $d$ much larger than the instanton radius $a$, the non-abelian energy per instanton has general form
\begin{multline}
E_{\rm NA}\ =\ N_c\lambda M\bigg(1\ +\ (M^{\prime2}+\frac12\, M^2)\times a^2\ +\ M^2\times \epsilon^2 \ +  \\
\ +\ M^2\,{a^4\over d^2}\times{\cal A}\ +\ M^{\prime2}{a^4\over d^2}\times{\cal B}\
+\ O(M^{\prime2}a^6/d^4)\biggr)\,,
\label{ENAgeneric}
\end{multline}
where $\epsilon$ is the zigzag amplitude and $\cal A$, $\cal B$ are some $O(1)$ functions of the  $SU(2)$ twists between instantons and also of the $\epsilon/d$ ratio, \emph{cf.} equations~(\ref{NAenergy}) and~(\ref{ENAmod}). Likewise, the Coulomb energy per instanton of a chain or zigzag has general form
\be
E_{\rm C}\ =\ {N_c\over\lambda M}\left( {1\over 5a^2}\
+\ {{\cal C}\over d^2}\ +\ O(a^2/d^4)\right),
\label{ECgeneric}
\ee
for some function $\cal C$ of the inter-instanton twists and $\epsilon/d$, \emph{cf.}~(\ref{ECstraight}), (\ref{ECsmallMod}). Now let us assume  that $M'\gg M$ and a  lattice spacing $d\sim (\sqrt{\lambda}M)^{-1}\gg a$. With these assumptions, we may approximate the net energy as
\begin{align}
E_{\rm net}\ =\ N_c\lambda M\ &
+\ N_c M'\left( \lambda MM' a^2\,+\,{1\over5\lambda MM' a^2}\right)\\
&+\ N_cM\left(
	 (\lambda M^2 d^2)\times(\epsilon/d)^2\,
	+\,{(\lambda MM' a^2)^2\over(\lambda M^2d^2)}\times{\cal B}\,
	+\,{1\over(\lambda M^2d^2)}\times{\cal C}\right)\\
&+\ O(N_c M^2/M')\,.
\end{align}
Minimizing this net energy with respect to the instanton radius we find
\be
\lambda MM'\times a^2\ =\ {1\over\sqrt{5}}\ +\ O(M/M'),
\ee
similar to a standalone instanton, and
\be
E_{\rm net}\ =\ E_B\ +\ N_c M\left( (\lambda M^2 d^2)\times(\epsilon/d)^2\, +\,{{\cal B}+5{\cal C}\over 5(\lambda M^2d^2)} \right)\ +\ O(N_c M^2/M')\,,
\label{ENgeneric}
\ee
where $E_B$ is the energy of a standalone instanton, \emph{cf.} equation~(\ref{StandAloneE}). Note that the lattice spacing $d$ and the zigzag amplitude $\epsilon$ enter this formula via the ratio $\epsilon/d$ and the combination $\lambda M^2d^2$, so the natural scale for $\epsilon$ and $d$ -- at which we might have some phase transition(s) -- is
\be
d,\epsilon\ \sim\ {1\over\sqrt{\lambda}M}\,.
\ee
For $M'\gg M$ this scale is much larger than the equilibrium instanton radius~(\ref{StandAloneA}), which justifies the perturbative expansion of the non-abelian and Coulomb energies in powers of
\be
{a^2\over d^2}\ \sim\ {M\over M'}\ \ll\ 1.
\ee

This concludes our general analysis, but the devil is in details. To work out the phase structure of the instanton zigzag, we need to specify the instanton orientations in $SU(2)$ in terms of some moduli, solve the ADHM equations for the non-diagonal matrix elements of the $\Gamma^\mu_{mn}$ matrices, calculate the instanton number density $I(X)$, then use it to obtain the non-abelian and Coulomb energies per instanton. Writing those energies in the form~(\ref{ENAgeneric}) and~(\ref{ECgeneric}) for some functions $\cal A$, $\cal B$, and $\cal C$ of $\epsilon/d$ and the moduli of the instanton orientations, we plug them into equation~(\ref{ENgeneric}) and minimize the combination
\be
{\cal F}\
\equiv\ (\lambda M^2 d^2)\times(\epsilon/d)^2\ +\ {{\cal B}+5{\cal C}\over 5(\lambda M^2d^2)}\,.
\label{Fdef}
\ee


\subsubsection{ADHM data}
\label{sssecNAdata}

As before we are interested in infinite zigzag configurations with periodicity $d$ in $x_4$ direction and amplitude $\epsilon$ in $x_3$ direction. Therefore instanton positions (diagonal elements $ X^\mu_n\equiv \Gamma^\mu_{nn}$) are given by
\be
X^4_n\ =\ n d,\quad X^3_n\ =\ (-1)^n \epsilon,\quad X_n^1\ =\ X_n^2\ =\ 0,\quad n\in{\bf Z}\,.
\label{Centers}
\ee
In general we expect $\epsilon,d=O(1/\sqrt{\lambda}M)$, also we allow for a straight chain with $\epsilon=0$. We assume that the radii and the orientations of the instantons share the geometric symmetries of the zigzag: all instantons have the same radius $a$, while the relative $SU(2)$ orientations between neighboring instantons depend on parity of $n$:
\be
\label{genTwist}
Y_{2n+1}\ =\ U_1\times Y_{2n}\,,\quad Y_{2n}\ =\ U_2\times Y_{2n-1}\,,\quad
U_1,U_2\in SU(2).
\ee
In $SO(3)$ terms, $U_1$ and $U_2$ are rotations through the same angle $\alpha$ but around different axes $\bn_1$ and $\bn_2$,
\be
\label{genU}
U_1\ =\ \exp\bigl(i(\alpha/2){\vec{\bf n}_1\cdot\vec{\tau}}\bigr),\quad
U_2\ =\ \exp\bigl(i(\alpha/2){\vec{\bf n}_2\cdot\vec{\tau}}\bigr).
\ee
We find it convenient to reparameterize these $SU(2)$ matrices in terms of two angles $\phi$ and $\beta$ and two unit vectors $\bp$ and $\bq$ that are perpendicular to each other:
\be
U_{1,2}\ =\ \exp\bigl(\mp i(\beta/2)\bq\cdot\vec\tau\bigr)\times
\exp\bigl(i(\phi/2)\bp\cdot\vec\tau\bigr)\times\exp\bigl(\mp i(\beta/2)\bq\cdot\vec\tau\bigr)
\label{Reparam}
\ee
so that the individual instantons' $Y_n$ parameters get a rather symmetric form
\be
Y_n\ =\ a\times\exp\bigl(in(\phi/2)\bp\cdot\vec\tau\bigr)
\times\exp\bigl(i(-1)^n(\beta/2)\bq\cdot\vec\tau\bigr).
\label{GeneralInstantons}
\ee
The parametrization of~ (\ref{Reparam}) is redundant, so we limit the range of angles to $0\le\phi\le\pi$ and $0\le\beta\le\pi/2$. But even in this range some redundancies remain: For $\phi=0$ the $\bp$ vector is irrelevant, while for $\phi=\pi$  we have $U_1=U_2=i\bp\cdot\vec\tau$ regardless of the $\beta$ and $\bq$ moduli. Also, thanks to the sign redundancy of ADHM data $Y_n\to(\pm)_n Y_n$, for $\beta=\pi/2$ the exchange
\be
(\phi,\bp,\bq)\ \leftrightarrow\ (\pi-\phi,\bq,\bp)
\label{Symmetry}
\ee
has no physical effect. We should keep these redundancies in mind when we seek the lowest-energy configurations of the instanton zigzag.

The instanton centers~(\ref{Centers}) govern the diagonal matrix elements of the quaternion-valued symmetric matrix $X$, which we expand in the basis of $\tau$-matrices as in equation~(\ref{GammaDef}). The non-diagonal matrix elements of $X$ follow from the ADHM equations~(\ref{ADHMeqs}). We can solve these equations in the matrix form, but we find it easier to work in a different basis where the coefficient matrices $\Gamma^{\mu}$, $Y^\mu\otimes Y^\mu$ become operators acting on two-component wavefunctions of a periodic variable as we defined in equation~(\ref{TwoComponent}). In this basis, the $T_{mn}={\rm Re}(Y_m^\dagger Y^{}_n)$ matrix becomes a singular local operator, \emph{cf.}~(\ref{Tbeta0}),
\be
T(\theta)\ =\ {\pi a^2\over2}\left(1+\Sigma_1\cos\beta\right)\times
\left(\delta(\theta+\frac{\phi}{2})+\delta(\theta-\frac{\phi}{2})\right).
\label{Toperator}
\ee
Likewise, the imaginary quaternionic matrix ${\rm Im} (Y_m^\dagger Y^{}_n)=i\vec\tau\cdot\vec{\bf t}_{mn}$ becomes a singular local operator $i\vec\tau\cdot\vec{\bf t}(\theta)$ where
\begin{multline}
\vec{\bf t}(\theta)\ =\ {\pi a^2\over2}\biggl[
\Bigl(\bp(\Sigma_1+\cos\beta)\,+\,(\bp\times\bq)\Sigma_3\sin\beta\Bigr)
	 \times\Bigl(-i\delta(\theta+\frac{\phi}{2})+i\delta(\theta-\frac{\phi}{2})\Bigr) \ +\cr
\ +\ \bq\,\Sigma_2\sin\beta\times
	 \Bigl(-i\delta(\theta+\frac{\phi}{2})-i\delta(\theta-\frac{\phi}{2})\Bigr)\biggr].
\label{Jumps}
\end{multline}
As to the $\Gamma^\mu_{mn}$ matrices, the $\Gamma^4$ is the same differential operator as before
\be
\Gamma^4\ =\ -id\,\frac{\partial}{\partial\theta}\,,
\ee
while the 3--vector $\vec\Gamma$ becomes ($2\times2$)-matrix-valued local function of $\theta$ satisfying ADHM equations~(\ref{GammaRel}) in 3--vector form
\be
-id\,\frac{\partial}{\partial\theta}\,\vec\Gamma(\theta)\
+\ \vec\Gamma(\theta)\times\vec\Gamma(\theta)\
+\ \vec{\bf t}(\theta)\ =\ 0\, ,
\label{ADHM}
\ee
subject to boundary conditions
\be
\Gamma(\theta=-\pi/2)\ =\ \Sigma_3\,\Gamma(\theta=+\pi/2)\,\Sigma_3\,.
\label{BC}
\ee
These equations may be solved exactly in terms of an auxiliary variable related to $\theta$ via an elliptic integral, or rather two auxiliary variables, one for $-\phi/2<\theta<\phi/2$ and the other for $\theta<-\phi/2$ or $\theta>\phi/2$ --- the solutions are discontinuous at $\theta=\pm \phi/2$ because of the $\delta$-functions in the $\vec{\bf t}(\theta)$ term~(\ref{Jumps}). However the exact solutions are too messy to work with and we will focus on the $a\ll d$ limit of the $\vec\Gamma(\theta)$.

Instead of expanding the exact solution in powers of $a^2/d^2$, it is easier to solve the ADHM equations~(\ref{ADHM}) perturbatively, starting from zero-order zigzag solution~(\ref{SD3mat}) considered in the previous section.
\be
\bigl(\vec\Gamma_0\bigr)_{mn}\ =\ \epsilon\,\bn_3\times\delta_{mn}(-1)^n\ \mapsto\
\vec\Gamma_0(\theta)\ \equiv\ \epsilon\,\bn_3\times\Sigma_3\,,
\ee
where $\bn_3$ is the unit vector in the $x_3$ direction. The first-order term in $\vec\Gamma(\theta)$ satisfies the linearized equations~(\ref{ADHM}), namely
\be
-id\frac{\partial}{\partial\theta}\,\vec\Gamma_1(\theta)\ +\ 2\epsilon\,\bn_3\times\left[\Sigma_3,\vec\Gamma_1(\theta)\right]\
+\ \vec{\bf t}(\theta)\ =\ 0 \,.
\label{gammaeqs}
\ee
The diagonal 11 and 22 matrix elements of the solution to these equations are piecewise-constant functions of $\theta$,
\be
\bigl(\vec{\Gamma}_1\bigr)_{\rm diag}(\theta)\
=\ {\pi a^2\over2d}\Bigl(\bp\,\cos\beta\,+\,(\bp\times\bq)\Sigma_3\sin\beta\Bigr)
\times\left({\phi\over\pi}-\Theta(-\frac{\phi}{2}<\theta<\frac{\phi}{2})\right),
\label{SolDiag}
\ee
where $\Theta$ is the step-function, while the off-diagonal 12 and 21 matrix elements are more complicated: for $-\phi/2<\theta<\phi/2$
\begin{align}
\nonumber
\bigl(\vec{\Gamma}_1\bigr)^{}_{12}(\theta)\ & =\ \bigl(\vec{\Gamma}_1\bigr)^*_{21}(\theta)
\\ \nonumber
&=\ {\pi a^2\over2d}\Bigl(-p_3\,\bn_3\ +\ \bv\,\cosh(2(\epsilon/d)\theta)\ - \ i(\bn_3\times\bv)\,\sinh(2(\epsilon/d)\theta)\Bigr),\\
{\rm where}\quad\bv\ & =\ {1\over\cosh(\pi \epsilon/d)}\Bigl( 	 -\bp_\perp\,\cosh((\pi-\phi)\epsilon/d)\,  +(\bn_3\times\bq_\perp)\sin\beta\sinh ((\pi-\phi)\epsilon/d)
	\Bigr)\,,
\label{gammain}
\end{align}
while for $\phi/2<\theta<\pi/2$, or $-\pi/2<\theta<-\phi/2$
\begin{align}
\nonumber
\bigl(\vec{\Gamma}_1\bigr)^{}_{12}(\theta)\ & =\ \bigl(\vec{\Gamma}_1\bigr)^*_{21}(\theta)
\\ \nonumber
&=\ \pm{\pi a^2\over2d}\left(\eqalign{
	-iq_3\sin\beta\,\bn_3\ &+\ \bv'\,\cosh((\epsilon/d)(\pi-2|\theta|))\cr
	&\pm\ i(\bn_3\times\bv')\,\sinh ((\epsilon/d)(\pi-2|\theta|))\cr
	}\right),\\
{\rm where}\quad\bv'\ &
=\ {i\over\cosh(\pi \epsilon/d)}\Bigl(
	(\bn_3\times\bp)\sinh(\phi \epsilon/d)\,
	-\,\bq_\perp\,\sin\beta\cosh (\phi \epsilon/d)
	\Bigr).
\label{gammaout}
\end{align}
In our notations, $p_3$ and $q_3$ are the $x_3$ components of the $\bp$ and $\bq$ unit vectors while $\bp_\perp$ and $\bq_\perp$ are their components perpendicular to the $x_3$ direction; the $\pm$ sign in equation~(\ref{gammaout}) is the sign of $\theta$. The next order in perturbation theory is rather complicated, so for the present purposes we simply let
\be
\vec\Gamma(\theta)\ =\ \epsilon\bn_3\,\Sigma_3\ +\ \vec\Gamma_1(\theta)\
+\ O(a^4/d^3).
\label{GammaExpansion}
\ee

In summary we have derived the ADHM data up to the first non-trivial order in $a/d$. Now we are going to present the results for the energy of the zigzag configuration as a function of the twist parameters. Details of this calculation can be found in the appendix~\ref{appGenTwist}. The energy needs to be minimized at given density to find the stable zigzag configuration.


\subsubsection{Phase space of the zigzag solution}
\label{sssecNAtwist}

To find the minimum energy zigzag configuration of instantons we need to evaluate the energy of the zigzag per instanton and read off the coefficients $\mathcal{B}$ and $\mathcal{C}$ defined by the general form of the net energy~(\ref{ENgeneric}) assuming $M'\gg M$ limit. Recall that in this limit the small instanton expansion is justified: $a\ll d$. As before, one first computes the determinant of matrix $L$~(\ref{Ldef}), then uses equations~(\ref{Idensity}) and~(\ref{A0eq2}) to find the instanton density and gauge field $\hat{A}_0$, and plugs the result in the formula for the instanton energy~(\ref{ActionInstanton}). Expansion of the expression for $\log\det L$ for the generalized ADHM data~(\ref{Toperator}) and (\ref{SolDiag})-(\ref{gammaout}) is presented in the appendix~\ref{appGenTwist}. Calculation of the net energy per instanton for the optimal instanton radius in the $M'\gg M$ limit gives, \emph{cf.} equations~(\ref{ENgeneric}) and~(\ref{Fdef}),
\begin{align}
E_{\rm net}\ =\ E_B\ &
+\ {\pi^2\,N_c M\over80(\lambda M^2d^2)}\times{\cal F}({\rm moduli})\
+\ O(N_c M^2/M'),\\
{\rm where}\quad{\cal F}\ =\ 3\ &
-\ {4\phi(\pi-\phi)\over\pi^2}\times
	 \Bigl(2\,-\,\bp_\perp^2\,\cos^2\beta\,-\,(\bp\times\bq)^2_\perp\,\sin^2\beta\Bigr)\cr
&+\ {\tanh(\pi \epsilon/d)\over(\pi \epsilon/d)}\times
	 \Bigl(5\,-\,4\sin^2\beta\,+\,2\bp_\perp^2\,+\,2\bq_\perp^2\,\sin^2\beta\Bigr)\cr
&+\ {\sinh((2\phi-\pi)\epsilon/d)\over (\pi \epsilon/d)\cosh (\pi \epsilon/d)}\times
	 \Bigl(-4\,+\,4\sin^2\beta\,+\,2\bp_\perp^2\,-\,2\bq_\perp^2\,\sin^2\beta\Bigr)\cr
&+\ {80\over\pi^4}\,(\lambda M^2d^2)^2\times(\pi \epsilon/d)^2.
\label{NetEnergy}
\end{align}
To find the phase structure of the instanton zigzag, we now need to minimize this energy with respect to the moduli $\epsilon/d$, $\phi$, $\beta$, $\bp$, and $\bq$. Let's start with the $\beta$, $\bp$, and $\bq$ moduli while keeping the $\pi \epsilon/d$ ratio and the $\phi$ angle fixed. Since the energy~(\ref{NetEnergy}) depends linearly on $\sin^2\beta$ and $\cos^2\beta$, its minimum with respect to $\beta$ must be at either $\beta=0$ or $\beta=\pi/2$. For $\beta=0$ we have an abelian zigzag, the $\bq$ parameter is irrelevant, and the energy dependence on the other moduli has form
\begin{multline}
{\cal F}(\beta=0)\ =\ 3\
-\ {4\phi(\pi-\phi)\over\pi^2}\times\bigl[2-\bp_\perp^2\bigr]\
+\ {\tanh(\pi \epsilon/d)\over(\pi \epsilon/d)}\times\bigl[5+2\bp_\perp^2\bigr]\cr
+\ {\sinh((2\phi-\pi)\epsilon/d)\over (\pi \epsilon/d)\cosh (\pi \epsilon/d)}\times\bigl[-4+2\bp_\perp^2\bigr]
+\ {80\over\pi^4}\,(\lambda M^2d^2)^2\times(\pi \epsilon/d)^2.
\end{multline}
This function depends on the $\bp$ unit vector as ${\cal F}=A+B\bp_\perp^2$ where the $B$ coefficient is always positive (except for $\phi=0$ when $\bp$ is irrelevant). Consequently, the abelian zigzag always prefer $\bp_\perp=0$ --- the $\bp$ vector should be parallel to the $X^3$ axis. For this choice, the abelian zigzag's energy as a function of the remaining moduli $\epsilon/d$ and $\phi$ is
\begin{multline}
{\cal F}_{\rm abelian}(\phi,\epsilon/d)\ =\ 3\
-\ {8\phi(\pi-\phi)\over\pi^2}\ +\ 5\times{\tanh(\pi \epsilon/d)\over(\pi \epsilon/d)}
-\ 4\times {\sinh((2\phi-\pi)\epsilon/d)\over (\pi \epsilon/d)\cosh (\pi \epsilon/d)}\cr
+\ {80\over\pi^4}\,(\lambda M^2d^2)^2\times(\pi \epsilon/d)^2.
\label{Fabelian}
\end{multline}

Now consider the non-abelian zigzags with $\beta=\pi/2$. For such zigzags, the energy depends on the other moduli as
\begin{align}
{\cal F}(\beta=\pi/2)\ =\ 3\ &
-\ {4\phi(\pi-\phi)\over\pi^2}\times
	 \Bigl(2\,-\,(\bp\times\bq)^2_\perp\,=\,\bp_\perp^2\,+\,\bq_\perp^2\Bigr)\cr
&+\ {\tanh(\pi \epsilon/d)\over(\pi \epsilon/d)}\times
	\Bigl(1\,+\,2\bp_\perp^2\,+\,2\bq_\perp^2\Bigr)\ +\ {\sinh((2\phi-\pi)\epsilon/d)\over (\pi \epsilon/d)\cosh (\pi \epsilon/d)}\times
	\Bigl(2\bp_\perp^2\,-\,2\bq_\perp^2\Bigr)\cr
&+\ {80\over\pi^4}\,(\lambda M^2d^2)^2\times(\pi \epsilon/d)^2.
\end{align}
As a function of the vectors $\bp$ and $\bq$, this energy has form ${\cal F}=A+B\bp_\perp^2+C\bq_\perp^2$, where the coefficients $B$ and $C$ can be both positive and negative, so at the minimum the $\bp$ and $\bq$ vectors must be either parallel or perpendicular to the $X^3$ axis. Since $\bp\perp\bq$, this gives us three choices:
\begin{align}
{\rm(N1)}:\quad \bp\parallel X^3,& \quad \bq\perp X^3,\cr
{\cal F}_{\rm N1}\ =\ 3\ &
-\ {4\phi(\pi-\phi)\over\pi^2}\ +\ 3\times {\tanh(\pi \epsilon/d)\over(\pi \epsilon/d)}\
-\ 2\times {\sinh((2\phi-\pi)\epsilon/d)\over (\pi \epsilon/d)\cosh (\pi \epsilon/d)}\cr
&+\ {80\over\pi^4}\,(\lambda M^2d^2)^2\times(\pi \epsilon/d)^2,\cr
{\rm(N2)}:\quad \bq\parallel X^3,& \quad \bp\perp X^3,\cr
{\cal F}_{\rm N2}\ =\ 3\ &
-\ {4\phi(\pi-\phi)\over\pi^2}\ +\ 3\times {\tanh(\pi \epsilon/d)\over(\pi \epsilon/d)}\
+\ 2\times {\sinh((2\phi-\pi)\epsilon/d)\over (\pi \epsilon/d)\cosh (\pi \epsilon/d)}\cr
&+\ {80\over\pi^4}\,(\lambda M^2d^2)^2\times(\pi \epsilon/d)^2,\cr
{\rm(N3)}:\quad\! \bp\perp X^3& \quad{\rm and}\quad \bq\perp X^3,\cr
{\cal F}_{\rm N3}\ =\ 3\ &
-\ {8\phi(\pi-\phi)\over\pi^2}\ +\ 5\times {\tanh(\pi \epsilon/d)\over(\pi \epsilon/d)}\
+\ 0\times {\sinh((2\phi-\pi)\epsilon/d)\over (\pi \epsilon/d)\cosh (\pi \epsilon/d)}\cr
&+\ {80\over\pi^4}\,(\lambda M^2d^2)^2\times(\pi \epsilon/d)^2,
\label{FNA}
\end{align}
but the first two choices (N1) and (N2) are physically equivalent in light of the redundancy~(\ref{Symmetry}) (which exchanges $\bp$ with $\bq$ and $\phi$ with $\pi-\phi$ for $\beta=\pi/2$), so we shall focus on (N1) and (N3) only and drop (N2) from further consideration.

At this point, we have three possibilities --- the abelian zigzag, and the non-abelian zigzags (N1) and (N3) --- so let's compare their energies at some $\phi_1<\pi/2$ and $\phi_2=\pi-\phi_1>\pi/2$. It is easy to see that
\be
{\cal F}_{\rm abelian}(\phi_2)\ <\ {\cal F}_{\rm(N3)}(\phi_2)\
=\ {\cal F}_{\rm(N3)}(\phi_1)\ <\ {\cal F}_{\rm abelian}(\phi_1),
\ee
so the abelian zigzag with $\phi=\phi_2>\pi/2$ always wins over the (N3) zigzag or the abelian zigzag with $\phi=\phi_1<\pi/2$. The only exception to this rule is $\phi=\pi/2$ for which the (N3) and the abelian zigzag have equal energy, but that point is not the global minimum of energy --- or even a local minimum since $\partial{\cal F}_{\rm abelian}/\partial\phi<0$ at $\phi=\pi/2$. As to the non-abelian zigzag (N1) it has lower energy at $\phi_2>\pi/2$ than at $\phi_2=\pi-\phi_1<\pi/2$, but for any particular $\phi>\pi/2$ the choice between the (N1) and the abelian zigzag depends on the $\epsilon/d$ ratio.

\begin{figure}[htb]
\begin{minipage}[b]{0.5\linewidth}
\begin{center}

\includegraphics[width=7.8cm]{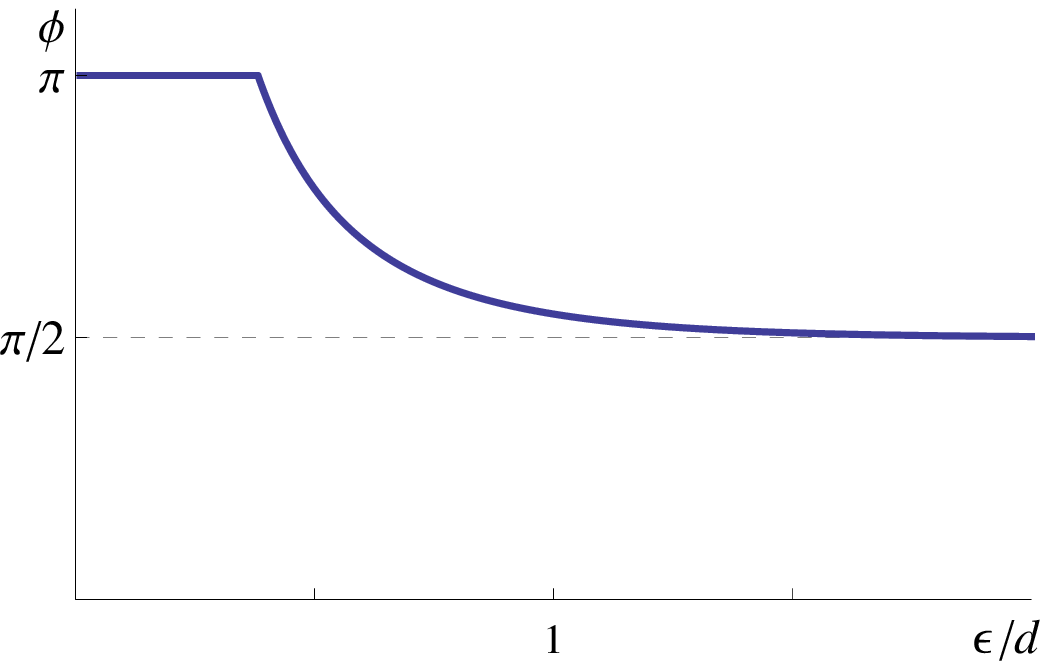}

(a)
\end{center}
\end{minipage}
\begin{minipage}[b]{0.5\linewidth}
\begin{center}

\includegraphics[width=8.cm]{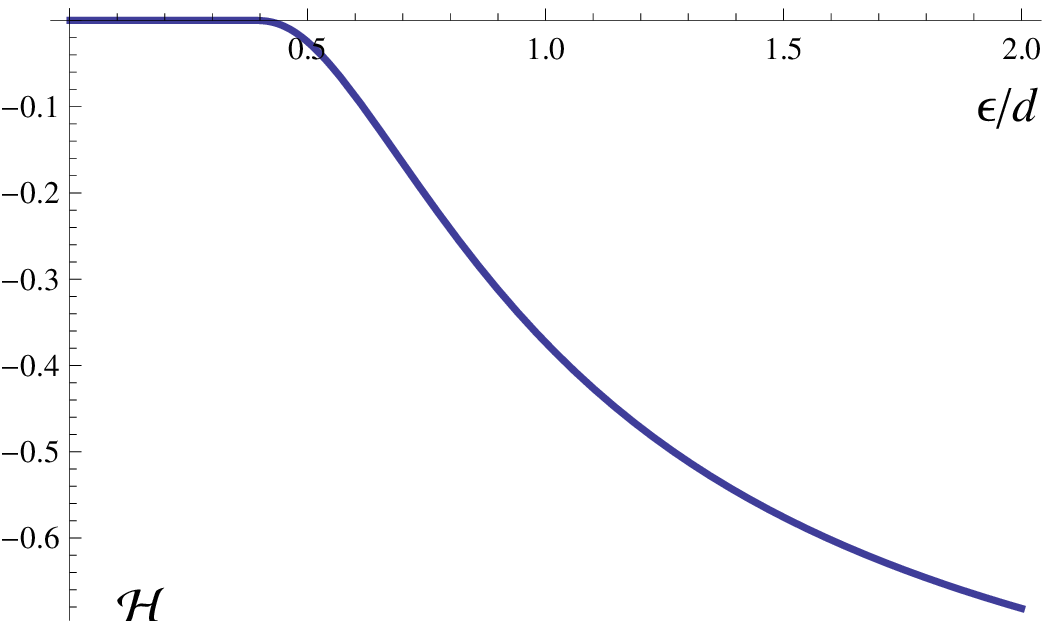}

\vspace{0.3cm}
(b)
\end{center}
\end{minipage}
\vspace{-0.6cm}
\caption{(a) Twist angle as a function of the $\epsilon/d$ ratio for an abelian or non-abelian (N1) zigzags. (b) ${\cal H}$ at optimal twist angle $\phi$ as a function of the zigzag ratio $\epsilon/d$.}
\label{figPHIplot}
\end{figure}

Thus far, we have narrowed  the choice of zigzags to the abelian and the (N1). Their energies depend on the $\phi$ angle in exactly the same way, up to an overall factor of 2. Indeed, we may write the $\mathcal{F}$ functions for the two zigzags as
\begin{align}
{\cal F}_{\rm abelian}(\phi,\epsilon/d;d)\ & = {\cal G}(\epsilon/d;d)\ +\ 2{\cal H}(\phi,\epsilon/d)\,, \\
{\cal F}_{\rm (N1)}(\phi,\epsilon/d;d)\ & = {\cal G}(\epsilon/d;d)\ +\ {\cal H}(\phi,\epsilon/d)\,, \\
{\rm where}\quad  {\cal G}(\epsilon/d;d)\ &
=\ 3 \ + \ \frac{\tanh(\pi \epsilon/d)}{(\pi \epsilon/d)}\ +\ \frac{80}{\pi^4}\,(\lambda M^2d^2)^2\times(\pi \epsilon/d)^2 \qquad{\rm independent \ of }\ \phi \,, \\
{\rm and}\quad {\cal H}(\phi,\epsilon/d)\ &
= \ -\ 4\times\frac{\phi(\pi-\phi)}{\pi^2}\ + \ 2\times\frac{\tanh(\pi \epsilon/d)}{(\pi \epsilon/d)} \
- \ 2\times\frac{\sinh((2\phi-\pi)\epsilon/d)}{(\pi \epsilon/d)\cosh (\pi \epsilon/d)}\,; \\
{\rm note}\quad {\cal F}_{\rm abelian} \ - \ {\cal F}_{\rm (N1)} \ & = \ {\cal H}(\phi,\epsilon/d)\,.
\label{DeltaF}
\end{align}
For a fixed $\epsilon/d$ ratio, both zigzags favor the same angle $\phi$ that minimizes the $\cal H$ function. Solving the minimum equation
\be
\frac{\pi}{4}\,\frac{\partial{\cal H}}{\partial\phi}\
=\ {2\phi-\pi\over\pi}\ -\ {\cosh((2\phi-\pi)\epsilon/d)\over\cosh(\pi \epsilon/d)}\ =\ 0
\ee
gives us $\phi$ as a function of the zigzag ratio $\epsilon/d$ shown in figure~\ref{figPHIplot}(a).

For small zigzag ratios $\epsilon/d$, the preferred angle is $\phi=\pi$ for which ${\cal H}=0$ and both zigzags have the same energy. Indeed, at that point the two zigzags become indistinguishable since for $\phi=\pi$ the $\beta$ angle is irrelevant. However, for larger zigzag ratios the minimum moves to $\phi<\pi$ and the two zigzags become distinct; for very large $\epsilon/d$, the minimum asymptotes to $\pi/2$. The transition between $\phi=\pi$ and $\phi<\pi$ regimes occurs at
$(\epsilon/d)\approx 0.382$ where the second derivative
\be
\left.\frac{\partial^2{\cal H}}{\partial\phi^2}\right|_{\phi=\pi}\
=\ {8\over\pi^2}\Bigl(1\,-\,(\pi \epsilon/d)\tanh(\pi \epsilon/d)\Bigr)
\ee
changes its sign.

To find which of the two zigzags is preferable in the regime with an optimal angle $\phi<\pi$ we plot in figure~\ref{figPHIplot}(b) the difference $\Delta\mathcal{F} = {\cal H}$~(\ref{DeltaF}) as a function of $\epsilon/d$. We see that $\mathcal{H}$ is either zero (when the minimum is at $\phi=\pi$), or negative (when the minimum is at $\phi<\pi$), but never becomes positive. Consequently, whenever the (N1) and the abelian zigzags become distinct, the abelian zigzag has lower energy.

\begin{figure}[t]
\begin{center}
\vspace{3ex}
\includegraphics[width= 100mm]{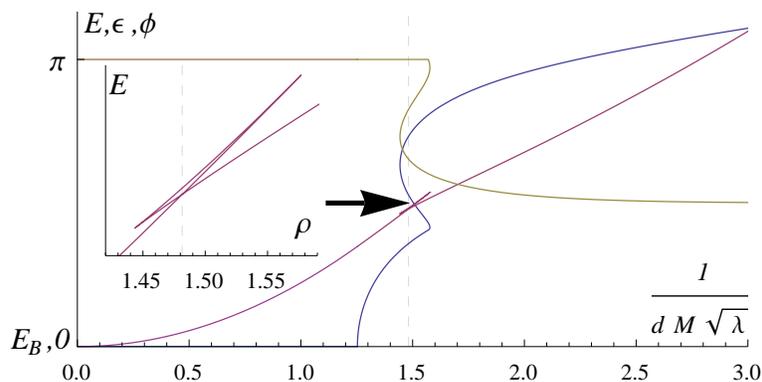}
\caption{The Phase diagram of the zigzag configuration. Three curves correspond to the free energy per instanton (magenta), zigzag amplitude (blue) and twist angle (yellow) as functions of density. The plot exhibits a second order (straight to zigzag) and a first order (orientation flip) phase transitions. The second order transition is only distinguishable in the behavior of $\epsilon$, which becomes non-zero above some critical density. The plot of the free energy has a ``butterfly" typical for a first order transition with stable, metastable and unstable parts. The energy is measured with reference to the stand-alone instanton energy $E_B$.}
\label{figZigzagPD}
\end{center}
\vspace{-0.5cm}
\end{figure}

We summarize the results of this section in figure~\ref{figZigzagPD}. In this figure we plot the free energy (per instanton) of the chain configuration, zigzag amplitude $\epsilon$ and twist angle $\phi$ as functions of the density in units of $\sqrt{\lambda}M$. At low densities the chain of instantons is straight ($\epsilon=0$) and the twist between adjacent instantons is $\phi=180^\circ$. At density $d^{-1}\simeq 1.25$ the straight configuration changes to the zigzag in a second order phase transition. The twist angle remains $180^\circ$ in this transition. At density $d^{-1}\simeq 1.48$ the relative twist in the zigzag changes from $\phi=180^\circ$ to $\phi\simeq 117^\circ$ in an apparent first order transitions. The zigzag amplitude also jumps across the transition. When the density is further increased the twist angle smoothly decreases and asymptotes to $90^\circ$ reflecting the fact that each of the two zigzag layers separately favors an anti-ferromagnetic orientation of neighboring instantons.

This is the picture of the phase diagram at the level when only the straight and zigzag configurations are taken into account. At the next step three and more layer configurations must be considered similarly to the analysis performed in section~\ref{secPointCharges}. It may happen that the transition to three layers occurs prior to the change-of-twist transition discovered above. It is also natural to expect that multidimensional ($D>1$) instanton arrays carry a non-abelian configuration of twists. Abelian configurations are always preferred by the straight and zigzag configurations, so it would be interesting to see when exactly the twists start respecting the increased dimensionality of the instanton chain.

\section{Summary}
\label{secSummary}

In this work we have modeled baryonic crystals of large $N_c$ nuclear matter phase via instantons of the effective flavor gauge field theory on probe D8 anti-D8 branes in the geometry of near extremal D4-branes. The effective theory was derived as a natural generalization of the  Sakai-Sugimoto \cite{SakaiSugimoto2004} setup and must be applicable to any holographic model of quenched flavor, which enjoys the limit of large 't Hooft coupling $\lambda$ and admits instanton description of baryons. The holographic nature of the model immediately implies that the most general baryon configurations are not the usual 3D crystals, but rather effectively 3+1D structures (including the holographic dimension). We have qualitatively analyzed the phase space of such structures. Since working with full-fledged 3D arrays of instantons is generally a challenging task, where even the pertinent flat space solutions are not known, we have resorted to several toy-models: point-like instantons and 1D lattices. Despite the simplification we believe that these models grasp the main features of the 3D instanton lattices.


We first applied the point-instanton approximation to a 1D chain. To stabilize the chain in 1D we introduced an \emph{ad hoc} curvature $M'$ in transverse spatial dimensions~(\ref{anisotropic0}). Assuming $M'\gg M$ (curvature in the holographic direction) we analyzed the free energy of the zigzag configuration of point charges with amplitude $\epsilon$ (figure~\ref{figZigzag}), which is the most apparent instability mode of the straight chain ($\epsilon=0$). We found that below critical density $\rho_c = { 2\cdot 3^{1/4} M\sqrt{\lambda}}/{\pi}$ the straight chain is the preferable configuration, while for $\rho>\rho_c$ a zigzag with a density dependent amplitude (figure~\ref{figzigzagpar0}) has lower energy. The transition between the straight chain and the zigzag is second order. Considering other configurations of point charges with up to four layers in the holographic dimension lead to the phase diagram on figure~\ref{figZigzagPD}. The phase diagram demonstrates that in general larger densities favor configurations with more layers, although precise sequence of transitions, \emph{e.g.} (\ref{1-4sequence}), is not necessarily a consecutive increase of number of layers.

After the 1D example we considered a simple cubic (sc) lattice of point-like instantons. For the sc lattice the leading instability mode must correspond to a splitting into two fcc layers. This is the 3D analog of the zigzag configuration. We found that for the densities $\rho\gtrsim 3.04(M\sqrt{\lambda})^3$ 2-layer configuration is favored over the original 1-layer sc lattice. Similarly bcc lattice must first split into 2 sc layers, while for the fcc lattice the double-layer splitting is probably a second order transition, which breaks cubic symmetries. The quantitative analysis of the 3D lattices of point instantons beyond two layers is similar to that of the 1D chains, albeit technically more challenging.

Next we analyzed the 1D chain of full-fledged instantons. Flat space instanton solution for a periodic 1D array of instantons was known before from the work of Kraan and van Baal\cite{Kraan:1998pm}. We found that in the holographic baryon model the minimum energy of the straight chain is attained for the anti-ferromagnetic (twist angle $\phi=\pi$) orientation of instantons. To investigate the stability of the straight chain we constructed a new flat space instanton solution corresponding to the zigzag of finite size instantons. The derivation was based on the ADHM construction and an analog of Nahm transform similarly to the derivation of Kraan and van Baal. Note that to have a clear interpretation of the ADHM data as instanton positions the  instanton separation must be larger than their size. This was achieved by taking $M'\gg M$ -- the same condition as for keeping the 1D chain stable in spatial dimension. The analysis of the free energy of the zigzag instanton solution lead to the phase diagram in figure~\ref{figZigzagPD}.  At low densities the instanton chain is straight. For $\rho\gtrsim1.25 M\sqrt{\lambda}$ ($5^{1/4}$ times larger than the critical density $\rho_c$ of point instanton lattice) the zigzag with $\epsilon=\epsilon(\rho)$ is preferable. The transition to the zigzag is second order as in the case of point charges.  As for the instanton orientation it is anti-ferromagnetic immediately after the phase transition, but for $\rho\simeq1.48M\sqrt{\lambda}$ the twist angle changes from $\phi=\pi$ to $\phi<\pi$ in a first order phase transition. For larger densities $\phi$ is a decreasing function of the density, which asymptotes to $\phi=\pi/2$. To see the orientation flip we constructed a zigzag instanton solution for general (non-abelian) twist in the first non-trivial order in a perturbative small instanton expansion $a\ll d$. Remarkably in the case of the zigzag the abelian twist are always favored over the non-abelian ones.


Analysis of the 3D instanton lattices is more challenging and was not attempted in this work. However the study of the above toy-models provides us with a qualitative picture of the phase space of the 3D lattices of holographic baryons. Namely, when the lattice is squeezed it gets stratified into 3D layers stacked one upon each other in the holographic dimension. Once the density is varied gradually from zero until some large values at which it still makes sense to talk about individual baryons the lattice undergoes a sequence of phase transitions with increasing number of layers. Such a behavior can already be seen at the level of point-like instantons. The phase space of the finite size instantons is reacher. In particular the energy of the instanton configuration depends on the mutual orientation of instantons. Apart from the layer splitting  transitions one will have transitions with a change of the orientation. Another consequence of the finite instanton size is an effectively weaker interaction as compared to the point charges due to the screening in the instanton cores.

The interpretation of the above results may be an interesting physical problem. The main idea have been already suggested by Rozali \emph{et al.} in the case of uniform instanton distribution~\cite{Rozali:2007rx}. The stratification of the lattice in the fourth dimension must be a holographic analog of the emerging Fermi sea. Indeed the holographic coordinate plays the role of an energy scale. The energy scale associated with finite density of baryons must be the chemical potential. So baryons piling-up on top of each other and acquiring higher energy scale are filling the Fermi sea of available quantum states. In this case the Fermi sea cannot be baryonic: the latter are classical in this picture and carry no spin, but rather a Fermi sea of quarks. In a realistic setup (with no additional scale $M'$) the critical density of the stratifying phase transitions and the instanton size will be defined by the same scale $M\sqrt\lambda$. Therefore the transitions will happen when the baryons start overlapping. Overlapping baryons will loose their identities: the constituent quarks will no longer know which baryons they belong to.  In the large $N_c$ QCD this kind of matter is believed to be quarkyonic, that is to have a quark Fermi sea but baryonic Fermi surface. The above phase transitions must then signal the onset of the quarkyonic phase.\footnote{Similar conclusion has been made in the case of skyrmions in~\cite{Lee:2009dpa}. The transition from the skyrmion to half-skyrmion matter is believed to signify the onset of the quarkyonic phase as well. Following the ideas of~\cite{Rho:2009ym} we expect that the holographic popcorn transition is related to the half-skyrmion transition in Skyrme lattices.}

For large enough densities the lattice will have many layers in the holographic dimension, \emph{i.e.} it will become mesoscopic. In this regime baryons will be well overlapping, quarks will form a kind of a liquid and the uniform baryon density will be a good description. In such a limit the baryon popcorn picture should match the story of the expanding Fermi sea of Rozali \emph{et al.}


Let us finish the discussion by summarizing interesting open questions.
\begin{itemize}

\item The analysis of this paper has not been generalized to the case of 3D baryon lattices. It will be interesting to construct new 3D instanton solutions and study the corresponding phase space.

\item Finite density baryon lattices were previously extensively studied within the framework of Skyrme model. Since skyrmions have a natural connection to holographic baryons there should be a way to reproduce the results from holographic baryon point of view. In particular one can try to understand the realization of the skyrmion-half-skyrmion phase transition along the lines suggested by Rho \emph{et al.}~\cite{Rho:2009ym}.

\item Finite temperature nuclear matter transition was studied in~\cite{Bergman} and~\cite{Rozali:2007rx} in the approximation of a uniform instanton density. It would be interesting to incorporate finite temperature in the above analysis.

\item The generalization of the Sakai-Sugimoto model used in this paper is applicable for a wide class of holographic models of baryons. Since baryon interactions are repulsive in the Sakai-Sugimoto model it is desirable to make a similar analysis with attractive baryons, \emph{e.g.} in the Klebanov-Strassler geometry~\cite{Dymarsky:2010ci}. It is also interesting to discuss baryons in the setup of non-critical models with $\lambda\sim 1$.

\end{itemize}

\vspace{0.5cm}

\paragraph{Acknowledgements.} We would like to thank O.~Bergman, M.~Strassler, A.~Vainstein and I.~Zahed for useful conversations. This work was supported in part by the US-Israel Binational Science Foundation (V.~K. and J.~S.); by the US National Science Foundation grant \#PHY--0969020 (V.~K.); by a PVE fellowship of CAPES Foundation -- Brazil, Brazilian Ministry of Science and Technology, the contract 14.740.11.0081 with the Ministry of Education and Science of the Russian Federation and the Grant of the President of the Russian Federation for Support of Scientific Schools NSh-3035.2008.2  (D.~M.); by a Centre of Excellence supported by the Israel Science Foundation (grant number 1468/06) and a grant (DIP H52) of the German Israel Project Cooperation (J.~S.). D.~M. would also like to thank theoretical and mathematical physics group at Universit\'e Libre de Bruxelles and the LPTHE at Universit\'e Pierre et Marie Curie, for hospitality at different stages of this project.


\begin{appendix}


\section{Straight chain. Analytical analysis at small and large densities}
\label{secSLDensity}

In section~\ref{secStraightChainSol} we described the flat space instanton solution that corresponds to a periodic straight chain of instantons. The flat space solution is the 0$^{\frac{\rm th}{}}$ order approximation of the holographic baryon lattice in the large $\lambda$ limit. In section~\ref{secStraightChainEn} we used it to compute the leading $1/\lambda$ corrections to the baryon free energy. Provided that we have a solution of the determinant of matrix $L$~(\ref{BigFla}) it is straightforward to compute the instanton density using~(\ref{Idensity}) and the solution of the abelian field $\hat{A}_0$~(\ref{Asoln}). Formulae obtained this way are too bulky to present in the paper. In fact, the explicit expressions are not required for the evaluation of the density moments~(\ref{vevx2}) in the non-abelian part of the energy. In the case of the abelian Coulomb energy~(\ref{Cenergy}) however, one still needs to take an integral of a bulky expression. In this work we evaluate the integral analytically in the limits of small $a\ll d$ and large $a\gg d$ instantons.

For large lattice spacing $d\gg a$ we can expand the expression for $\det(L)$ in powers of $a/d$.  In this case the instantons in the chain are approximately spherically symmetric in 4D, so at distances $R\sim a\ll d$ from the center we have
\be
\log\det(L)\ \approx\ {\rm const}\ +\ \log(a^2+R^2)\
+\ \frac{a^2-R^2\cos(2\psi)}{6d^2}\
+\ \frac{a^4}{d^2(a^2+R^2)}\times\frac{\pi^2+(\pi-\phi)^2}{12\pi^2}\
+\ O\Bigl( a^4/d^4\Bigr)\,,
\label{LDsmall}
\ee
where $R$ is the 4D radius and $\psi$ is the 4D `latitude' angle ($i.\,e.,\ x_4=R\cos\psi$). In this approximation, the instanton density becomes
\be
I(x)\ \approx\ \frac{6a^2}{\pi^2(R^2+a^2)^4}\times\left[
	 1\,+\,\frac{a^2}{d^2}\times\frac{\pi^2+(\pi-\phi)^2}{6\pi^2}\times\frac{R^2-a^2}{R^2+a^2}\,
	+\,O(a^4/d^4)\right]
\ee
and consequently the Coulomb energy per instanton~(\ref{Cenergy}) evaluates to
\be
E_{\rm C}\ \approx\ \frac{N_c}{\lambda M}\left[
\frac{1}{5a^2}\ +\ \frac{4\pi^2+3(\pi-\phi)^2}{30d^2}\ +\ O(a^2/d^4)\right].
\ee
The first term in this expression is the 4D Coulomb self-energy of the instanton, which diverges in the limit of zero size. The second, $a$-independent term is a modification of point charge energy~(\ref{AbelianEnergy}) to include the dependence on the twist angle. In the $d \gg a$ regime the Coulomb energy apparently prefers the instantons to be oriented in an anti-parallel way $\phi=\pi$, while the non-abelian energy~(\ref{NAenergy}) favors $\phi=0$. Combining the two energies together in section~\ref{secStraightChainEn} we found that the Coulomb energy wins and the equilibrium twist angle is $\phi=\pi$.

In the opposite limit $d\ll a$ of densely packed overlapping instantons, the instanton density becomes independent of the $x_4$ coordinate, while in the 3 transverse dimensions it becomes concentrated at two widely separated points
\be
\vec X^{(1)}\ =\ \Bigl(0,0,\frac{a^2}{2d}(2\pi-\phi)\Bigr)\qquad
{\rm and}\qquad\vec X^{(2)}\ =\ \Bigl(0,0,\frac{a^2}{2d}(-\phi)\Bigr)\,.
\ee
Indeed expanding $\log\det(L)$ in powers of $d/a$ gives
\be
\log\det(L)\ \approx\ \frac{\phi}{d}\times r_1\ +\ \frac{2\pi-\phi}{d}\times r_2\
+\ \log\frac{(r_1+r_2+(\pi a^2/d))^2}{8r_1r_2}\, ,
\label{DLlarge}
\ee
where $r_1=|\vec x-\vec X^{(1)}|$ and $r_2=|\vec x-\vec X^{(2)}|$, {\it cf.}\ equations~(\ref{Radii}). Taking the double Laplacian one can obtain the corresponding 3D instanton density as
\be
I_{3D}(\vec x)\ =\ d\times I_{4D}(\vec x)\
\approx\ \frac{\phi}{2\pi}\times\delta^{(3)}(\vec x-\vec X^{(1)})\
+\ \frac{2\pi-\phi}{2\pi}\times\delta^{(3)}(\vec x-\vec X^{(2)})\
+\ O\bigl(e^{-2\phi r/d}\bigr).
\label{Ilarge}
\ee
Consequently, the abelian gauge field has 3D Coulomb form
\be
\label{A0LargeInst}
\hat A_0\ =\ \frac{4\pi^2}{\lambda M d}\left(\frac{\phi/2\pi}{4\pi r_1}\,
	+\,\frac{1-\phi/2\pi}{4\pi r_2}\right).
\ee
Evaluating again the Coulomb energy using~(\ref{Cenergy}) gives
\be
E_{\rm C}\ =\frac{N_c}{8\pi\lambda M d}\left(
	\frac{\phi^2}{2\rho_1}\,+\,\frac{(2\pi-\phi)^2}{2\rho_2}\,
	+\,\frac{\phi(2\pi-\phi)}{b=\pi a^2/d}\right),
\label{GenCoulapp}
\ee
where $\rho_1$ and $\rho_2$ are effective radii of the instanton density concentrations that appear $\delta$-like in equation~(\ref{Ilarge}). Those $\delta$-functions are artefacts of the approximation~(\ref{DLlarge}) that becomes inaccurate near $\vec X^{(1)}$ and $\vec X^{(2)}$. A better approximation  for $r_1\sim d\ll b$ is given by
\be
\log\det(L)\ \approx\ \frac{2\pi-\phi}{d}\times r_2\
+\ \log\left[\frac{1}{r_1}\sinh\frac{\phi r_1}{d}\right]\
+\ \frac{r_1 d}{\pi a^2}\,\ctanh\frac{\phi r_1}{d}\
+\ O(r_1^2 d^2/a^4)\,,
\ee
which leads to instanton density near $\vec X^{(1)}$
\be
I_{3D}\
\approx\  \frac{\phi^4}{d^3}\left(\left(\frac{\d}{\d x}\,+\,\frac{2}{x}\right)
    \frac{\d}{\d x}\right)^2 \log\frac{\sinh(x)}{x}\
+\ \frac{\phi^3}{\pi a^2 d}\left(\left(\frac{\d}{\d x}\,
+\,\frac{2}{x}\right)
    \frac{\d}{\d x}\right)^2\frac{x}{\tanh(x)} \  +\ O(d/a^4)\,,
\ee
where $x\equiv{\phi r_1}/{d}$. The Coulomb energy of this charge density interacting with itself (but not with the other charge at the $\vec X^{(1)}$) corresponds to
\be
\label{rho1app}
{1\over2\rho_1}\ \approx\ \frac{\phi}{8d}\times C_1\
+\ \frac{d}{4\pi a^2}\times C_2\ +\ O(d^3/a^4)\,,
\ee
where $C_1$ and $C_2$ are numerical values of the dimensionless integrals:
\begin{align}
C_1\ &
= \intop\limits_0^\infty\!\d x\,x^2\left[
	\frac{\d}{\d x}\left(\frac{\d}{\d x}+\frac{2}{x}\right)\frac{\d}{\d x}
	\log\frac{\sinh(x)}{x}\right]^2\nonumber\\
&\approx\ 1.174,\label{C1defapp}\\
C_2\ &
= \intop\limits_0^\infty\!\d x\,x^2\left[
	\frac{\d}{\d x}\left(\frac{\d}{\d x}+\frac{2}{x}\right)\frac{\d}{\d x}
	\log\frac{\sinh(x)}{x}\right]\times\left[
	\frac{\d}{\d x}\left(\frac{\d}{\d x}+\frac{2}{x}\right)\frac{\d}{\d x}
	\frac{x}{\tanh(x)}\right]\nonumber\\
&\approx\ 1.761.\label{C2defapp}
\end{align}
Likewise near $\vec{X^{(2)}}$,
\be
\label{rho2Aapp}
{1\over2\rho_2}\ \approx\ \frac{2\pi-\phi}{8d}\times C_1\
+\ \frac{d}{4\pi a^2}\times C_2\ +\ O(d^3/a^4).
\ee
Plugging these charge radii into equation~(\ref{GenCoulapp}), we write the Coulomb energy
of large instantons as
\be
E_{\rm C}\ =\ \frac{N_c}{\lambda M}\left(
	\frac{C_1(\pi^2+3(\pi-\phi)^2)}{32d^2}\
	+\ \frac{(C_2+2)\pi^2+(C_2-2)(\pi-\phi)^2}{16\pi^2 a^2}\
	+\ O(d^2/a^4)\right).
\label{EClargeapp}
\ee
Two terms in the above expression favor different twist angles, but the first term is apparently stronger in the $d\ll a$ regime. Combining this Coulomb energy with the non-abelian energy (\ref{NAenergy}) one can find that $\phi=\pi$ is a preferred angle for the overlapping instantons as well. Below in the appendix~\ref{appMinEnergy} we argue that anti-parallel orientation is favored for any values of $a$ and $d$.


\section{Straight chain. Equilibrium configuration}
\label{appMinEnergy}

In section~\ref{secStraightChainEn} and appendix~\ref{secSLDensity} we computed the free energy of the straight twisted instanton chain in the limits of small and large instantons. Here we are going to argue that the equilibrium value of the twist angle is always $\phi=\pi$, no matter how is  the instanton size compared to the lattice spacing. First, thanks to the invariance of the gauge fields under $Z_2\subset SU(2)$, the instanton chain has an exact $\phi\to2\pi-\phi$, $x_3\to-x_3$ symmetry. Indeed the combination of the latter transformations leave the ADHM data $Y$ invariant up to an overall sign, which does not affect the gauge field. In principle, this symmetry could be spontaneously broken, but it does not happen for either large or small $d$, and it is extremely unlikely that such spontaneous breakdown would happen only at the intermediate $d\sim a_0$. Consequently, for any given $d$ and $a$, the minimum of the net energy with respect to the twist angle $\phi$ should lie at one of the fixed points of the transformation $\phi\to2\pi-\phi$, \emph{i.e.} at $\phi=\pi$ or $\phi=0$.

We are going to show that $\phi=0$ is a local maximum of the net energy $E_{\rm C}+E_{\rm NA}$ for the straight chain instanton solution derived from the following expression for determinant of the matrix $L$:
\begin{align}
\det(L)\ =\ &
\left(\cosh\frac{\phi r_1 }{d}\,+\,\frac{\pi a^2}{dr_1}\,\sinh\frac{\phi r_1 }{d}\right)
\left(\cosh\frac{(2\pi-\phi) r_2 }{d}\,+\,\frac{\pi a^2}{dr_2}\,\sinh\frac{(2\pi-\phi) r_2 }{d}\right)\nonumber \\
&+\ \frac{r_1^2+r_2^2-(\pi a^2/d)^2}{2r_1r_2}\,\sinh\frac{\phi r_1 }{d}\,\sinh\frac{(2\pi-\phi)r_2}{d} \ -\ \cos \frac{2 \pi x_4}{d}\,,
\label{BigFlaApp}
\end{align}
where
\begin{align}
r_1^2\ &=\ x_1^2\ +\ x_2^2\ +\ \left(x_3+\frac{a^2(\phi-2\pi)}{2d}\right)^2,\nonumber\\
r_2^2\ &=\ x_1^2\ +\ x_2^2\ +\ \left(x_3+\frac{a^2\phi}{2d}\right)^2.
\label{RadiiApp}
\end{align}
The non-abelian energy $E_{\rm NA}$ is given by equation~(\ref{NAenergy}), while the Coulomb energy must be computed from the integral~(\ref{Cenergy}).

Suppose the radius $a$ lies at the minimum of the total energy (for fixed $d$ and $\phi=0$). In this case
\be
\label{EnetAtMina}
\frac{\partial E_{\rm C}}{\partial a}\ +\ \frac{\partial E_{\rm NA}}{\partial a}\ =\ 0\,.
\ee
We will see below   that
\be
{\rm for}\ \phi\to0,\quad
\frac{\partial E_{\rm C}}{\partial\phi}\
=\ \frac{2\pi a^3}{d^2}\times\frac{\partial E_{\rm C}}{\partial a}\,.
\label{derrel}
\ee
From this relation at $\phi=0$ and equation~(\ref{EnetAtMina}) one derives that
\be
\left.\frac{\partial E_{\rm C}}{\partial\phi}\right|_{\phi=0}\
=\ -\frac{2\pi a^3}{d^2}\times\left.\frac{\partial E_{\rm NA}}{\partial a}\right|_{\phi=0}\
=\ -\frac{6\pi a^4 M^3\lambda N_c}{d^2}\Bigl(1+O(a^2M^2)\Bigr)\,,
\ee
where the second equality follows from equation~(\ref{NAenergy}). At the same time,
\be
\left.\frac{\partial E_{\rm NA}}{\partial\phi}\right|_{\phi=0}\
=\ +\frac{2\pi a^4 M^3\lambda N_c}{4d^2}\Bigl(1+O(a^2M^2)\Bigr)
\ee
and therefore
\be
\left.\frac{\partial E_{\rm net}}{\partial\phi}\right|_{\phi=0}\
=\ \frac{2\pi a^4 M^3\lambda N_c}{d^2}\left(+\frac14\,-\,3\,+\,O(a^2M^2)\right)\
<\ 0.
\ee
The derivative of the net energy with respect to $\phi$ is always negative for the equilibrium value of the radius $a$. Thus, $\phi=0$ is always a local maximum of the net energy. Since we do not expect the symmetry $\phi\to2\pi-\phi$ to be spontaneously broken, the only remaining choice for the minimum is to lie at $\phi=\pi$.

To complete this argument it remains to derive the relation~(\ref{derrel}). For this let us calculate the derivative $\partial E_{\rm C}/\partial\phi$ for $\phi=0$. The Coulomb energy~(\ref{Cenergy}) is quadratic in $\log\det(L)$, so we need both the derivative $\partial\det(L)/\partial\phi$ and the determinant $\det(L)$ itself in the $\phi\to0$ limit. Note that in this limit, the determinant~(\ref{BigFlaApp}) becomes spherically symmetric in 3D,
\be
\left.\det(L)\right|_{\phi=0}\
=\ \cosh\frac{2\pi r}{d}\ +\ \frac{\pi a^2}{rd}\sinh\frac{2\pi r}{d}\
-\ \cos\frac{2\pi x_4}{d}\,,
\ee
where $r = x_1^2+x_2^2+x_3^2$. Also d'Alembertians of this expression are spherically symmetric. In such a case only the spherically symmetric part of the derivative $\partial\det(L)/\partial\phi$ contributes to the integral
\be
\left.\frac{\partial E_{\rm C}}{\partial\phi}\right|_{\phi=0}\ =\ -\frac{N_c}{128\pi^2\lambda M} \int_0^d\!\!\d x_4\!\int\!\!\d^3x\ \square\square\square\log\det(L)\times\frac{1}{\det(L)}\, \frac{\partial\det(L)}{\partial\phi}\,.
\ee
Extracting the spherically symmetric part from the expression for the derivative one gets
\be
\left.\frac{\partial\det(L)}{\partial\phi}\right|_{\phi=0}\ =\ \frac{\pi^2 a^4}{rd^3}\sinh\frac{2\pi r}{d}\
+\ \cos\Theta\times[\mbox{some function of}\ r]\,,
\label{phider}
\ee
where $\Theta$ is the azimuthal angle in 3D. The term proportional to $\cos\Theta$ will vanish after integration over the spherical coordinates. Thus, we obtain the following relation between the $\phi$ derivative of $\det(L)$ and its derivative with respect to instanton radius $a$,
\be
\int\!\frac{\d^2\Omega}{4\pi} \left.\frac{\partial\det(L)}{\partial\phi}\right|_{\phi=0}\
=\ \frac{\pi^2 a^4}{rd^3}\sinh\frac{2\pi r}{d}\
=\ \frac{2\pi a^3}{d^2}\times\frac{\partial\det(L)}{\partial a}\,.
\ee
Equation (\ref{derrel}) immediately follows from the above relation between the derivatives. This completes the argument. With the only assumption that the $Z_2$ symmetry related to the exchange of $\phi\to 2\pi - \phi$ is not spontaneously broken (which is highly unlikely since it is not broken in both $a\ll d$ and $a\gg d$ limits) we conclude that at equilibrium the straight chain always favors the anti-ferromagnetic ($\phi=\pi$) orientation of instantons.


\section{Zigzag - Stability at high densities}
\label{appHighDensity}

In section~\ref{secZigzag} we found that for large enough densities the straight chain of instantons becomes unstable against the zigzag deformation (figure~\ref{figZigzag}). This result was derived in the approximation of small instanton size $a\ll d$. The latter condition can be ensured by the high anisotropy limit $M'\gg M$. For $M'\to M$ the critical spacing of the zigzag transition $d_c\to a$, which is outside of the small instanton approximation. The analysis of the stability needs to be done numerically. It is possible that the phase diagram is qualitatively different for $M'\simeq M$. By analyzing the opposite limit of the instanton radius $a\gg d$ we are going to show that as far as only the zigzag and the straight chain are compared the former has lower free energy only in some intermediate density regime, while the straight chain is a preferable configuration also at asymptotically large densities.

In the case of the zigzag with the antiparallel orientation of instantons we derived the following solution for the determinant of the matrix $L$ in section~\ref{secZigzag}.
\begin{align}
\frac{\det(L)}{\rm const}\ =\ &
\left( \cosh\frac{\pi r_1}{d}\,+\,\frac{\pi a^2}{dr_1}\,\sinh\frac{\pi r_1}{d}\right)\times
\left( \cosh\frac{\pi r_2}{d}\,+\,\frac{\pi a^2}{dr_2}\,\sinh\frac{\pi r_2}{d}\right)\nonumber\\
&\qquad+\ \frac{r_1^2+r_2^2-(\pi a^2/d)^2}{2r_1 r_2}\,
	\sinh\frac{\pi r_1}{d}\,\sinh\frac{\pi r_2}{d}\
-\ \cos\frac{2\pi x_4}{d}\nonumber\\
&+\ \sin\nu\times\cos\frac{\pi x_4}{d}\times\left(
	\left( 2\cosh\frac{\pi r_1}{d}\,+\,\frac{\pi a^2}{dr_1}\,\sinh\frac{\pi r_1}{d}\right)\,
	-\,\left( 2\cosh\frac{\pi r_2}{d}\,+\,\frac{\pi a^2}{dr_2}\,\sinh\frac{\pi r_2}{d}\right)
	\right)\nonumber\\
&+\ \sin^2\nu\times\left(
	-1\,+\,\cosh\frac{\pi r_1}{d}\,\cosh\frac{\pi r_2}{d}\,
	-\,\frac{r_1^2+r_2^2}{2r_1r_2}\sinh\frac{\pi r_1}{d}\,\sinh\frac{\pi r_2}{d}
	\right),
\label{DetLmodApp}
\end{align}
where
\begin{align}
r_{1,2}^2\ &=\ x_1^2\ +\ x_2^2\ +\ (x_3\,\mp\,\tfrac12 b_\epsilon)^2,\\
\label{beApp}
b_\epsilon\ &=\ \sqrt{4\epsilon^2+(\pi a^2/d)^2},\\
\nu\ &=\ \arctan\frac{2\epsilon}{\pi a^2/d}\ =\ \arcsin\frac{2\epsilon}{b_\epsilon}\,.
\label{nuApp}
\end{align}
In the regime of densely overlapping instantons ($d\ll a$) equation~(\ref{DetLmodApp}) yields
\be
\log\det(L)\ \approx\ \frac{\pi(r_1+r_2)}{d}\
+\ \log\frac{(r_1+r_2+(\pi a^2/d)^2)-\sin^2\nu(r_1-r_2)^2}{r_1r_2}\,.
\label{LDzigzagapp}
\ee
Hence the  instanton density
\be
I_{4d}(x)\ \approx\ \frac{1}{2d}\,\delta^{(3)}(\vec X-\vec X^{(1)})\
+\ \frac{1}{2d}\,\delta^{(3)}(\vec X-\vec X^{(2)})\ +\ O(1/b_\epsilon^4).
\label{Izigzagapp}
\ee
Similar to the straight chain with $\phi=\pi$, half of the zigzag's net instanton number is concentrated at $\vec X^{(1)}=(0,0,+b_\epsilon/2)$, and the other half at $\vec X^{(1)}=(0,0,-b_\epsilon/2)$, where the separation $b_\epsilon$ is given by~(\ref{beApp}). Consequently, the abelian Coulomb energy of the zigzag  has  a 3D form
\be
E_{\rm C}\ =\ \frac{\pi N_c}{8\lambda M d}\left(2\times\frac{1}{2\rho}\,
	+\,\frac{1}{b_\epsilon}\,+\,O(d/b_\epsilon^2)\right),
\label{ECzigzagapp}
\ee
where $\rho\sim d$ is the effective charge radius of the instanton concentration at $\vec X^{(1)}$ or $\vec X^{(2)}$, which appears $\delta$-like in equation~(\ref{Izigzagapp}).
Note that both $\rho$ and $b_\epsilon$ depend on the zigzag deformation~$\epsilon$. To find $\rho$, we need to replace (\ref{LDzigzagapp}) with a better approximation
for $r_1\sim d\ll b_\epsilon$, namely
\begin{align}
\log\det(L)\ &
=\ \frac{\pi r_2}{d}\ +\ \log\left[\frac{1}{r_1}\,\sinh\frac{\pi r_1}{d}\right]\\
&\qquad+\frac{d^2}{\pi^2a^2}\,\frac{\pi r_1/d}{\sinh(\pi r_1/d)}\times\left[
	(2-\cos\nu)\,\cosh\frac{\pi r_1}{d}\,
	-\, \frac{\sin\nu(2+\cos\nu)}{1+\cos\nu}\,\cos\frac{\pi x_4}{d}
	\right]
+\ O(d^4/a^4).\nonumber
\end{align}
Consequently, the instanton density near $\vec X^{(1)}$ comes out to be
\begin{align}
I(x)\ \approx\ &
\frac{\pi^2}{8d^4}\times\frac{1}{\sinh^4(\pi r_1/d)}\left\{
	2\,+\,\cosh(2\pi r_1/d)\,+\,\frac{\sinh^4(\pi r_1/d)}{(\pi r_1/d)^4}\,
	-\,\frac{4\sinh (2\pi r_1/d)}{(2\pi r_1/d)}
	\right\}\label{ModulatedDensityapp}\\
&+\ \frac{(2-\cos\nu)}{a^2 d^2}\times\frac{1}{\sinh^4(\pi r_1/d)}
\left\{\eqalign{
	2\,+\,\cosh(2\pi r_1/d)\,&
	-\,\frac32\times\frac{(\pi r_1/d)}{\tanh(\pi r_1/d)}\cr
	&+\,\frac{3+(\pi r_1/d)^2}{2}\times\frac{\sinh(2\pi r_1/d)}{(2\pi r_1/d)}\cr
	}\right\}\nonumber\\
&-\ \frac{\sin\nu(2+\cos\nu)}{16a^2d^2(1+\cos\nu)}
	\times\frac{\cos(\pi x_4/d)}{\sinh^5(\pi r_1/d)}
\left\{\eqalign{
	-9\,&
	+\,16 (\pi r_1/d)^2\,+\,\cosh (4\pi r_1/d)\,\cr
	&+\,8[1+ (\pi r_1/d)^2]\times\cosh (2\pi r_1/d)\,\cr
	&-\,24 (\pi r_1/d)\times\sinh (2\pi r_1/d)\cr
	}\right\}.\nonumber
\end{align}
Note the $x_4$--dependent subleading term on the last line here. Its periodicity in $x_4$ is $2d$ rather than $d$, which is a characteristic of a zigzag rather than a straight-chain arrangement of the instantons. Indeed, the  instanton density near $\vec X^{(2)}$ has a similar term but with the opposite phase.

The Coulomb self-energy of the instanton density~(\ref{ModulatedDensityapp}) near $\vec X^{(1)}$ corresponds in 3D terms to a charge radius
\be
\frac{1}{2\rho}\ =\ \frac{\pi C_1}{8d}\ +\ \frac{C_2 d}{4\pi a^2}\times(2-\cos\nu)\,,
\ee
where $C_1$ and $C_2$ were introduced in  equations~(\ref{C1defapp}) and~(\ref{C2defapp}). Overall the instanton density is concentrated in two tubes parallel to the $x_4$ direction. The tubes are separated by a distance $b_\epsilon$ along the $x_3$ direction. The 3D radius of the tubes scales as $\rho\sim d\ll b_\epsilon$. The density is independent of $x_4$ in the leading approximation, but it has a subleading modulation with the period $2d$ which goes in opposite phase for  the two tubes.

The net Coulomb energy~(\ref{ECzigzagapp}) of the whole zigzag is
\begin{align}
E_{\rm C}\ &
=\ \frac{N_c}{8\lambda M d}\left[
	2\times\frac{\pi C_1}{8d}\,
	+\,2\times\frac{C_2 d}{4\pi a^2}\times(2-\cos\nu)\,
	+\,\frac{\cos\nu}{\pi a^2/d}
	\right]\nonumber\\
&=\ \frac{N_c}{\lambda M}\left[
	\frac{\pi^2 C_1}{32 d^2}\
	+\ \frac{2C_2\,+\,(2-C_2)\cos\nu}{16a^2}
	\right]\nonumber\\
&=\ \frac{N_c}{\lambda M}\left[ \frac{\pi^2 C_1}{32 d^2}\
    +\ \frac{C_2+2}{16a^2}\ -\ \frac{(2-C_2)d^2}{8\pi^2 a^6}\times\epsilon^2\
    +\ \frac{(2-C_2)d^4}{24\pi^4 a^8}\times\epsilon^4\ +\ \cdots\ \right] .
\label{EClargeModapp}
\end{align}

Combining it with the non-abelian energy~(\ref{ENAmod}), we find the net energy cost of the zigzag deformation
\begin{align}
\Delta E_{\rm net}(\epsilon)\ &
\equiv\ E_{\rm net}[\mbox{zigzag}]\ -\ E_{\rm net}[\mbox{straight chain}]\nonumber\\
&=\ \left( N_c\lambda M^3\,-\,\frac{(2-C_2)N_cd^2}{8\pi^2\lambda M a^6}\right)\times\epsilon^2\
+\ \frac{(2-C_2)N_cd^4}{24\pi^4\lambda M a^{10}}\times\epsilon^4\ +\ \cdots
\label{ZigzagLarge}
\end{align}
The coefficient of $\epsilon^4$ is always positive, but the coefficient of $\epsilon^2$ can be either positive or negative, depending on the lattice spacing~$d$ and the instanton's formal radius~$a$, which suggests another second-order phase transition between a straight chain and a zigzag, this time for $d\ll a$. Note that the coefficient of $\epsilon^2$ is positive in the limit $d\to 0$, that is the straight chain becomes stable again at large densities. For the moment let us take this transition seriously and analyze it in more detail.

To study the zigzag phase, it is convenient to express the energy in terms of $a$ and
\be
p\ =\ \frac{a^2}{\cos\nu}\ =\ \sqrt{a^4\,+\,(2d/\pi)^2\times\epsilon^2}\,.
\ee
In the zigzag phase $\epsilon\neq 0$, the variables $p$ and $a$ are independent, albeit $p>a^2$, while for the straight chain there is a constraint $p=a^2$. In terms of these variables the non-abelian energy reads
\be
E_{\rm NA}\ =\ {\rm const}\ +\ N_c\lambda M\left(
	(M^{\prime2}+\tfrac12 M^2)\times a^2\ +\ \frac{\pi^2\lambda^2M^4}{4d^2}\times p^2\right),
\ee
and hence the net energy of the zigzag is
\be
E_{\rm net}\ =\ {\rm const}\
+\ \frac{N_c}{\lambda M}\left( \lambda^2M^2(M^{\prime2}+\tfrac12 M^2)\times a^2\,
	+\,\frac{C_2}{8a^2}\right)\
+\ \frac{N_c}{\lambda M}\left(\frac{\pi^2\lambda^2M^4}{4d^2}\times p^2\,
	+\,\frac{2-C_2}{16p}\right).
\ee
Minimizing this energy with respect to the independent variables $p$ and $a$, we obtain
\begin{align}
p\ &
=\ \root 3\of{2-C_2\over 8\pi^2}\,\left(\frac{d}{\lambda M^2}\right)^{2/3},
\label{Pvalue}\\
a\ &
\equiv\ a_z\ =\ \frac{\root 4\of{C_2/8}}{[\lambda^2M^2(M^{\prime2}+\tfrac12 M^2)]^{1/4}}\
=\ \root 4\of{5C_2/8}\times a'_0\ \approx\ 1.024\,a'_0.
\label{Avalue}
\end{align}
Thus in the zigzag phase the formal radius $a\equiv a_z$ of an instanton does not depend on the lattice spacing $d$. Also, despite the $d\ll a$ regime in which individual instantons merge into two continuous lines in  the $x_4$ direction, $a_z$ is remarkably close to the radius $a'_0$~(\ref{InstSize}) of a standalone instanton!

The zigzag amplitude $\epsilon$ is obtained as follows
\be
\epsilon\ =\ \frac{2}{\pi d}\sqrt{p^2-a^4}\
=\ \frac{\pi a_z^2}{2d}\times\sqrt{\left(\frac{d}{D}\right)^{4/3}\,-\,1}\,,
\label{ZigzagAmp}
\ee
where $a_z$ is given in equation~(\ref{Avalue}) and
\be
D\ =\ \frac{\sqrt{C_2}\pi}{\sqrt{2-C_2}}\,
\sqrt{M^2\over M^{\prime2}+\frac12 M^2}\times a_z\
\approx\ 8.5\,\frac{M}{M'}\times a_z\quad({\rm for}\ M'\gg M).
\label{Ddef}
\ee
For $d<D$ formula~(\ref{ZigzagAmp}) breaks down because the solutions~(\ref{Pvalue}) and~(\ref{Avalue}) fail to satisfy the constraint $p\ge a^2$.  For smaller lattice spacings $d<D$, the instantons form a straight chain, $\epsilon\equiv0$, $p\equiv a^2$, and the net energy should be minimized subjected to this constraint. This leads to a cubic equation for $a^2$, namely
\be
(2-C_2){D^2\over d^2}\times a^6\ +\ 2C_2 a_z^2\times a^4\ -\ (2+C_2)a_z^6\ =\ 0.
\label{Aequation}
\ee
Note that the solution is smaller than $a_z^2$ for any $d<D$ --- in the high-density straight-chain phase, the formal instanton radius $a$ shrinks with the lattice spacing $d$; asymptotically, for $d\ll D$,
\be
\frac{a}{a_z}\ \approx\ \root 6\of{2+C_2\over 2-C_2}\times\left(\frac{d}{D}\right)^{1/3}.
\ee

\begin{figure}[h]
\begin{center}
\vspace{3ex}
\includegraphics[width= 0.95\linewidth]{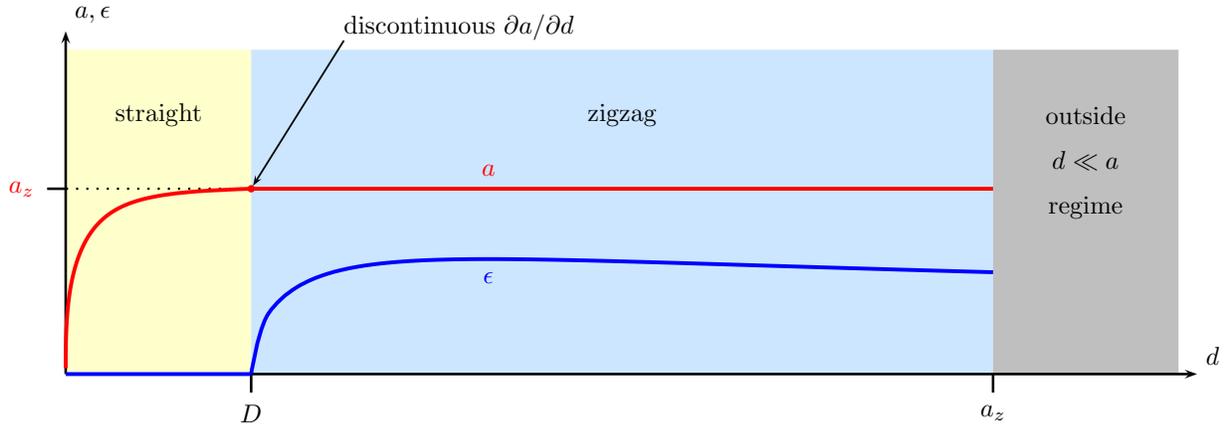}
\end{center}
\caption{Equilibrium instanton radius $a$ and the zigzag amplitude $\epsilon$ in the regime of large density $a\gg d$ in the case we allow only the straight chain and zigzag configurations with $\phi=\pi$ in the phase diagram.}
\label{figDenseRegime}
\end{figure}
The situation is summarized on figure~\ref{figDenseRegime}, where we plot the dependence of the $a$ parameter and the zigzag amplitude $\epsilon$ on the lattice spacing in the $d\ll a $ regime. Note that this whole picture --- the curves $a(d)$ and $\epsilon(d)$, the transition point $D$, and even the existence of the transition from zigzag for $d>D$ to a stable straight chain for $d<D$  --- is based on the Coulomb energy calculated in the $d\ll a$ approximation. Consistency of this analysis requires $D\ll a_z$, which in light of equation~(\ref{Ddef}) means $M'\gg 8.5 M$. Again, as in the case of small instantons, consistency requires $M'\gg M$.

\begin{figure}[h]
\begin{center}
\vspace{3ex}
\includegraphics[width= \linewidth]{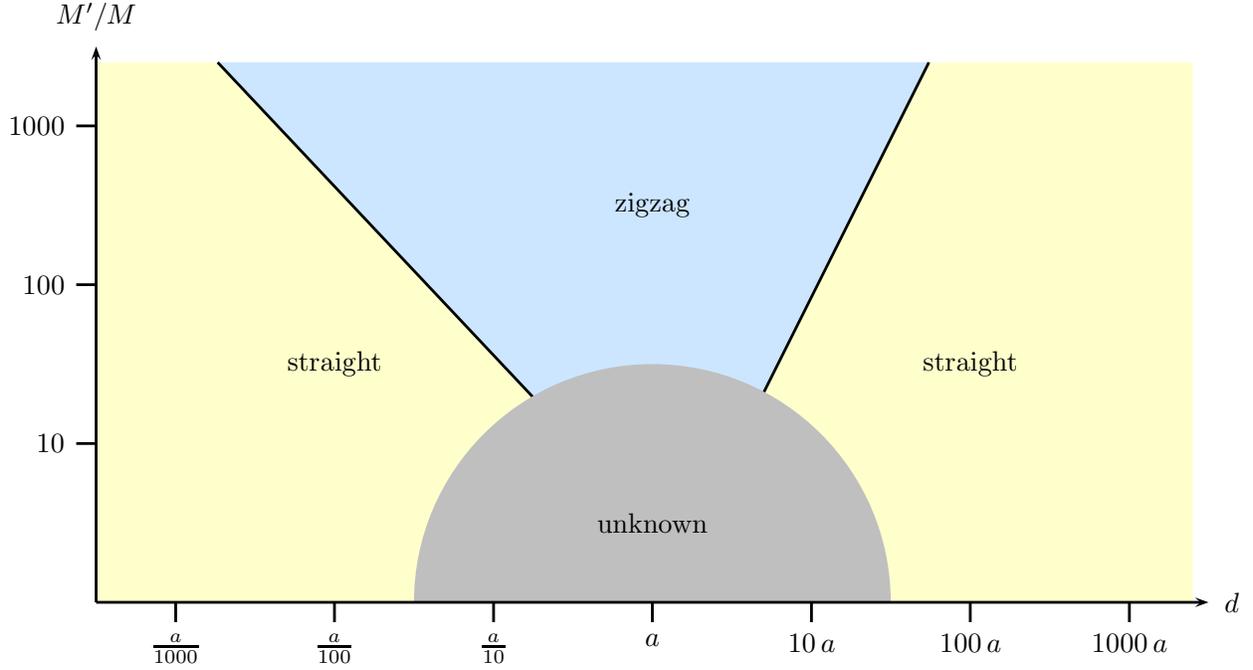}
\end{center}
\caption{Phase diagram at different anisotropy ratios $M'/M$ in the case we ignore multi-layer instanton configurations other than the zigzag with $\phi=\pi$.}
\label{fig1Dphase1}
\end{figure}

Thus as far as just the straight chain and the zigzag with $\phi=\pi$ are concerned one can have the following possibilities. For the highly anisotropic setups with $M'\gg M$, the instantons form a stable straight chain when the lattice spacing is either very large or very small. At the intermediate lattice spacings, the straight 1D lattice is unstable and the instantons form a zigzag. For less anisotropic setups, the zigzag region shrinks. For $M'\to M$, both transition points between straight-chain and zigzag phases move into the $d\sim a$ region where our analytical calculations become unreliable. These observations are summarized in figure~\ref{fig1Dphase1}. The big gray blob in the center indicates the region, where we do not have analytical control. Logically there are two possibilities for what is happening in this region. Perhaps there is similar phase structure for all $M'>M$: stable straight chain for either large or small lattice spacing, but a zigzag instability for intermediate~$d$. Alternatively, for small enough anisotropy $M'/M$, the zigzag phase may disappear altogether, and the straight chain is stable for all lattice spacings.

In principle one can be concerned about the meaning of the straight chain configuration at asymptotically large densities. In reality however we cannot just restrict to the zigzag and straight chain and must consider other relevant instanton configurations and compare their free energy. In particular favored configurations at large densities must be multi-layer 2D instanton lattices. In fact as we show in section~\ref{secGeneralTwist} assuming $M'\gg M$, at high density the more stable configuration is the zigzag with $\phi<\pi$ rather than the straight chain, see also figure~\ref{figZigzagPD}. Notice that the transition to the phase with the twist angle $\phi<\pi$ occurs at critical spacing $d\sim 1/(\sqrt{\lambda}M)$, while the scale for the straight chain transition considered in this section is
\be
D\simeq \left(\frac{M}{M'}\right)^{3/2}\times \frac{1}{\sqrt{\lambda}M} \ll \frac{1}{\sqrt{\lambda}M}.
\ee
One can thus be worried that for $M'\simeq M$, when both critical densities are of the same order, there is a transition from the zigzag back to the straight chain. Such a transition seems to be counterintuitive and would contradict the physical interpretation of the popcorn phase transitions given for example in section~\ref{secSummary}. Therefore it is still interesting to make a numerical analysis of the $M'\simeq M$ case and show that at high densities the straight chain is always dominated by multi-layer configurations.


\section{Zigzag. Details of calculations for general twist}
\label{appGenTwist}

In section~\ref{secGeneralTwist} we discussed the zigzag configuration  with a general (non-abelian) orientation twist~(\ref{genTwist}),~(\ref{genU}) between adjacent instantons.  For the general twist it is also a straightforward procedure to derive the full instanton solution, or at least the matrix $L$, which is related to the topological density~(\ref{Idensity}). However, unlike for the abelian limit, there is normally no way to present the results of such a calculation in a sensible compact manner. On the other hand we have to resort to approximations anyway when evaluating the Coulomb energy integral~(\ref{Cenergy}). Therefore for our purposes it is enough to calculate the instanton solutions up to few non-trivial orders in $a/d$, or its inverse. Here we will explain the calculation of the solution as an expansion in the case of $a\ll d$. We will find the expression for the determinant of matrix $L$ and compute the Coulomb potential $\hat{A}_0$, instanton density $I$ and finally the net energy of the zigzag.

The preliminaries of the calculation were explained in the beginning of section~\ref{sssecNAdata}. In particular we start from writing the matrix
\be
\label{Lapp}
L= (x^\mu - \Gamma^\mu(\theta))^2 + T(\theta)
\ee
in terms of the ADHM data~(\ref{Centers}),~(\ref{GeneralInstantons}) defined as differential operator acting on functions $\psi(\theta)$ defined on a circle.  The expansion of the ADHM data in powers of $a/d$ in section~\ref{sssecNAdata} took the following form.The  $\Gamma^4$ component is exact:
\be
\label{G4app}
\Gamma^4(\theta) = i\,d\,\frac{\partial}{\partial\theta}\,.
\ee
The matrix $T$ becomes
\be
T(\theta)\ =\ {\pi a^2\over2}\left(1+\Sigma_1\cos\beta\right)\times
\left(\delta(\theta+\frac{\phi}{2})+\delta(\theta-\frac{\phi}{2})\right).
\label{ToperatorApp}
\ee
while the remaining 3 components of $\Gamma^\mu$ take the form
\be
\vec\Gamma(\theta)\ =\ \epsilon\bn_3\,\Sigma_3\ +\ \vec\Gamma_1(\theta)\
+\ O(a^4/d^3)\, .
\label{GammaExpansionApp}
\ee
Here we wrote the operators in a basis of two-component wavefunctions
\be
\label{PsiApp}
\Psi(\theta)\ =\,\begin{pmatrix} \psi(\theta)+\psi(\theta+\pi)\\ \psi(\theta)-\psi(\theta+\pi)\end{pmatrix},\quad
-\frac{\pi}{2}\le\theta\le\frac{\pi}{2},\quad \Psi(\pi/2)\,=\,\Sigma_3\Psi(-\pi/2)\,,
\ee
and $\Sigma_{1,2,3}$ are Pauli matrices acting on the two components. The first term in~(\ref{GammaExpansionApp}) encodes the zeroth order abelian zigzag solution from section~\ref{secZigzag}. The second term $\vec{\Gamma}_1(\theta)$ is an $O(a^2/d)$ leading correction, which one obtains from solving equation~(\ref{gammaeqs}). The diagonal 11 and 22 matrix elements of the solution are piecewise-constant functions of $\theta$, from (\ref{SolDiag})
\be
\bigl(\vec{\Gamma}_1\bigr)_{\rm diag}(\theta)\
=\ {\pi a^2\over2d}\Bigl(\bp\,\cos\beta\,+\,(\bp\times\bq)\Sigma_3\sin\beta\Bigr)
\times\left({\phi\over\pi}-\Theta(-\frac{\phi}{2}<\theta<\frac{\phi}{2})\right),
\nonumber
\ee
where $\Theta$ is the step-function, while the off-diagonal 12 and 21 matrix elements are more complicated: from~(\ref{gammain}) for $-\phi/2<\theta<\phi/2$
\begin{align}
\nonumber
\bigl(\vec{\Gamma}_1\bigr)^{}_{12}(\theta)\ & =\ \bigl(\vec{\Gamma}_1\bigr)^*_{21}(\theta)
\\ \nonumber
&=\ {\pi a^2\over2d}\Bigl(-p_3\,\bn_3\ +\ \bv\,\cosh(2(\epsilon/d)\theta)\ - \ i(\bn_3\times\bv)\,\sinh(2(\epsilon/d)\theta)\Bigr),\\
{\rm where}\quad\bv\ & =\ {1\over\cosh(\pi \epsilon/d)}\Bigl( 	 -\bp_\perp\,\cosh((\pi-\phi)\epsilon/d)\,  +(\bn_3\times\bq_\perp)\sin\beta\sinh ((\pi-\phi)\epsilon/d)
	\Bigr)\,,
\nonumber
\end{align}
while for $\phi/2<\theta<\pi/2$, or $-\pi/2<\theta<-\phi/2$, equation~(\ref{gammaout}) tells that
\begin{align}
\nonumber
\bigl(\vec{\Gamma}_1\bigr)^{}_{12}(\theta)\ & =\ \bigl(\vec{\Gamma}_1\bigr)^*_{21}(\theta)
\\ \nonumber
&=\ \pm{\pi a^2\over2d}\left(\eqalign{
	-iq_3\sin\beta\,\bn_3\ &+\ \bv'\,\cosh((\epsilon/d)(\pi-2|\theta|))\cr
	&\pm\ i(\bn_3\times\bv')\,\sinh ((\epsilon/d)(\pi-2|\theta|))\cr
	}\right),\\
{\rm where}\quad\bv'\ &
=\ {i\over\cosh(\pi \epsilon/d)}\Bigl(
	(\bn_3\times\bp)\sinh(\phi \epsilon/d)\,
	-\,\bq_\perp\,\sin\beta\cosh (\phi \epsilon/d)
	\Bigr).
\nonumber
\end{align}
In our notations, $p_3$ and $q_3$ are the $x_3$ components of the $\bp$ and $\bq$ unit vectors while $\bp_\perp$ and $\bq_\perp$ are their components perpendicular to the $x_3$ direction; the $\pm$ sign in equation~(\ref{gammaout}) is the sign of $\theta$.

In summary we need to compute determinant of the differential operator $L(\theta)$ acting on the two component functions $\Psi$~(\ref{PsiApp}).The   operator $L$ is defined through~(\ref{Lapp}) with $T$ and $\Gamma^4$ given by~(\ref{ToperatorApp}) and~(\ref{G4app}) respectively, while the $\Gamma^{1,2,3}$ components are given by the expansion~(\ref{GammaExpansionApp}) with the leading order correction $\vec{\Gamma}_1$ specified by equations~(\ref{SolDiag}),~(\ref{gammain}) and~(\ref{gammaout}). Consequently we will compute $L$ as an expansion up to first few non-trivial corrections:
\be
\label{LExpApp}
L(x)\ =\ L_0(x)\ +\ L_1(x)\ +\ L_2(x)\ +\ \cdots,\
\ee
where
\begin{align}
L_0\ &
=\ \left(x_4-id\frac{\partial}{\partial\theta}\right)^2\ +\ \left(\vec{\bf x}-\epsilon\Sigma_3\bn_3\right)^2\
+\ {\pi a^2\over2}\times\left(\delta\bigl(\theta+\frac{\phi}{2}\bigr)+\delta\bigl(\theta-\frac{\phi}{2}\bigr)\right),\cr
L_1\ &
=\ -2\vec{\bf x}\cdot\vec\Gamma_1\ +\ \epsilon\bigl\{\Sigma_3,\bn_3\cdot\vec\Gamma_1\}\
+\ {\pi a^2\over2}\times\Sigma_1\cos\beta\times
	 \left(\delta\bigl(\theta+\frac{\phi}{2}\bigr)+\delta\bigl(\theta-\frac{\phi}{2}\bigr)\right),\cr
L_2\ &
=\ \left(\vec\Gamma_1\right)^2\ -\ 2\vec{\bf x}\cdot\vec\Gamma_2\
+\ \epsilon\bigl\{\Sigma_3,\bn_3\cdot\vec\Gamma_2\}, \qquad \cdots
\label{Lexp}
\end{align}
Notice that in the leading order we have removed $a/d$ terms from $\Gamma^3$, \emph{cf.}~(\ref{Gamma3beta0}). Also the last term in~(\ref{LExpApp}) is only $O(a^2)$ and naively should be dropped. Careful analysis shows however that this term is required for the consistency of our expansion in the regions close to instanton centers. Although the expression for $L_2$ depends on the subleading corrections $\Gamma_2$ to the ADHM matrices, only the first term, quadratic in $\vec{\Gamma}_1$, will be important.  The leading $L_0$ operator has determinant\footnote{In the notations of this section, `Tr' and `Det' are the trace and the determinant in the Hilbert space of $\Psi(\theta)$, while `tr' and `det' are the trace and the determinant of the $2\times2$ matrices only.}
\begin{align}
\Det(L_0)\ =\ {\rm const}
\times\Bigl(\cosh(\pi R_1/d)\,&
-\,\cos(\pi x_4/d)\,+\,\frac{\pi a^2}{ 2d R_1}\,\sinh(\pi R_1/d)\cr
&+\,\frac{\pi^2 a^4}{8d^2R_1^2}\,\sinh(\phi R_1/d)\sinh((\pi-\phi)R_1/d)\Bigr)\times{}\cr
\times\Bigl(\cosh(\pi R_2/d)\,&
+\,\cos(\pi x_4/d)\,+\,\frac{\pi a^2}{ 2d R_2}\,\sinh(\pi R_2/d)\cr
&+\,\frac{\pi^2 a^4}{ 8d^2R_2^2}\,\sinh(\phi R_2/d)\sinh((\pi-\phi)R_2/d)\Bigr)\,,
\label{LZeroDet}
\end{align}
where
\be
R^2_1\ =\ x_1^2\ +\ x_2^2\ +\ (x_3-\epsilon)^2\quad{\rm and}\quad
R^2_2\ =\ x_1^2\ +\ x_2^2\ +\ (x_3+\epsilon)^2,
\ee
while the effect of the subleading $L_1$ and $L_2$ operators on $\log\det(L)$ may be calculated perturbatively as
\begin{align}
\log\Det(L)\ =\ \log\Det(L_0)\ &
+\ \Tr\left(L_0^{-1}L_1\right)\ +\ \Tr\left(L_0^{-1}L_2\right)\
-\ \frac12\,\Tr\left(L_0^{-1}L_1L_0^{-1}L_1\right)\ +\ \cdots\cr
{}=\ \log\Det(L_0)\ &
+\!\int\limits_{-\pi/2}^{+\pi/2}\frac{\d\theta}{2\pi}\,\tr\Bigl(G_0(\theta,\theta) L_1(\theta)\Bigr)\
+\!\int\limits_{-\pi/2}^{+\pi/2}\frac{\d\theta}{2\pi}\,\tr\Bigl(G_0(\theta,\theta) L_2(\theta)\Bigr)\cr
&-\ \frac{1}{2}\int\!\!\!\int\frac{\d\theta_1\,\d\theta_2}{(2\pi)^2}\,\tr\Bigl(
	G_0(\theta_1,\theta_2) L_1(\theta_2)G_0(\theta_2,\theta_1) L_1(\theta_1)
	\Bigr)\ +\ \cdots,
\label{PertExp}
\end{align}
where $G_0(\theta_1,\theta_2)$ is the Green's function of the operator $L_0$. At generic points in space this Green's function is rather messy, but for small instantons (of radius much smaller than the lattice spacing, $a\ll d$) the instanton density is concentrated in small volumes around the instanton centers where $G_0$ has a much simpler form. Specifically, at $O(a)$ 4D distance $\rho$ from the center $X^\mu=(0,0,(-1)^n \epsilon,nd)$ of the instanton at position $n$,
\begin{align}
G_0(\theta_1,\theta_2)\ &
=\ e^{-in\pi(\theta_1-\theta_2)}
\begin{pmatrix}
g_1&0
\\ 0 &g_2\\
\end{pmatrix}
\qquad{\rm for\ even}\ n,\cr
G_0(\theta_1,\theta_2)\ &
=\ e^{-in\pi(\theta_1-\theta_2)}\begin{pmatrix}g_2&0\cr 0&g_1\cr\end{pmatrix}\qquad{\rm for\ odd}\ n,\cr
{\rm where}\quad
g_1(\theta_1,\theta_2)\ &
=\ {2\over \rho^2+a^2}\,
	\langle\!\langle\,\rm{independent\ of}\ \theta_1,\theta_2\,\rangle\!\rangle\
    +\ O(1/d^2)\cr
{\rm and}\quad
g_2(\theta_1,\theta_2)\ &
=\ {\pi\over 2\epsilon d}\times{\sinh((\pi-2|\theta_1-\theta_2|)\epsilon/d)\over\cosh(\pi \epsilon)}\
	+\ O(a^2/d^4)\,.
\label{Gsimp}
\end{align}
Note that the $g_1$ is $O(1/a^2)$ while the $g_2$ is $O(1/d^2)$. Consequently, near instanton centers,  both the first and the second orders of the formal perturbative expansion~(\ref{PertExp}) may contribute $O(a^2/d^2)$ terms to the determinant~(\ref{LZeroDet}). In terms of matrix elements of $L_1$ and $L_2$, at distance $\rho\sim a$ from the center of an even-numbered instanton we have
\begin{align}
\log\frac{\Det(L)}{\Det(L_0)}\ =\ & \frac{2}{\rho^2+a^2}\times
	\!\!\int\frac{\d\theta}{2\pi}\,(L_1)_{11}(\theta)\
-\ \frac{2}{ (\rho^2+a^2)^2}\times
	 \left[\int\frac{\d\theta}{2\pi}\,(L_1)_{11}(\theta)\right]^2\cr
&+\ \frac{\pi\tanh(\pi \epsilon)}{ 2\epsilon d}\times\!\!\int\frac{\d\theta}{2\pi}\,(L_1)_{22}(\theta)\cr
&+\ \frac{2}{\rho^2+a^2}\times\!\!\int\frac{\d\theta}{2\pi}\,(L_2)_{11}(\theta)\cr
&-\ \frac{2}{\rho^2+a^2}\times{\pi\over 2ED\,\cosh(\pi \epsilon/d)}\times{}\cr
&\qquad\times\!\!\int\!\!\!\int\frac{\d\theta_1\,\d\theta_2}{(2\pi)^2}\,
	\sinh((\pi-2|\theta_1-\theta_2|)\epsilon/d)\times
	(L_1)_{12}(\theta_1)\times(L_1)_{21}(\theta_2)\cr
&+\ O(a^4/d^4)\,;
\label{PertSum}
\end{align}
near the center of an odd-numbered instanton we have a similar expression with the 22 and 11 matrix elements exchanging their roles.

The diagonal matrix elements of the $L_1$ follow from those of the $\vec\Gamma_1$; in light of equation~(\ref{SolDiag}),
\be
(L_1)_{11,22}\ =\ -2\bigl(\vec{\bf x}\mp \epsilon\bn_3\bigr)\cdot
\bigl(\bp\,\cos\beta\mp(\bp\times\bq)\sin\beta\bigr)\times
\left({\phi\over\pi}-\Theta(-\frac{\phi}{2}<\theta<\frac{\phi}{2})\right),
\ee
hence
\be
\int\limits_{-\pi/2}^{+\pi/2}\frac{\d\theta}{2\pi}\,(L_1)_{11}(\theta)\
=\ \int\limits_{-\pi/2}^{+\pi/2}\frac{\d\theta}{2\pi}\,(L_1)_{22}(\theta)\
=\ 0\,,
\ee
which eliminates the first three terms on the right hand side of the expansion~(\ref{PertSum}). Moreover, in the remaining two terms we need only the leading $O(a^4/d^2)$ terms in $(L_2)_{\rm 11\,or\,22}$ and the leading $O(a^2)$ terms in $(L_1)_{\rm 12\,or\,21}$. Since we are working at $O(a)$ distance from an instanton center, terms involving to $\vec{\bf x}\mp \epsilon\bn_3$ carry extra powers of $a$ so that they may be neglected at this order of perturbation theory. In other words, we may approximate $\vec{\bf x}\approx\pm \epsilon\bn_3$, thus
\be
\bigl(L_2\bigr)_{\rm 11\,or\,22}\ \approx\ \bigl({\vec\Gamma}_1^2\bigr)_{\rm 11\,or\,22}
\ee
regardless of the second-order $\vec\Gamma_2(\theta)$ while
\be
(L_1)^{}_{12}\ =\ (L_1)_{21}^*\ \approx\
{\pi a^2\over2}\,\cos\beta\times
	 \bigl(\delta(\theta+\frac{\phi}{2})+\delta(\theta-\frac{\phi}{2})\bigr)\
\mp\ 2\epsilon\times(\bn_3\cdot\vec\Gamma_1)_{12}\,.
\ee
From equations~(\ref{SolDiag}), (\ref{gammain}), and~(\ref{gammaout}) for the $\vec\Gamma_1(\theta)$,
we obtain
\be
(L_2)_{11}\ =\ (L_2)_{22}\ =\ {\pi^2 a^4\over 4d^2}\times \begin{cases}
	 (1-\phi/\pi)^2\,+\,p_3^2\,+\,\bv^2\times\cosh(4(\epsilon/d)\theta) &
	\text{for } |\theta|<\phi/2,\cr
	 (\phi/\pi)^2\,+\,q_3^2\sin^2\beta\,+\,\bv^{\prime2}\times\cosh(2(\epsilon/d)(\pi-2|\theta|)) &
	\text{for } |\theta|>\phi/2,
	\end{cases}
\ee
hence
\begin{align}
\!\!\int\frac{\d\theta}{2\pi}\,(L_2)_{\rm11\,or\,22}\
=\ \frac{\pi^2a^4}{16d^2}\biggl( \frac{2\phi(\pi-\phi)}{\pi^2}\,&
+\,\frac{2\phi}{\pi}\,p_3^2\,+\,\frac{2(\pi-\phi)}{\pi}\,q_3^2\sin^2\beta\cr
&+\,\frac{\tanh(\pi \epsilon/d)}{(\pi \epsilon/d)}\bigl(\bp_\perp^2+\bq_\perp^2\sin^2\beta\bigr)\cr
&+\,\frac{\sinh((2\phi-\pi)\epsilon/d)}{(\pi \epsilon/d)\cosh(\pi \epsilon/d)}
	\bigl(\bp_\perp^2-\bq_\perp^2\sin^2\beta\bigr)\biggr)\,.
\label{FirstIntegral}
\end{align}
Similarly,
\begin{align}
(L_1)^{}_{12}\ =\ (L_1)_{21}^*\ &
=\ {\pi a^2\over2}\,\cos\beta\times
	 \bigl(\delta(\theta+\frac{\phi}{2})+\delta(\theta-\frac{\phi}{2})\bigr)\cr
&\qquad\pm\ {\pi a^2 \epsilon\over d}\times\begin{cases}
	p_3 & \text{for } -\frac{\phi}{2}<\theta<\frac{\phi}{2},\cr
	+iq_3\,\sin\beta & \text{for } \frac{\phi}{2}<\theta<\frac{\pi}{2},\cr
	-iq_3\,\sin\beta & \text{for } -\frac{\pi}{2}<\theta<-\frac{\phi}{2},\cr
	\end{cases}
\end{align}
leads to
\begin{align}
\!\!\!\int\!\!\!\int\frac{\d\theta_1\,\d\theta_2}{(2\pi)^2}\,&
\sinh\bigl((\pi-2|\theta_1-\theta_2|)\epsilon/d\bigr)
\times(L_1)_{12}(\theta_1)\times(L_1)_{21}(\theta_2)\ ={}\cr
{}=\ \frac{a^4}{ 8}\biggl[ &
\cos^2\beta\times\Bigl(\sinh(\pi \epsilon/d)\,-\,\sinh((2\phi-\pi)\epsilon/d)\Bigr)\cr
&+\ p_3^2\times\Bigl( 2\phi(\epsilon/d)\cosh(\pi \epsilon/d)\,
	-\,\sinh(\pi \epsilon/d)\,-\,\sinh ((2\phi-\pi)\epsilon/d)\Bigr)\cr
&+\ q_3^2\sin^2\beta\times\Bigl( 2(\pi-\phi)(\epsilon/d)\cosh(\pi \epsilon/d)\,
	-\,\sinh(\pi \epsilon/d)\,+\,\sinh ((2\phi-\pi)\epsilon/d)\Bigr)\cr
&\pm\ 2p_3\times\Bigl(\cosh(\pi \epsilon/d)\,-\,\cosh((2\phi-\pi)\epsilon/d)\Bigr)\biggr]\,,
\label{SecondIntegral}
\end{align}
where the last line changes sign between odd-numbered and even-numbered instantons. We are interested in the average energy per instanton, so in the first order of the perturbation theory that line does not contribute.

Substituting the integrals~(\ref{FirstIntegral}) and~(\ref{SecondIntegral}) into equation~(\ref{PertSum}) and making use of $\bp_\perp^2+p_3^2=\bq_\perp^2+q_2^2=1$, we obtain
\begin{align}
\log{\Det(L)\over\Det(L_0)}\ =\ &
{\pi^2 a^2\over 8d^2}\times{a^2\over a^2+\rho^2}\times\left(\eqalign{
	{\phi(\pi-\phi)\over\pi^2}\,&
	+\,{\tanh(\pi \epsilon/d)\over(\pi \epsilon/d)}\times\sin^2\beta\cr
	&+\,{\sinh((2\phi-\pi)\epsilon/d)\over (\pi \epsilon/d)\cosh (\pi \epsilon/d)}\times\cos^2\beta\cr
	}\right)\cr
&+\ O(a^4/d^4)\,.
\label{DeltaLog}
\end{align}
Somehow, the dependence of  the intermediate expressions such as~(\ref{FirstIntegral}) or~(\ref{SecondIntegral}) on the unit vectors $\bp$ and $\bq$ cancels out from this formula, at least at the $O(a^2/d^2)$ level.

To obtain the instanton number density $I(x)$ and the abelian Coulomb potential $\hat A_0(x)$ it produces, we need to add the perturbative correction~(\ref{DeltaLog}) to $\log\det(L_0)$ and take d'Alembertians. Since we are working at distances $\rho\sim a$ from an instanton center, we start by expanding the logarithm of $\det(L_0)$ from equation~(\ref{LZeroDet}) in powers of $a/d$ and $\rho/d$,
\begin{align}
\log\Det(L_0)\ =\ &
{\rm const}\ +\ \log(\rho^2+a^2)
\pm\ {\pi\over 2d}\,\tanh(\pi \epsilon/d)\times\Delta x_3\cr
&+\ {\pi^2\over 12d^2}\left(
	\bigl(4-3\tanh^2(\pi \epsilon/d)\bigr)\bigl((\Delta x_3)^2-(\Delta x_4)^2\bigr)\ + a^2\right)\cr
&+\ {\pi^2\over 12d^2}\left(1+{3\tanh(\pi \epsilon/d)\over(\pi \epsilon/d)}\right)\times
	\bigl((\Delta x_1)^2+(\Delta x_2)^2\bigr)\cr
&+\ {3\phi\pi-3\phi^2-\pi^2\over 12d^2}\times{a^4\over a^2+\rho^2}
+\ O(a^4/d^4)\,,
\label{OrigLog}
\end{align}
where $\Delta x_\mu$ is the displacement away from the instanton center. Combining the two logarithms~(\ref{OrigLog}) and~(\ref{DeltaLog}) and taking the d'Alembertians, we find
\begin{align}
\hat A_0(x)\ =\ \frac{1}{4\lambda M}&
\,\Box\left( \log\Det(L)\,=\,\log\Det(L_0)\,+\,\log \frac{\Det(L)}{\Det(L_0)}\right)\cr
{}=\ \frac{1}{4\lambda M}&\biggl[
\frac{8a^2+4\rho^2}{(a^2+\rho^2)^2}\
	+\ \frac{\pi^2}{d^2}\left(\frac{1}{3}\,+\,{\tanh(\pi \epsilon/d)\over(\pi \epsilon/d)}\right)
\ +\ {2\pi^2-12\phi\pi+12\phi^2\over 3d^2}\times{a^6\over(a^2+\rho^2)^3}\cr
&\quad-\ {2\pi^2\over d^2}\times{a^6\over(a^2+\rho^2)^3}\times
	{\cos^2\beta\sinh((2\phi-\pi)\epsilon/d)+\sin^2\beta\sinh(\pi \epsilon/d)\over(\pi \epsilon/d)\cosh(\pi \epsilon/d)}\cr
&\quad+\ O(a^2/d^4)\biggr]
\end{align}
and the instanton density itself
\begin{align}
I(x)\ &=\ -\,{1\over(4\pi)^2}\,\Box\Box\,\log\Det(L)\cr
&=\ {6\over\pi^2}\,{a^4\over(a^2+\rho^2)^4}\,
+\, {\pi^2-6\phi\pi+6\phi^2\over d^2}\times{a^6(a^2-\rho^2)\over(a^2+\rho^2)^5}\cr
&\qquad-\ {3\over d^2}\times{a^6(a^2-\rho^2)\over(a^2+\rho^2)^5}\times
	{\cos^2\beta\sinh((2\phi-\pi)\epsilon/d)+\sin^2\beta\sinh(\pi \epsilon/d)\over(\pi \epsilon/d)\cosh(\pi \epsilon/d)}\cr
&\qquad+\ O(1/d^4)\,.
\end{align}
Consequently, the Coulomb energy (per instanton) of the zigzag configuration is
\begin{align}
E_{\rm C}\ & =\ \frac{N_c}{4}\!\!\!\int\limits_{\rm near~center}\!\!\!\!\d^4 x\,\hat A_0(x)\times I(x)\cr
&=\ \frac{N_c}{\lambda M}\biggl[ {1\over 5a^2} \
+\ \frac{\pi^2}{80 d^2}\biggl(1\
+\ 2\left(1-{2\phi\over\pi}\right)^2\,
+\,{5\sinh(\pi \epsilon/d)\,-\,4\sinh((2\phi-\pi)\epsilon/d)\over (\pi \epsilon/d)\cosh(\pi \epsilon/d)}\cr
& \qquad -\,4\sin^2\beta\times{\sinh(\pi \epsilon/d)\,-\,\sinh((2\phi-\pi)\epsilon/d)\over (\pi \epsilon/d)\cosh(\pi \epsilon/d)}
	\biggr) \ +\ O(a^2/d^4)\biggr]\,,
\end{align}
or in terms of the generic formula~(\ref{ECgeneric}) for the Coulomb energy of the instanton configuration in section~\ref{sssecNAgeneral}
\begin{align}
{\cal C}(\phi,\beta,\epsilon/d)\ =\ \frac{\pi^2}{80}\biggl(1\,&
+\,2\left(1-\frac{2\phi}{\pi}\right)^2\,
  +\,\frac{5\sinh(\pi \epsilon/d)\,-\,4\sinh((2\phi-\pi)\epsilon/d)}{(\pi \epsilon/d)\cosh(\pi \epsilon/d)}\cr
&-\,4\sin^2\beta\times\frac{\sinh(\pi \epsilon/d)\,-\,\sinh((2\phi-\pi)\epsilon/d)}{(\pi \epsilon/d)\cosh(\pi \epsilon/d)}
	\biggr)\,.
\label{Cone}
\end{align}
Note that this Coulomb energy depends on the twist angles $\phi$ and $\beta$ but it does not depend on the directions of the twist axes $\bp$ and $\bq$!

Now consider the non-abelian energy per instanton
\be
E_{\rm NA}\ =\ \lambda MN_c\int\limits_0^d\d x^4\!\int\d^3x\,
\Bigl(1\,+\,M^2(x_3)^2\,+\,{M'}^2\bigl((x_1)^2+(x_2)^2\bigr)\Bigr)\times I(x).
\ee
Since the instanton density $I(x)$ is a total derivative, we can take this integral by parts and obtain an exact formula for the non-abelian energy in terms of the $T$ and $\vec\Gamma$ operators, namely
\begin{align}
E_{\rm NA}\ =\ N_c\lambda M\biggl(1\ +\ \bigl({M'}^2+ \frac12 M^2\bigr)\times\Tr(T)\ &
+\ M^2\times\Tr\left(\bigl(\bn_3\cdot\vec\Gamma\bigr)^2\right)\cr
&+\ {M'}^2\times\Tr\left({\vec\Gamma}^2-\bigl(\bn_3\cdot\vec\Gamma\bigr)^2\right)
	\biggr.\,.
\label{ENAgeneral}
\end{align}
The $T$ operator has a twist-independent trace
\be
\Tr\,(T)\ \equiv\int\frac{\d\theta}{2\pi}\,\tr\,(T)\ =\ a^2,
\ee
so our equation~(\ref{ENAgeneral}) agrees with the generic formula~(\ref{ENAgeneric}) for the non-abelian energy, provided we identify
\begin{align}
E^2\ +\ \frac{a^4}{d^2}\times{\cal A}\ +\ O(a^6/d^4)\ &
=\ \Tr\left(\bigl(\bn_3\cdot\vec\Gamma\bigr)^2\right)\
\equiv\int\frac{\d\theta}{2\pi}\,\tr\left(\bigl(\bn_3\cdot\vec\Gamma(\theta)\bigr)^2\right),\cr
\frac{a^4}{d^2}\times{\cal B}\ +\ O(a^6/d^4)\ &
=\ \Tr\left({\vec\Gamma}^2-\bigl(\bn_3\cdot\vec\Gamma\bigr)^2\right)\
\equiv\int\frac{\d\theta}{2\pi}\,
	\tr\left({\vec\Gamma}^2(\theta)-\bigl(\bn_3\cdot\vec\Gamma(\theta)\bigr)^2\right).
\label{ENAident}
\end{align}
In perturbation theory, the bare zigzag $\vec\Gamma_0=\epsilon\Sigma_3\,\bn_3$ is in the $x_3$ direction, hence in the transverse directions
\be
\Tr\left({\vec\Gamma}^2-\bigl(\bn_3\cdot\vec\Gamma\bigr)^2\right)\
=\ \Tr\left({\vec\Gamma}_1^2-\bigl(\bn_3\cdot\vec\Gamma_1^{}\bigr)^2\right)\
+\ O(a^6/d^4)
\ee
and therefore, in light of equations~(\ref{SolDiag}), (\ref{gammain}), and (\ref{gammaout}),
\begin{align}
{\cal B}\ &
=\ {d^2\over a^4}\times\Tr\left({\vec\Gamma}_1^2-\bigl(\bn_3\cdot\vec\Gamma_1^{}\bigr)^2\right)\cr
&=\ {\phi(\pi-\phi)\over4}\times\Bigl(\bp_\perp^2\,\cos^2\beta\,
	+\,(\bp\times\bq)_\perp^2\,\sin^2\beta\Bigr)\cr
&\qquad+\ {\pi^2\over 8(\pi \epsilon/d)}\times\Bigl( \bv^2\,\sinh(2\phi \epsilon/d)\,
	+\,\bv^{\prime2}\,\sinh(2(\pi-\phi)\epsilon/d\Bigr)\cr
&=\ {\phi(\pi-\phi)\over4}\times\Bigl(\bp_\perp^2\,\cos^2\beta\,
	+\,(\bp\times\bq)_\perp^2\,\sin^2\beta\Bigr)\cr
&\qquad+\ {\pi^2\over8}\times{\tanh(\pi \epsilon/d)\over(\pi \epsilon/d)}
	\times\bigl(\bp_\perp^2\,+\,\bq_\perp^2\sin^2\beta\bigr)\cr
&\qquad+\ {\pi^2\over8}\times{\sinh((2\phi-\pi)\epsilon/d)\over(\pi \epsilon/d)\cosh(\pi \epsilon/d)}
	\times\bigl(\bp_\perp^2\,-\,\bq_\perp^2\sin^2\beta\bigr)\,.
\end{align}
In the $x_3$ direction the perturbation theory is more involved:
\begin{align}
\Tr\left(\bigl(\bn_3\cdot\vec\Gamma\bigr)^2\right)\
=\ \epsilon^2\,\Tr\left(\bigl(\Sigma_3\bigr)^2\right)\ &
+\ 2\epsilon\Tr\left(\Sigma_3\bigl(\bn_3\cdot\vec\Gamma_1\bigr)\right)\
+\ \Tr\left(\bigl(\bn_3\cdot\vec\Gamma_1\bigr)^2\right)\cr
&+\ 2\epsilon\Tr\left(\Sigma_3\bigl(\bn_3\cdot\vec\Gamma_2\bigr)\right)\
+\ O(a^6/d^4)\,,
\end{align}
where
\begin{align}
\Tr\left(\bigl(\Sigma_3\bigr)^2\right)\ &
\equiv\int\frac{\d\theta}{2\pi}\,\tr\left(\bigl(\Sigma_3\bigr)^2\right)\
=\ 1,\cr
{\rm and}\quad
\Tr\left(\Sigma_3\bigl(\bn_3\cdot\vec\Gamma_1\bigr)\right)\ &
\equiv\int\frac{\d\theta}{\pi}\,
	\tr\left(\Sigma_3\bigl(\bn_3\cdot\vec\Gamma_1\bigr)\right)\cr
&=\ 2(\bp\times\bq)_3\,\sin\beta\times \int\frac{\d\theta}{2\pi}\,
	\left(\frac{\phi}{\pi}-\Theta(-\frac{\phi}{2}<\theta<+\frac{\phi}{2})\right)\cr
&=\ 0\,.
\end{align}
Hence
\be
\frac{a^4}{d^2}\times{\cal A}\
=\ \Tr\left(\bigl(\bn_3\cdot\vec\Gamma_1\bigr)^2\right)\
+\ 2\epsilon\Tr\left(\Sigma_3\bigl(\bn_3\cdot\vec\Gamma_2\bigr)\right),
\ee
where the second term involves the second-order (in $a^2/d^2$) perturbative correction to the $\vec\Gamma$ operator that we have not calculated yet. Fortunately, the $\cal A$ function of the moduli gives a subleading contribution to the net energy in the $M'\gg M$ limit, \emph{cf.} equation~(\ref{ENgeneric}), so we do not have to work it out.

Altogether, in the $M'\gg M$ limit the net energy per instanton comes out to be
\begin{align}
E_{\rm net}\ =\ E_B\ &
+\ {\pi^2\,N_c M\over80(\lambda M^2d^2)}\times{\cal F}({\rm moduli})\
+\ O(N_c M^2/M'),\\
{\rm where}\quad{\cal F}\ =\ 3\ &
-\ {4\phi(\pi-\phi)\over\pi^2}\times
	 \Bigl(2\,-\,\bp_\perp^2\,\cos^2\beta\,-\,(\bp\times\bq)^2_\perp\,\sin^2\beta\Bigr)\cr
&+\ {\tanh(\pi \epsilon/d)\over(\pi \epsilon/d)}\times
	 \Bigl(5\,-\,4\sin^2\beta\,+\,2\bp_\perp^2\,+\,2\bq_\perp^2\,\sin^2\beta\Bigr)\cr
&+\ {\sinh((2\phi-\pi)\epsilon/d)\over (\pi \epsilon/d)\cosh (\pi \epsilon/d)}\times
	 \Bigl(-4\,+\,4\sin^2\beta\,+\,2\bp_\perp^2\,-\,2\bq_\perp^2\,\sin^2\beta\Bigr)\cr
&+\ {80\over\pi^4}\,(\lambda M^2d^2)^2\times(\pi \epsilon/d)^2\,.
\label{NetEnergyApp}
\end{align}

In this appendix we presented the derivation of the net energy per instanton of the zigzag configuration with general $SU(2)$ orientation twist in the limit of small instantons $a\ll d$ and high anisotropy $M'\gg M$. We have demonstrated that the result can be presented in the general form of~(\ref{ENgeneric}) and specified the generic function $\mathcal{F}$ in terms of zigzag amplitude, lattice spacing and twist parameters. In section~\ref{sssecNAtwist} we minimize the net energy with respect to parameters at given spacing $d$ to describe the phase space of zigzag configurations.

\end{appendix}



\begin{thebibliography}{99}

\bibitem{Rajagopal:2000wf}
  K.~Rajagopal and F.~Wilczek,
  ``The Condensed matter physics of QCD,''
  arXiv:hep-ph/0011333.

\bibitem{Schafer:2005ff}
  T.~Schafer,
  ``Phases of QCD,''
  arXiv:hep-ph/0509068.

\bibitem{Alford:2006wn}
  M.~G.~Alford,
  ``Color superconductivity in ultra-dense quark matter,''
  PoS {\bf LAT2006}, 001 (2006)
  [arXiv:hep-lat/0610046].

\bibitem{Stephanov:2007fk}
  M.~A.~Stephanov,
  ``QCD phase diagram: An Overview,''
  PoS {\bf LAT2006}, 024 (2006)
  [arXiv:hep-lat/0701002].

\bibitem{McLerran:2007qj}
  L.~McLerran and R.~D.~Pisarski,
  ``Phases of cold, dense quarks at large N(c),''
  Nucl.\ Phys.\  A {\bf 796}, 83 (2007)
  [arXiv:0706.2191 [hep-ph]].

\bibitem{AdS/CFT} J.~M.~Maldacena,
  ``The Large N limit of superconformal field theories and supergravity,''
  Adv.\ Theor.\ Math.\ Phys.\  {\bf 2} (1998) 231
  [Int.\ J.\ Theor.\ Phys.\  {\bf 38} (1999) 1113]
  [arXiv:hep-th/9711200].

  S.~S.~Gubser, I.~R.~Klebanov and A.~M.~Polyakov,
  ``Gauge theory correlators from noncritical string theory,''
  Phys.\ Lett.\  B {\bf 428} (1998) 105
  [arXiv:hep-th/9802109].

  E.~Witten,
  ``Anti-de Sitter space and holography,''
  Adv.\ Theor.\ Math.\ Phys.\  {\bf 2 } (1998)  253-291.
  [hep-th/9802150].

\bibitem{SakaiSugimoto2004}
  T.~Sakai and S.~Sugimoto,
  ``Low energy hadron physics in holographic QCD,''
  Prog.\ Theor.\ Phys.\  {\bf 113} (2005) 843
  [arXiv:hep-th/0412141].

\bibitem{WittensModel} E.~Witten,
  ``Anti-de Sitter space, thermal phase transition, and confinement in  gauge
  theories,''
  Adv.\ Theor.\ Math.\ Phys.\  {\bf 2} (1998) 505
  [arXiv:hep-th/9803131].

\bibitem{KK}
  A.~Karch and E.~Katz,
  ``Adding flavor to AdS/CFT,''
  JHEP {\bf 0206}, 043 (2002) [arXiv:hep-th/0205236].

\bibitem{Kruczenski:2003uq} M.~Kruczenski, D.~Mateos, R.~C.~Myers and D.~J.~Winters,
  ``Meson spectroscopy in AdS / CFT with flavor,''
  JHEP {\bf 0307} (2003) 049
  [hep-th/0304032].

  M.~Kruczenski, D.~Mateos, R.~C.~Myers and D.~J.~Winters,
  ``Towards a holographic dual of large N(c) QCD,''
  JHEP {\bf 0405} (2004) 041
  [arXiv:hep-th/0311270].

\bibitem{Erdmenger:2007cm}
  J.~Erdmenger, N.~Evans, I.~Kirsch and E.~Threlfall,
  ``Mesons in Gauge/Gravity Duals - A Review,''
  Eur.\ Phys.\ J.\ A {\bf 35} (2008) 81
  [arXiv:0711.4467 [hep-th]].

\bibitem{WittenBaryons}
  E.~Witten,
  ``Baryons and branes in anti de Sitter space,''
  JHEP {\bf 9807}, 006 (1998)
  [arXiv:hep-th/9805112].

\bibitem{Hata:2007mb}
H.~Hata, T.~Sakai, S.~Sugimoto and S.~Yamato,
  ``Baryons from instantons in holographic QCD,''
  Prog.\ Theor.\ Phys.\  {\bf 117}, 1157 (2007)
  [arXiv:hep-th/0701280].

\bibitem{Kim:2011ey}
  Y.~Kim and D.~Yi,
  ``Holography at work for nuclear and hadron physics,''
  arXiv:1107.0155 [hep-ph].

\bibitem{Horigome:2006xu}
  N.~Horigome and Y.~Tanii,
  ``Holographic chiral phase transition with chemical potential,''
  JHEP {\bf 0701} (2007) 072
  [arXiv:hep-th/0608198].

S.~Nakamura, Y.~Seo, S.~-J.~Sin and K.~P.~Yogendran,
  ``A New Phase at Finite Quark Density from AdS/CFT,''
  J.\ Korean Phys.\ Soc.\  {\bf 52} (2008) 1734
  [hep-th/0611021].

\bibitem{Yamada:2007} D.~Yamada,
  ``Sakai-Sugimoto model at high density,''
  JHEP {\bf 0810} (2008) 020
  [arXiv:0707.0101 [hep-th]].

\bibitem{Bergman}
  O.~Bergman, G.~Lifschytz and M.~Lippert,
  ``Holographic Nuclear Physics,''
  JHEP {\bf 0711} (2007) 056
  [arXiv:0708.0326 [hep-th]].

\bibitem{Rozali:2007rx}
  M.~Rozali, H.~H.~Shieh, M.~Van Raamsdonk and J.~Wu,
  ``Cold Nuclear Matter In Holographic QCD,''
  JHEP {\bf 0801}, 053 (2008)
  [arXiv:0708.1322 [hep-th]].

\bibitem{FiniteDensity}
  S.~Kobayashi, D.~Mateos, S.~Matsuura, R.~C.~Myers and R.~M.~Thomson,
  ``Holographic phase transitions at finite baryon density,''
  JHEP {\bf 0702} (2007) 016
  [arXiv:hep-th/0611099].

  S.~K.~Domokos and J.~A.~Harvey,
  ``Baryon number-induced Chern-Simons couplings of vector and axial-vector
  mesons in holographic QCD,''
  Phys.\ Rev.\ Lett.\  {\bf 99}, 141602 (2007)
  [arXiv:0704.1604 [hep-ph]].

  Y.~Kim, B.~H.~Lee, S.~Nam, C.~Park and S.~J.~Sin,
  ``Deconfinement phase transition in holographic QCD with matter,''
  Phys.\ Rev.\  D {\bf 76} (2007) 086003
  [arXiv:0706.2525 [hep-ph]].

  Y.~Kim, C.~H.~Lee and H.~U.~Yee,
  ``Holographic Nuclear Matter in AdS/QCD,''
  Phys.\ Rev.\  D {\bf 77}, 085030 (2008)
  [arXiv:0707.2637 [hep-ph]].

  S.~J.~Sin,
  ``Gravity back-reaction to the baryon density for bulk filling branes,''
  JHEP {\bf 0710} (2007) 078
  [arXiv:0707.2719 [hep-th]].

A.~Karch and A.~O'Bannon,
  ``Holographic thermodynamics at finite baryon density: Some exact results,''
  JHEP {\bf 0711} (2007) 074
  [arXiv:0709.0570 [hep-th]].

D.~Mateos, S.~Matsuura, R.~C.~Myers and R.~M.~Thomson,
  ``Holographic phase transitions at finite chemical potential,''
  JHEP {\bf 0711} (2007) 085
  [arXiv:0709.1225 [hep-th]].

  P.~Basu, F.~Nogueira, M.~Rozali, J.~B.~Stang and M.~Van Raamsdonk,
  ``Towards A Holographic Model of Color Superconductivity,''
  arXiv:1101.4042 [hep-th].

\bibitem{Rho:2009ym}
  M.~Rho, S.~J.~Sin and I.~Zahed,
  ``Dense QCD: a Holographic Dyonic Salt,''
  Phys.\ Lett.\  B {\bf 689}, 23 (2010)
  [arXiv:0910.3774 [hep-th]].

\bibitem{Skyrme}
  T.~H.~R.~Skyrme,
  ``A Nonlinear field theory,''
  Proc.\ Roy.\ Soc.\ Lond.\  A {\bf 260} (1961) 127.

\bibitem{Klebanov} I.~R.~Klebanov,
  ``Nuclear Matter In The Skyrme Model,''
  Nucl.\ Phys.\  B {\bf 262} (1985) 133.

\bibitem{Goldhaber:1987pb}
  A.~S.~Goldhaber and N.~S.~Manton,
  ``Maximal Symmetry Of The Skyrme Crystal,''
  Phys.\ Lett.\  B {\bf 198}, 231 (1987).

\bibitem{KuglerShtrikman}
  M.~Kugler and S.~Shtrikman,
  ``A new skyrmion crystal,''
  Phys.\ Lett.\  B {\bf 208}, 491 (1988).

  M.~Kugler and S.~Shtrikman,
  ``Skyrmion crystals and their symmetries,''
  Phys.\ Rev.\  D {\bf 40}, 3421 (1989).

\bibitem{DKS} A.~Dymarsky, S.~Kuperstein and J.~Sonnenschein,
  ``Chiral Symmetry Breaking with non-SUSY D7-branes in ISD backgrounds,''
  JHEP {\bf 0908} (2009) 005
  [arXiv:0904.0988 [hep-th]].

\bibitem{KS}
  I.~R.~Klebanov and M.~J.~Strassler,
  ``Supergravity and a confining gauge theory: Duality cascades and
  chiSB-resolution of naked singularities,''
  JHEP {\bf 0008}, 052 (2000)
  [arXiv:hep-th/0007191].

\bibitem{tHooftLargeN}
  G.~'t Hooft,
  ``A Planar Diagram Theory for Strong Interactions,''
  Nucl.\ Phys.\  B {\bf 72} (1974) 461.

\bibitem{Witten:1979kh}
  E.~Witten,
  ``Baryons In The 1/N Expansion,''
  Nucl.\ Phys.\  B {\bf 160}, 57 (1979).

E.~Witten,
  ``Current Algebra, Baryons, And Quark Confinement,''
  Nucl.\ Phys.\  B {\bf 223}, 433 (1983).

G.~S.~Adkins, C.~R.~Nappi and E.~Witten,
  ``Static Properties Of Nucleons In The Skyrme Model,''
  Nucl.\ Phys.\  B {\bf 228}, 552 (1983).

\bibitem{Casher}  A.~Casher,
  ``Chiral Symmetry Breaking in Quark Confining Theories,''
  Phys.\ Lett.\  B {\bf 83} (1979) 395.

  T.~Banks and A.~Casher,
  ``Chiral Symmetry Breaking in Confining Theories,''
  Nucl.\ Phys.\  B {\bf 169} (1980) 103.

\bibitem{Aharony:2006da} O.~Aharony, J.~Sonnenschein and S.~Yankielowicz,
  ``A Holographic model of deconfinement and chiral symmetry restoration,''
  Annals Phys.\  {\bf 322}, 1420 (2007)
  [arXiv:hep-th/0604161].

\bibitem{Kaplan} D.~B.~Kaplan and A.~V.~Manohar,
  ``Nucleon nucleon potential in the 1/N(c) expansion,''
  Phys.\ Rev.\  C {\bf 56} (1997) 76
  [arXiv:nucl-th/9612021].

\bibitem{Kaplunovsky:2010eh}
  V.~Kaplunovsky and J.~Sonnenschein,
  ``Searching for an Attractive Force in Holographic Nuclear Physics,''
  JHEP {\bf 1105} (2011) 058
  [arXiv:1003.2621 [hep-th]].

\bibitem{Bonanno:2011yr}
  L.~Bonanno and F.~Giacosa,
  ``Does nuclear matter bind at large $N_c$?,''
  Nucl.\ Phys.\ A {\bf 859} (2011) 49
  [arXiv:1102.3367 [hep-ph]].

\bibitem{DecChiSplitting} J.~J.~M.~Verbaarschot and T.~Wettig,
  ``Random matrix theory and chiral symmetry in QCD,''
  Ann.\ Rev.\ Nucl.\ Part.\ Sci.\  {\bf 50}, 343 (2000)
  [arXiv:hep-ph/0003017].

  J.~C.~Osborn, K.~Splittorff and J.~J.~M.~Verbaarschot,
  ``Chiral symmetry breaking and the Dirac spectrum at nonzero chemical
  potential,''
  Phys.\ Rev.\ Lett.\  {\bf 94}, 202001 (2005)
  [arXiv:hep-th/0501210].

  L.~Y.~Glozman and R.~F.~Wagenbrunn,
  ``Chirally symmetric but confining dense and cold matter,''
  Phys.\ Rev.\  D {\bf 77}, 054027 (2008)
  [arXiv:0709.3080 [hep-ph]].

\bibitem{Rubakov} D.~V.~Deryagin, D.~Y.~Grigoriev and V.~A.~Rubakov,
  ``Standing wave ground state in high density, zero temperature QCD at large
  N(c),''
  Int.\ J.\ Mod.\ Phys.\  A {\bf 7}, 659 (1992).

\bibitem{SonShuster} E.~Shuster and D.~T.~Son,
  ``On finite density QCD at large N(c),''
  Nucl.\ Phys.\  B {\bf 573}, 434 (2000)
  [arXiv:hep-ph/9905448].

\bibitem{Kutschera:1984zm}
  M.~Kutschera, C.~J.~Pethick and D.~G.~Ravenhall,
  ``Dense Matter In The Chiral Soliton Model,''
  Phys.\ Rev.\ Lett.\  {\bf 53} (1984) 1041.

 N.~S.~Manton and P.~J.~Ruback,
  ``Skyrmions In Flat Space And Curved Space,''
  Phys.\ Lett.\ B {\bf 181} (1986) 137.

\bibitem{AtiyahManton} M.~F.~Atiyah and N.~S.~Manton,
  ``Skyrmions from instantons,''
  Phys.\ Lett.\  B {\bf 222} (1989) 438.

\bibitem{Schroers:1993yk}
  B.~J.~Schroers,
  ``Dynamics of moving and spinning Skyrmions,''
  Z.\ Phys.\ C {\bf 61} (1994) 479
  [hep-ph/9308236].

C.~J.~Houghton, N.~S.~Manton and P.~M.~Sutcliffe,
  ``Rational maps, monopoles and Skyrmions,''
  Nucl.\ Phys.\ B {\bf 510} (1998) 507
  [hep-th/9705151].

\bibitem{Manton87} N.~S.~Manton,
  ``Is the B=2 skyrmion axially symmetric?,''
  Phys.\ Lett.\  B {\bf 192}, 177 (1987).

\bibitem{Wust}  E.~Wuest, G.~E.~Brown and A.~D.~Jackson,
  ``Topological chiral bags in a baryonic environment,''
  Nucl.\ Phys.\  A {\bf 468}, 450 (1987).

\bibitem{Jackson:1988bd}
  A.~D.~Jackson, A.~Wirzba and N.~S.~Manton,
  ``New skyrmion solutions on a three sphere,''
  Nucl.\ Phys.\  A {\bf 495} (1989) 499.

\bibitem{ParityDoubling} H.~Forkel, A.~D.~Jackson, M.~Rho, C.~Weiss, A.~Wirzba and H.~Bang,
  ``Chiral symmetry restoration and the Skyrme model,''
  Nucl.\ Phys.\  A {\bf 504} (1989) 818.

\bibitem{Sutcliffe}
  P.~Sutcliffe,
  ``Skyrmions, instantons and holography,''
  JHEP {\bf 1008} (2010) 019
  [arXiv:1003.0023 [hep-th]].

\bibitem{Nawa} K.~Nawa, H.~Suganuma and T.~Kojo,
  ``Baryons in holographic QCD,''
  Phys.\ Rev.\ D {\bf 75} (2007) 086003
  [hep-th/0612187].

\bibitem{Brandhuber:1998xy} D.~J.~Gross and H.~Ooguri,
  ``Aspects of large N gauge theory dynamics as seen by string theory,''
  Phys.\ Rev.\  D {\bf 58} (1998) 106002
  [arXiv:hep-th/9805129].

  A.~Brandhuber, N.~Itzhaki, J.~Sonnenschein and S.~Yankielowicz,
  ``Baryons from supergravity,''
  JHEP {\bf 9807}, 020 (1998)
  [arXiv:hep-th/9806158].

  C.~G.~Callan, A.~Guijosa, K.~G.~Savvidy and O.~Tafjord,
  ``Baryons and flux tubes in confining gauge theories from brane actions,''
  Nucl.\ Phys.\  B {\bf 555}, 183 (1999)
  [arXiv:hep-th/9902197].

\bibitem{Seki:2008mu}
  S.~Seki and J.~Sonnenschein,
  ``Comments on Baryons in Holographic QCD,''
  JHEP {\bf 0901} (2009) 053
  [arXiv:0810.1633 [hep-th]].

\bibitem{Dbrane-instanton}
  E.~Witten,
  ``Sigma Models And The Adhm Construction Of Instantons,''
  J.\ Geom.\ Phys.\  {\bf 15}, 215 (1995)
  [arXiv:hep-th/9410052].

  M.~R.~Douglas,
  ``Branes within branes,''
  arXiv:hep-th/9512077.

  M.~R.~Douglas,
  ``Gauge Fields and D-branes,''
  J.\ Geom.\ Phys.\  {\bf 28}, 255 (1998)
  [arXiv:hep-th/9604198].

\bibitem{ADHM}  M.~F.~Atiyah, N.~J.~Hitchin, V.~G.~Drinfeld and Yu.~I.~Manin,
  ``Construction of instantons,''
  Phys.\ Lett.\  A {\bf 65} (1978) 185.

\bibitem{Kuperstein:2004yf}
  S.~Kuperstein and J.~Sonnenschein,
  ``Non-critical, near extremal AdS(6) background as a holographic laboratory
  of four dimensional YM theory,''
  JHEP {\bf 0411}, 026 (2004)
  [arXiv:hep-th/0411009].

\bibitem{Dymarsky:2010ci}
  A.~Dymarsky, D.~Melnikov and J.~Sonnenschein,
  ``Attractive Holographic Baryons,''
  JHEP {\bf 1106} (2011) 145
  [arXiv:1012.1616 [hep-th]].

\bibitem{Pomarol} A.~Pomarol and A.~Wulzer,
  ``Baryon Physics in Holographic QCD,''
  Nucl.\ Phys.\ B {\bf 809} (2009) 347
  [arXiv:0807.0316 [hep-ph]].

G.~Panico and A.~Wulzer,
  ``Nucleon Form Factors from 5D Skyrmions,''
  Nucl.\ Phys.\ A {\bf 825} (2009) 91
  [arXiv:0811.2211 [hep-ph]].

\bibitem{Cherman:2009gb}
  A.~Cherman, T.~D.~Cohen and M.~Nielsen,
  ``Model Independent Tests of Skyrmions and Their Holographic Cousins,''
  Phys.\ Rev.\ Lett.\  {\bf 103} (2009) 022001
  [arXiv:0903.2662 [hep-ph]].

\bibitem{Cherman:2011ve}
  A.~Cherman and T.~Ishii,
  ``Long-distance properties of baryons in the Sakai-Sugimoto model,''
  Phys.\ Rev.\ D {\bf 86} (2012) 045011
  [arXiv:1109.4665 [hep-th]].


\bibitem{Kuperstein:KW}
S.~Kuperstein and J.~Sonnenschein,
  ``A New Holographic Model of Chiral Symmetry Breaking,''
  JHEP {\bf 0809}, 012 (2008)
  [arXiv:0807.2897 [hep-th]].

\bibitem{KW}
  I.~R.~Klebanov and E.~Witten,
  ``Superconformal field theory on threebranes at a Calabi-Yau singularity,''
  Nucl.\ Phys.\  B {\bf 536} (1998) 199
  [arXiv:hep-th/9807080].

\bibitem{Kim:2006gp}
  K.~Y.~Kim, S.~J.~Sin and I.~Zahed,
  ``Dense hadronic matter in holographic QCD,''
  arXiv:hep-th/0608046.

\bibitem{Lee:1997vp}
  K.~M.~Lee and P.~Yi,
  ``Monopoles and instantons on partially compactified D-branes,''
  Phys.\ Rev.\  D {\bf 56}, 3711 (1997)
  [arXiv:hep-th/9702107].

\bibitem{Harland} D.~Harland and R.~S.~Ward,
  ``Chains of Skyrmions,''
  JHEP {\bf 0812} (2008) 093
  [arXiv:0807.3870 [hep-th]].

\bibitem{Kraan:1998pm}
  T.~C.~Kraan and P.~van Baal,
  ``Periodic instantons with non-trivial holonomy,''
  Nucl.\ Phys.\  B {\bf 533}, 627 (1998)
  [arXiv:hep-th/9805168].

\bibitem{Harrington:1978ve}
  B.~J.~Harrington and H.~K.~Shepard,
  ``Periodic Euclidean Solutions And The Finite Temperature Yang-Mills Gas,''
  Phys.\ Rev.\  D {\bf 17} (1978) 2122.

\bibitem{Rossi}  P.~Rossi,
  ``Propagation Functions In The Field Of A Monopole,''
  Nucl.\ Phys.\  B {\bf 149} (1979) 170.

\bibitem{Osborn} H.~Osborn,
  ``Calculation Of Multi - Instanton Determinants,''
  Nucl.\ Phys.\  B {\bf 159} (1979) 497.

\bibitem{Nahm} W.~Nahm,
  ``A Simple Formalism For The Bps Monopole,''
  Phys.\ Lett.\  B {\bf 90}, 413 (1980).

\bibitem{Lee:2009dpa}
  H.~K.~Lee and M.~Rho,
  ``Half-Skyrmion Hadronic Matter at High Density,''
  arXiv:0905.0235 [hep-ph].


\end{thebibliography}
\end{document}